\documentclass[journal,onecolumn]{IEEEtran}
\IEEEoverridecommandlockouts
\usepackage[utf8]{inputenc}
\usepackage{amsthm}
\usepackage{amsfonts,amssymb,latexsym,cite}
\usepackage[cmex10]{amsmath}
\usepackage{algorithmic}
\usepackage{array}
\usepackage{bbm}
\usepackage{verbatim}
\usepackage{color,xcolor}
\usepackage{graphicx}
\usepackage{epstopdf}
\usepackage{subcaption}
\usepackage[font={small,it}]{caption}
\usepackage[T1]{fontenc}
\usepackage{url}
\usepackage{enumerate}
\usepackage{hyperref} 
\graphicspath{{./figures/}} 
\usepackage{caption}
\usepackage[blocks]{authblk}
\usepackage{epsfig}
\usepackage{latexsym}
\usepackage{soul}
\usepackage{thmtools,thm-restate}
\usepackage[margin=1in]{geometry}

\newtheorem{definition}{Definition}
\newtheorem{theorem}{Theorem}

\newtheorem{lemma}{Lemma}
\newtheorem{claim}{Claim}

\newtheorem{remark}{Remark}
\allowdisplaybreaks
\newcommand{\upperRomannumeral}[1]{\uppercase\expandafter{\romannumeral#1}}

\newcommand\blfootnote[1]{%
  \begingroup
  \renewcommand\thefootnote{}\footnote{#1}%
  \addtocounter{footnote}{-1}%
  \endgroup
}

\newcommand{\C}{\mathcal{C}} 
\newcommand{\T}{T} 
\newcommand{\That}{\widehat{T}} 
\newcommand{\N}{N}
\newcommand{\PP}{\mathbb{P}}
\newcommand{\M}{M}
\newcommand{\D}{\Delta(n)}

\newcommand{\Mhat}{\widehat{M}}

\newcommand{\fa}{f^{xz}_{10}}
\newcommand{\fb}{f^{xz}_{11}}
 
\newcommand{\n}{n}

\newcommand{\wy}{W_{Y|X}}

\newcommand{\x}{\mathbf{x}}

\newcommand{\X}{\mathbf{X}} 
 
\newcommand{\y}{\mathbf{y}}
\newcommand{\Y}{\mathbf{Y}} 
\newcommand{\z}{\mathbf{z}}
\newcommand{\Z}{\mathbf{Z}}
\newcommand{\s}{\mathbf{s}}

\newcommand{\txz}{\mathcal{T}_{\X|\z}(f^{xz}_{10}, f^{xz}_{11})}
\newcommand{\axz}{\mathcal{A}_{\X|\z}}
\newcommand{\az}{\mathcal{A}_{\Z}}
\newcommand{\ax}{\mathcal{A}_{\X}}

\newcommand{\E}{\mathbb{E}}
\newcommand{\pzx}{W_{\mathbf{Z}|\mathbf{X}}}

\begin{document}
\title{Covert Communication over Adversarially Jammed Channels}

\author{
	Qiaosheng (Eric) Zhang\IEEEauthorrefmark{1}\thanks{
		\IEEEauthorrefmark{1}%
		Department of Electrical and Computer Engineering,
		National University of Singapore, elezqiao@nus.edu.sg}
	\and Mayank Bakshi\IEEEauthorrefmark{2}\thanks{
		\IEEEauthorrefmark{2}%
		Department of Information Engineering,
		The Chinese University of Hong Kong, jaggi@ie.cuhk.edu.hk, mayank@inc.cuhk.edu.hk}
	\and Sidharth Jaggi\IEEEauthorrefmark{2}
}

\date{}
\maketitle
\thispagestyle{empty}

\begin{abstract}
	
\textbf{Suppose that a transmitter Alice potentially wishes to communicate with a receiver Bob over an adversarially jammed binary channel. An active adversary James eavesdrops on their communication over a binary symmetric channel (BSC($q$)), and may maliciously flip (up to) a certain fraction $p$ of their transmitted bits based on his observations. We consider a setting where the communication must be simultaneously covert as well as reliable, i.e., James should be unable to accurately distinguish whether or not Alice is communicating, while Bob should be able to correctly recover Alice’s message with high probability regardless of the adversarial jamming strategy. We show that, unlike the setting with passive adversaries, covert communication against active adversaries requires Alice and Bob to have a shared key (of length at least $\Omega(\log n)$) even when Bob has a better channel than James. We present lower and upper bounds on the information-theoretically optimal throughput as a function of the channel parameters, the desired level of covertness, and the amount of shared key available. These bounds match for a wide range of parameters of interest. We also develop a computationally efficient coding scheme (based on concatenated codes) when the amount of shared key available is $\Omega(\sqrt{n} \log n)$, and further show that this scheme can be implemented with much less amount of shared key when the adversary is assumed to be computationally bounded.}

\end{abstract}

\blfootnote{A preliminary version of this work was presented in IEEE Information Theory Workshop (ITW), 2018.}

\setcounter{page}{1}

\section{Introduction} \label{sec:introduction}
The security of our communication schemes is of significant concern --- Big Brother is often watching! While much attention focuses on schemes that aim to hide the \emph{content} of communication, in many scenarios, the \emph{fact} of communication should also be kept secret. For example, a secret agent being caught communicating with an accomplice is of potentially drastic consequences ---  merely ensuring secrecy does not guarantee undetectability. This observation has drawn attention to the problem of \emph{covert communication}. In a canonical information-theoretic setting for this problem, a transmitter Alice {\it may} wish to transmit messages to a receiver Bob over a noisy channel, and remains \emph{silent} otherwise. James  eavesdrops on her transmission through another noisy channel. The communication goals are twofold. Firstly,  the communication should be \emph{covert}, i.e., James should be unable to reliably distinguish whether or not Alice is transmitting.  Simultaneously, it should also be \emph{reliable}, i.e., Bob should be able to correctly estimate Alice's transmission with a high probability of success. Recent literature~\cite{BasGT:12a,bash2015quantum,7407378,tahmasbi2018first,7447769,CheBJ:13} has quite successfully characterized the information-theoretic aspects of this problem, in terms of characterizing the fundamental limits on the total amount of covert communication possible from Alice to Bob. Specifically, it turns out that no more than $c_{p,q}\sqrt{n}$ bits may be covertly transmitted from Alice to Bob over $n$ channel uses, where $c_{p,q}$ is an explicitly characterizable constant depending on the channels from Alice to Bob and James. This sub-linear throughput (as opposed to the linear throughput in most communication settings) results from the stringent requirement on Alice's transmissions imposed by the need to remain covert --- she must ``whisper'', so to speak (pun intended). Indeed, most of her transmitted codewords must have low Hamming weight (${\cal O}(\sqrt{n})$).

Prior information-theoretic work on covert communication largely focuses on random  noise to both Bob and James. While such channel models are appropriate for passive eavesdroppers, a truly malicious adversary might wish to also actively disrupt any potential communication even when he is unable to detect if transmission has indeed taken place. To model this scenario, in this work, we take a somewhat {\it coding-theoretic} view --- we let the channel from Alice to James be probabilistic, but we allow James to try to {\it jam} the channel to Bob adversarially, as a function of his noisy observations of Alice's potential transmissions.

Semi-formally, in our setting, Alice's channel input is an $n$-length binary vector $\X$. The channel from Alice to James is a binary symmetric channel with transition probability $q$ (i.e., BSC($q$)). James uses his observation $\Z$ in two ways --- to detect if communication is being attempted via an estimator $\Phi$, and to choose a binary jamming vector $\textbf{S}$ of Hamming weight at most  $pn$ --- Bob receives the vector $\Y=\X\oplus\textbf{S}$. We denote the channel from Alice to Bob as ADVC($p\vert q$).  When Alice is silent, $\X$ {\it must} be the all-zeros vector $\mathbf{0}$; when Alice is \emph{active}, the $\X$ she transmits may be a function of the message she wishes to transmit. Alice and Bob's encoding/decoding procedures are known to all parties.  We measure covertness via a {\it hypothesis-testing metric} --- we say that the communication is $(1-\epsilon_d)$-covert if irrespective of James' estimator $\Phi$, his probability of {\it false alarm} plus his probability of {\it missed detection} is always lower-bounded by $1-\epsilon_d$. Secondly, we require reliability --- Bob should be able to reconstruct Alice's transmission with high probability (w.h.p.) regardless of James' jamming strategy. 

Unfortunately, in our setting this turns out to be impossible --- it turns out (as we show in our first main result) that the noise $\textbf{S}$ on Bob's channel is adversarially chosen (rather than randomly as in the classical setting, e.g.~\cite{BasGT:12a,7407378,tahmasbi2018first,7447769}) implies James can ensure {\it any} such communication protocol must be either non-covert or unreliable. This is true even if James has computational restrictions, or is required to behave causally~\cite{chen2015characterization}. This is in stark contrast to the probabilistic channel setting wherein covert communication is possible for a wide range of parameters. 
Hence, we mildly relax our problem --- prior to transmission, we allow Alice and Bob to secretly share a $\Delta(n)$-bit  randomly generated \emph{shared key} that is unknown to James.

\subsection{Main contribution}
The main contributions of this paper can be summarized as follows.
\begin{enumerate}
\item We show in Theorem~\ref{thm:converse1} that to ensure the communication to be reliable and covert simultaneously, the size of the shared key $\Delta(n)$ should be at least $\frac{1}{2}\log(n)$.\footnote{All logarithms in this paper are binary.} 

\item On the other hand, Theorem~\ref{thm:upperbound} provides an information-theoretic upper bound on the throughput which holds regardless of the amount of shared key available --- in particular, the throughput is restricted to zero when the adversarial channel model ADVC($p\vert q$) satisfies $p \ge q$. 

\item We then provide an achievability scheme for ADVC($p\vert q$) with $p < q$ and different values of $\Delta(n)$ in Theorem~\ref{thm:achievable}.  
\begin{itemize}
	\item For a wide range of $(p,q)$-regime, a modest value of $\Delta(n) = 6\log(n)$ suffices\footnote{Using a finer analytical technique provided by a recent work~\cite{Shared2019}, one may show that even $\Delta(n) = (2.5+\delta)\log(n)$ (where $\delta >0$ can be made arbitrarily small) suffices.} for our achievability scheme to match the upper bound derived in Theorem~\ref{thm:upperbound}. Hence, the amount of shared key $\Delta(n) \in {\cal O}(\log(n))$ required to initiate reliable and covert communication scales much more gracefully than the amount of communication ${\cal O}(\sqrt{n})$ thereby instantiated. 
	
	\item For the remaining $(p,q)$-regime in which our scheme does not match the upper bound with $\Delta(n) = 6\log(n)$, a larger amount of shared key in general yields a better achievability bound when the shared key is in the moderate-sized regime $\Delta(n) \in (\Omega(\log(n),{\cal O}(\sqrt{n}))$, but further increasing $\Delta(n)$ leads to diminishing returns in the large-sized regime $\Delta(n) \in \omega(\sqrt{n})$. 
	
	\item When $\Delta(n) \in \omega(\sqrt{n})$, our achievability scheme matches the upper bound for all the values of $p < q$. This leads to a full information-theoretic characterization of the optimal throughput.
	
	\item Our achievability schemes make no computational or causality assumptions on James.

\end{itemize}

\item When the shared key $\Delta(n) \in \Theta(\sqrt{n}\log(n))$, in Theorem~\ref{thm:efficient} we demonstrate a {\it computationally efficient} communication scheme with polynomial encoding and decoding complexities. It achieves within a constant factor of the information-theoretically optimal throughput, and also makes no computational or causality assumptions on James. If the computational complexity of James is further restricted to polynomial time, then the scheme can be implemented with a much smaller amount of shared key, as shown in Theorem~\ref{cor:prg}.

\end{enumerate}

\subsection{Related Work \& Comparisons}
\noindent{\bf Covert Communication:} Bash et al.~\cite{BasGT:12a} were the first to study covert communication for additive white Gaussian noise (AWGN) channels in an information-theoretic setting and demonstrate a {\it square-root law} --- communication that is simultaneously covert and reliable is possible when the message length is $\mathcal{O}(\sqrt{n})$ bits and shared key is available. Subsequently, Che et al. showed that for BSCs, as long as James has a noisier channel than Bob, no shared key is necessary~\cite{CheBJ:13,CheBCJ:14a,CheBCJ:14b,CheSBCJA:14}. Bloch et al.~\cite{7407378,bloch2017optimal} and Wang et al.~\cite{7447769} then derived tight capacity characterizations for general discrete memoryless channels (DMCs) and AWGN channels. The work in~\cite{7407378} also showed that the amount of shared key needed when Bob has a noisier channel than James is ${\cal O}(\sqrt{n})$. While prior work on covert communication focuses on random noise channels (e.g., BSCs, AWGNs, and DMCs), to the best of our knowledge, our work is the first to examine covert communication over adversarial channels. 

\noindent{\bf Random noise vs adversarial noise channels:} In the non-covert setting, much work has focused on two classes of noisy channels --- {\it random noise} channels and {\it adversarial noise} channels. 
 
The capacities of random noise channels have been fully characterized by Shannon in his seminal work~\cite{shannon2001mathematical}. On the contrary, though many upper and lower bounds (sometimes but not always matching) for a variety of special adversarial jamming models, a tight capacity characterization for general adversarial channels (also called Arbitrarily Varying Channels (AVCs) in the information theory literature --- see~\cite{lapidoth1998reliable} for an excellent survey) is still elusive. One way to classify adversarial models is via the adversary's knowledge level of the transmitted codeword $\X$.
Models of interest include {\it classical/omniscient} adversarial model~\cite{gilbert1952comparison, varshamov1957estimate, mceliece1977new} (full knowledge of $\X$), the {\it myopic} adversarial model~\cite{sarwate_avc_2012, sarwate_coding_2010, dey2015sufficiently,zhang2018quadratically} (noisy observations of $\X$), the {\it oblivious} adversarial model~\cite{lapidoth1998reliable, langberg2008oblivious, guruswami2010codes} (no knowledge of $\X$) and the {\it causal} adversarial model~\cite{chen2015characterization, dey2016bit, dey_improved_2012} (causal observations of $\X$). Also, the {\it computationally bounded} adversary model~\cite{gopalanerror, micali2005optimal} considers models wherein Alice/Bob/James are all computationally bounded. 

\noindent {\bf Arbitrarily Varying Channels (AVCs):} At a high level, reliable communication in the model considered in this work is closely related that of communication over an AVC~\cite{blackwell_capacities_1960, csiszar_capacity_1988, lapidoth1998reliable} with stringent input constraints. Indeed, the impossibility result we present in Theorem~\ref{thm:converse1} is motivated by the {\it symmetrizability} condition for AVCs. 
\begin{enumerate}
\item {\it Myopic adversaries with shared key:} These are AVC problems first explicitly considered by Sarwate~\cite{sarwate_coding_2010} wherein James only observes a noisy version $\Z$ of $\X$ (for instance through a BSC$(q)$) before deciding on his jamming vector $\textbf{S}$. Sarwate~\cite{sarwate_coding_2010} provided a tight characterization of the throughput in such settings over general DMCs in the presence of an unlimited-sized shared key --- as such, the model therein has strong connections to the problem we consider. Indeed, the converse we present in Theorem~\ref{thm:upperbound} relies heavily on the information-theoretic framework for impossibility results in AVCs in general and~\cite{sarwate_coding_2010} in particular. 
\item {\it Myopic adversaries without shared key:} Problems concerning myopic adversaries {\it without} shared key between Alice and Bob~\cite{dey2015sufficiently} are considerably more challenging than when shared key is available. However, if the adversary is {\it sufficiently myopic}, i.e., the noise $q$ on the BSC($q$) to James is strictly larger than the fraction $p$ of bit flips he can impose on the channel to Bob, and there are no constraints on Alice's transmissions, the capacity of such a channel has been shown to exactly equal that of a BSC$(p)$. 
\end{enumerate}
Despite similarities, the main focus of most work in the AVC literature differs from this work in the following aspects: (i) an unlimited-sized shared key between Alice and Bob is assumed, as opposed to the careful classification of achievabilities/converses obtained in our work pertaining to differing-sized shared keys; (ii) covertness is not considered as in this work; (iii) the stringent requirements in channel inputs enforced due to covertness imply that some of the analytical techniques used in the AVC literature do not translate to our setting; and (iv) no effort is made to consider computational restrictions on Alice/Bob/James, unlike in our work.\\
\noindent {\bf List decoding:} One of the primitives our achievability schemes rely heavily on is that of {\it list decoding}~\cite{elias_list_1957, guruswami2004list, sarwate_list-decoding_2012}. Results in this subset of the literature guarantee that even in the presence of omniscient adversaries, Bob is able to localize Alice's transmission to a small (often constant-sized) list at a communication rate approaching that of a corresponding random noise channel. However, we note that the ``usual'' list decoding model does not translate to our setting due to the severity of the constraint on Alice's transmissions imposed by covertness. Hence in our work we prove a novel version of list decoding for such input-constrained channels, in which we rely heavily on James' myopicity.

\noindent {\bf Usage of shared key:} One pathway to achievability schemes for AVCs (e.g.~\cite{langberg2004private}) is to ensure that Bob can list-decode to a small list, and then to use the key shared with Alice to disambiguate this list down to a unique message. There are multiple such schemes in the literature, including computationally efficient schemes~\cite{cramer2008detection}.

\noindent {\bf Permutation-based coding:} Another idea in the literature that has borne multiple dividends (e.g.~\cite{ahlswede_elimination_1978-1, langberg2004private}) in the context of code design for AVCs (especially computationally efficient codes, e.g.~\cite{lipton1994new, guruswami2010codes}) and even in covert communication from a source-resolvability perspective~\cite{bash2015hiding, 7407378} is that of {\it permutation-based coding}. Alice and Bob generate a small (polynomial-size) set $\Pi$ (known also to James) of randomly sampled permutations as part of code-design, and then use their shared key to pick a particular permutation $\pi$ that is unknown to James. Alice then transmits the codeword $\pi(\X)$, and Bob attempts to decode $\pi^{-1}(\Y)$. In several problems it can be shown that the effect of this permutation $\pi$ is to ``scramble'' James' jamming action, and hence makes him behave essentially like i.i.d. noise. In our work we show that similar ideas work even in the presence of a myopic and computationally unbounded jammer James, and results in a computationally efficient communication scheme for Alice and Bob.

\subsection{Organization}
The rest of this paper is organized as follows. We formally introduce the problem setup in Section~\ref{sec:model}. Section~\ref{sec:result} presents the main results (Theorems~\ref{thm:converse1}-\ref{cor:prg}) as well as some key ideas. The detailed proofs of Theorem 1 (the converse result on the shared key) and Theorem 2 (the upper bound on the covert capacity) are respectively provided in Sections~\ref{sec:converse} and~\ref{sec:upperbound}. In Section~\ref{sec:achievability}, we introduce and analyze a coding scheme that leads to the lower bound on the covert capacity (Theorem~\ref{thm:achievable}). Finally, Section~\ref{sec:efficient} provides a detailed description of our computationally efficient coding scheme which is briefly introduced in Theorem~\ref{thm:efficient}. In Section~\ref{sec:conclusion}, we conclude this work and propose several directions that are worthy exploring for future work.

\section{Model}\label{sec:model}
Random variables are denoted by uppercase letters, e.g., $X$, and their realizations are denoted by lowercase letters, e.g., $x$. Sets are denoted by calligraphic letters, e.g., $\mathcal{X}$. Vectors of length $n$ are denoted by boldface letters, e.g., $\X$ and $\x$. The $i$-th locations of $\X$ and $\x$ are denoted by $X_i$ and $x_i$ respectively. The $Q$-function takes the form 
\begin{align}
Q(x) = \frac{1}{2\pi} \int_{x}^{\infty}\exp\left(-\frac{u^2}{2}\right)du.
\end{align}

\noindent{\bf Encoder:} Let $n$ denote the blocklength (number of channel uses) of Alice's communication. Alice's {\it encoder} $\Psi(.,.,.)$ takes three inputs\footnote{In some scenarios in the literature, in addition to the three inputs below, the encoder also incorporates additional private randomness (known {\it a priori} only to Alice, but not to Bob or James). Indeed, in some communication scenarios~\cite{dey2016bit} it can be shown that the throughput in the presence of such private randomness is strictly higher than in its absence. However, since in this work such types of encoders do not help, we ignore this potential flexibility in code design.}:
(i) the single bit {\it transmission status} $T$: Alice's silence is denoted by $T=0$ whereas $T=1$ denotes that she is {\it active}.\footnote{Note that no assumptions are made about any probability distribution on $T$.} 
(ii) the {\it message} $M$, which is either $0$ (if Alice is silent), or uniformly distributed over $\{1,2,\ldots,N\}$ (if Alice is active).
(iii) the $\Delta(n)$-bit {\it shared key} $K$ distributed uniformly over $\{0,1\}^{\Delta(n)}$. 
Prior to transmission, only Alice knows the transmission status $T$ and message $M$, and both Alice and Bob know the key $K$ --- James is {\it a priori} ignorant of all three.

If $T=0$, then Alice's encoder $\Psi(0,.,.)$ {\it must} output $\X = {\mathbf 0}$, a length-$n$ vector comprising entirely of zeros.
On the other hand if $T = 1$, then Alice's encoder $\Psi(1,.,.)$ may output an arbitrary length-$n$ binary vector $\X$. The collection of all outputs of Alice’s encoder $\Psi(1,.,.)$ is called the {\it codebook}, denoted by $\C$. This encoder is known {\it a priori} to all parties (Alice, Bob, and James).
The {\it relative throughput} of the code is defined as $r \triangleq  (\log N )/ \sqrt{n}$. 

\begin{figure}
	\begin{center}
		\includegraphics[scale=0.55]{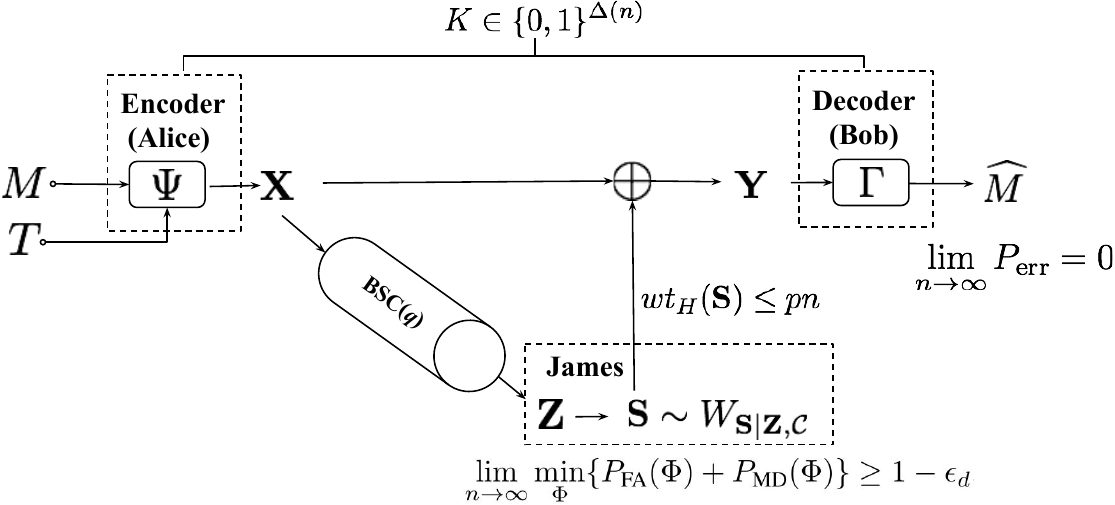}
		\caption{System diagram.} \label{fig:system}
	\end{center}
\end{figure}

\noindent {\bf James' observations:}
James receives the vector $\Z = \X \oplus \bar{\mathbf{S}}$, where $\bar{S}_i$ is a Bern$(q)$ random variable. Hence James' observed vector $\Z$ is the output of a BSC$(q)$ channel to which the input is Alice's transmission $\X$. 
On the basis of this observation $\Z$ and his knowledge of Alice's encoder $\Psi(.,.,.)$, James, as described below: (i) estimates Alice's transmission status $T$, and (ii) generates a jamming vector $\mathbf{S}$ to disrupt communication.

\noindent{\textbf{Estimator:}} James' {\it estimator} $\Phi(.): \left\{0, 1\right\}^{\n} \rightarrow \left\{0, 1\right\}$ estimates Alice's transmission status $T$ as $\That = \Phi(\Z)$. We use a hypothesis-testing metric (defined below) to measure covertness: 
\begin{definition}[\textbf{Covertness}]\label{def:covert}
Let $P_{\emph{FA}}(\Phi) \triangleq \PP_{K,\bar{\mathbf{S}}}(\That = 1| \T = 0)$ and $P_{\emph{MD}}(\Phi) \triangleq \PP_{M,K,\bar{\mathbf{S}}}(\That = 0| \T = 1)$ respectively be the probability of false alarm and the probability of missed detection of an estimator $\Phi$. The communication is said to be ($1-\epsilon_d$)-covert if\footnote{Note that even if James ignores the knowledge of $\Z$, a na\"ive estimator $\tilde{\Phi}$ (which always outputs $\widehat{\T} = 0$ or $\widehat{\T}=1$) also guarantees $P_{\text{FA}}(\tilde{\Phi}) + P_{\text{MD}}(\tilde{\Phi}) = 1$. Therefore, Definition~\ref{def:covert} implies that James' optimal estimator $\Phi^\ast$ cannot be much better than the na\"ive estimator $\widehat{\Phi}$. }
\begin{align}
\liminf_{n \to \infty} \min_{\Phi} \{ P_{\emph{FA}}(\Phi) + P_{\emph{MD}}(\Phi) \} \ge 1 - \epsilon_d,
\end{align}
where $\Phi$ is minimized over all possible estimators.
\end{definition}
For the optimal estimator $\Phi^\ast, P_{\text{FA}}(\Phi^\ast) + P_{\text{MD}}(\Phi^\ast) = 1 - \mathbb{V}(Q_0(\Z),Q_1(\Z))$, where $\mathbb{V}(Q_0(\Z),Q_1(\Z))$ is the variational distance between the two distributions (corresponding to $T=0$ and $T=1$, respectively) on James' observation $\Z$. In general the computational complexity of implementing the optimal estimator $\Phi^\ast$ is high (potentially $\exp(n)$); also, analyzing its performance can also be tricky. 


\noindent{\textbf{Jamming function:}} As a function of his observation $\Z$ and his knowledge of Alice's encoding function $\Psi(.,.,.)$ James chooses a {\it jamming function} to output a length-$n$ binary {\it jamming vector} $\mathbf{S}$ of Hamming weight at most $pn$. In general James' jamming function corresponds to a conditional probability distribution $W_{\mathbf{S}|\Z,\C}$ that stochastically maps James' observations to his jamming vector $\mathbf{S}$. Note that $W_{\mathbf{S}|\Z,\C}$ generates an $n$-letter distribution over length-$n$ binary sequences $\mathbf{S}$, given James’ length-$n$ observation $\Z$, and his knowledge of Alice and Bob's code $\C$.


\noindent{\textbf{Decoder:}}
Bob receives the length-$\n$ binary vector $\Y = \X \oplus \textbf{S}$, and then applies his {\it decoding function} $\Gamma(.,.): \{0, 1\}^n \times \{0, 1\}^{\Delta(n)} \rightarrow \{0\} \cup \{1, 2, \ldots , N\}$ to produce his {\it message reconstruction} $\widehat{M}$ from his observed vector $\Y$ and the shared key $K$.

\noindent{\textbf{Probability of decoding error:}} Bob's probability of error is defined as\footnote{The two terms correspond to Bob's decoder making an error in each of two scenarios: when Alice is silent, and when she is active. 
} 
\begin{align}
P_{\text{err}} \triangleq \max_{W_{\mathbf{S}|\Z,\C}} \left(
\PP_{K,\bar{\mathbf{S}},\mathbf{S}} (\widehat{M} \neq 0 | T = 0) +
\PP_{M,K,\bar{\mathbf{S}},\mathbf{S}} (\widehat{M} \neq M | T = 1) 
\right).  \label{eq:pe}
\end{align}
Note that the probability as defined in~\eqref{eq:pe} is maximized over the $n$-letter distribution $W_{\mathbf{S}|\Z,\C}$. This is to indicate that there may (or may not) be a stochastic component to the jamming function James uses to generate $\mathbf{S}$ from his observation $\Z$. Hence we include an averaging over $\mathbf{S}$.

\noindent{\textbf{Achievable relative throughput/covert capacity:}} For any $p,q \in (0,\frac{1}{2})$, $\Delta(n)\geq 0$, and $\epsilon_d \in (0,1)$, a relative throughput $r_{\Delta(n),\epsilon_d}(p,q)$ is said to be {\it achievable} if there exists an infinite sequence of codes with $\Delta(n)$ bits of shared key such that each of the codes in the sequence has relative throughput at least $r_{\Delta(n),\epsilon_d}(p,q)$, $\limsup_{n\to \infty }P_{\text{err}} = 0$, and ensures the communication is $(1-\epsilon_d)$-covert. 
Then the {\it covert capacity}\footnote{Note that the covert capacity defined here depends on the amount of shared key available.} $r^{\ast}_{\Delta(n),\epsilon_d}(p,q)$ is defined as the supremum over all possible achievable relative throughputs.

\noindent{{\bf Positive throughput region:}} For any $\Delta(n)$ and $\epsilon_d \in (0,1)$, the {\it positive throughput region} $\mathcal{R}^{+}_{\Delta(n),\epsilon_d}(p,q)$ is defined as a collection of values $(p,q)$ such that the covert capacity $r^{\ast}_{\Delta(n),\epsilon}(p,q)$ is positive.

\section{Main Results} \label{sec:result}
We now summarize the main contributions of this work. There are at least two types of estimators and jamming functions James can use, each of which results in a non-trivial restriction on the reliable and covert throughput obtainable from Alice to Bob. Perhaps surprisingly, there is a unified achievability scheme that Alice and Bob can use that meets these constraints for a wide range of parameters of interest, and thereby shows that these types of estimators/jamming functions are in some sense optimal from James' perspective.
\begin{itemize}
\item \textbf{Weight-detector:} This estimator (with computational complexity ${\cal O}(n)$) merely computes the Hamming weight of the observed $\Z$, and if this is significantly higher than expected ($qn+c_t\sqrt{n}$ for some constant $c_t$), then James estimates\footnote{Even though this estimator is a sub-optimal proxy to the Hypothesis-testing estimator, it has been shown in~\cite{CheBJ:13,7407378} to be ``good enough'' from James' perspective, in the sense that it constrains Alice's throughput to the same extent as does the Hypothesis-testing estimator, which is known~\cite{neyman1992problem} to be optimal.} $\widehat{T}=1$.

\item \textbf{Hypothesis-testing estimator:} James first computes two distributions $Q_0(\Z)$ and $Q_1(\Z)$, which respectively correspond to the distributions of $\Z$ when $T = 0$ and $T = 1$. This (optimal) estimator $\Phi^\ast$ outputs $\widehat{T}=1$ if $Q_1(\z) \ge Q_0(\z)$, and outputs $\widehat{T}=0$ if $Q_0(\z) > Q_1(\z)$. Note that this estimator potentially has computational complexity $\exp(n)$ for James.

\item \textbf{Oblivious jamming:} This jamming strategy ignores James' channel observations $\Z$, and chooses $\mathbf{S}$ as a binary addition of multiple (at most ${\cal O}(\sqrt{n})$) codewords from the codebook. Since Bob's observation is a sum of Alice's transmission $\X$ and James' jamming vector $\mathbf{S}$, this jamming strategy
	attempts to confuse Bob as to what Alice truly transmitted. Note that this jamming strategy can be implemented by James causally, with computational complexity at most $\sqrt{n}$ times the computational complexity of Alice's encoder. The converse with respect to this oblivous jamming strategy is presented in Theorem~\ref{thm:converse1}.
		
\item \textbf{Myopic jamming:} Even if Alice's transmission is covert (hence James is unsure whether or not Alice is transmitting), James can nonetheless use his observations $\Z$ to guess which channel uses correspond to potential $1$’s in Alice’s transmissions, and then preferentially flip these bits that are likeliest to be $1$. Specifically, if $Z_i =1$ then he flips the corresponding $X_i$ with probability about $p/q$, but if $Z_i = 0$ he does not flip $X_i$. Hence, James concentrates his bit-flip power in bits that are likelier to correspond to the actual transmissions from Alice. Note that this jamming strategy can be implemented by James causally, with computational complexity linear in $n$. The converse with respect to this myopic jamming strategy is presented in Theorem~\ref{thm:upperbound}.
\end{itemize}

For any channel parameters $q \in (0,\frac{1}{2})$ and covertness parameter $\epsilon_d \in (0,1)$, we first define the \emph{code weight parameter} $t(q,\epsilon_d)$ as 
\begin{align}
t(q,\epsilon_d)  \triangleq \frac{2 \sqrt{q (1-q)}}{1-2q}\cdot Q^{-1}\left(\frac{1-\epsilon_d}{2}\right). \label{eq:q}
\end{align}
The parameter $t(q,\epsilon_d)$ is independent of the blocklength $n$, and roughly speaking, ``most'' codewords have Hamming weight about $t(q,\epsilon_d)\sqrt{n}$. Following the techniques in~\cite{tahmasbi2017second}, the average Hamming weight $t(q,\epsilon_d)\sqrt{n}$ has been optimized to be as large as possible while still ensuring $(1-\epsilon_d)$-covertness. 

When the average Hamming weight of the channel inputs is $t(q,\epsilon_d)\sqrt{n}$, we then define the \emph{weight normalized mutual information} for Bob and James as follows. 
\begin{definition}[Weight normalized mutual information] For any $p,q \in (0,\frac{1}{2})$ such that $p \le q$, the weight normalized mutual information for Bob and James are respectively defined as
\begin{align}
&I_B(p,q) \triangleq \frac{p(q-1)}{q}\log \left(\frac{(q-p+pq)(1-p)}{p^2(1-q)}\right)  +\log\left(\frac{q-p+pq}{pq}\right),  \mathrm{ and} \\
&I_J(q) \triangleq (1-2q)\log \left(\frac{1-q}{q}\right),
\end{align}
\end{definition}
The quantity $I_J(q)t(q,\epsilon_d)\sqrt{n}$ denotes the mutual information corresponding to the BSC$(q)$ from Alice to James, derived by taking the appropriate Taylor series expansion of $I(\X,\Z)$. The quantity $I_B(p,q)t(q,\epsilon_d)\sqrt{n}$ denotes the mutual information of the worst i.i.d. channel from Alice to Bob, induced by an i.i.d. myopic jamming strategy employed by James (i.e., flipping $X_i$ with probability approximately $p/q$ only within the support of $\Z$). 

While the mutual information from Alice to Bob $I_B(p,q)t(q,\epsilon_d)\sqrt{n}$ (in the presence of such an i.i.d. myopic jamming strategy) clearly serves as an upper bound on Alice's achievable throughput, it is perhaps more surprising that this is also achievable by our codes in a wide range of parameter regimes (corresponding to the {\it achievable positive throughput region} presented in Theorem~\ref{thm:achievable} below).

\subsection{Impossibility of covert communication with $\Delta(n) < \frac{1}{2}\log (n)$}

When the amount of shared key is less than $\frac{1}{2}\log{n}$, if James employs a weight-detector with an appropriate threshold, combined with an oblivious jamming strategy, it turns out that he can ensure that the probability of decoding error is bounded away from zero. Roughly speaking, since Alice's codebook comprises mostly of low-weight codewords, James is able to confuse Bob by choosing a jamming vector that comprises of the binary addition of multiple potential codewords -- ``spoofs'' -- Bob is unable to disambiguate Alice's true $\X$ from among the cacophony of spoofs. The following theorem makes the above claim precise, and the proof of Theorem~\ref{thm:converse1} can be found in Section~\ref{sec:converse}.
\begin{theorem} \label{thm:converse1}
Let $\epsilon_d\in(0,1)$ and $\Delta(n)<\frac{1}{2}\log(n)$. For every sequence of codes $\{\C_n\}$ of blocklength $n$, message length $\log{\N} = nr$, and encoding complexity $f_{\C}(n)$, at least one of the following is true:
\begin{enumerate}
\item ($\C_n$ is not covert) There exists a detector $\Phi$ with computational complexity $\mathcal{O}(n)$ such that  $P_{\emph{FA}}(\Phi)+P_{\emph{MD}}(\Phi)<1-\epsilon_d$. In particular, $\Phi$ can be chosen to be the weight-detector $\Phi_{\rho}$ for an appropriately set threshold $\rho$.
\item ($\C_n$ is not reliable) There exists a constant $\eta=\eta(\epsilon_d,p,q)$ and causal jamming strategy $W_{\mathbf{S}|\Z,\C}$ with  computational complexity $\mathcal{O}(\sqrt{n} f_{\C}(n))$, such that the probability of error is bounded from below as \begin{equation} 
P_{\emph{err}} \ge \PP_{\M,K,\bar{\mathbf{S}},\mathbf{S}}(\M\neq\Mhat |T=1)\geq 1-\max\left\{\frac{2}{\N},\frac{2^{\Delta(n)}\eta}{\sqrt{n}}\right\}.\label{eq:errorsmallshared}\end{equation}
In particular, $W_{\mathbf{S}|\Z,\C}$ may be chosen as the oblivious jamming strategy $W^{\emph{(ob)}}_{\mathbf{S}|\Z,\C}$.
\end{enumerate}
\end{theorem}
\begin{remark} (a) The lower bound on the probability of error in~\eqref{eq:errorsmallshared} is valid for all values of $\Delta(n)$. However, it is non-vanishing only if $\Delta(n)<\frac{1}{2}\log(n)$.	
(b) The converse result in Theorem~\ref{thm:converse1} is also valid when all the parties (Alice, Bob, James) are assumed to be computationally bounded (i.e., the computational power is restricted to be polynomial in the blocklength $n$). This is because both  the weight-detector and oblivious jamming strategy $W^{\text{(ob)}}_{\mathbf{S}|\mathbf{Z},\C}$ can be employed by James efficiently.
\end{remark}

	
\begin{table}[]
\scriptsize
\centering
\caption{Summary of Main results}
\begin{tabular}{|l|l|l|l|l|l|}
\hline
Theorem               & length of $\D$ & Relative throughput $r$ & \begin{tabular}[c]{@{}l@{}}Enc/Dec\\ Complexity\end{tabular} & \begin{tabular}[c]{@{}l@{}}Complexity of\\ adversary's attack\end{tabular} \\ \hline
Thm~\ref{thm:converse1} (Converse)      & less than $\frac{1}{2}\log (n) $  & $0$      & N/A     &  $\mathcal{O}(\sqrt{n}f_{\C}(n))$, causal     \\ \hline
Thm~\ref{thm:upperbound} (Converse)  & arbitrary &  $t(q,\epsilon_d)I_B(p,q)$  & N/A    & $\mathcal{O}(n)$, causal     \\ \hline
Thm~\ref{thm:achievable} (Achievability) & arbitrary     &   $t(q,\epsilon_d)I_B(p,q)$    & $2^{\mathcal{O}(\sqrt{n})}$  &   arbitrary  \\ \hline
Thm~\ref{thm:efficient} (Achievability) & $\mathcal{O}(\sqrt{n}\log (n))$   & \begin{tabular}[c]{@{}l@{}}$\frac{t(q,\epsilon_d)}{\rho^\ast} C_{\text{BAC}}(p,q)$, where $\rho^{\ast}$, $C_{\text{BAC}}(p,q)$ are defined in Section III-E  \end{tabular}   &  poly$(n)$     &  arbitrary     \\ \hline

Thm~\ref{cor:prg} (Achievability) & $n^{\xi}$ for any $\xi > 0$   & \begin{tabular}[c]{@{}l@{}}$\frac{t(q,\epsilon_d)}{\rho^\ast} C_{\text{BAC}}(p,q)$  \end{tabular}   &  poly$(n)$     &  poly$(n)$     \\ \hline

\end{tabular}
\end{table}

\subsection{An upper bound on the covert capacity for any $\Delta(n)$}
\begin{figure}
	\begin{center}
		\includegraphics[scale=0.3]{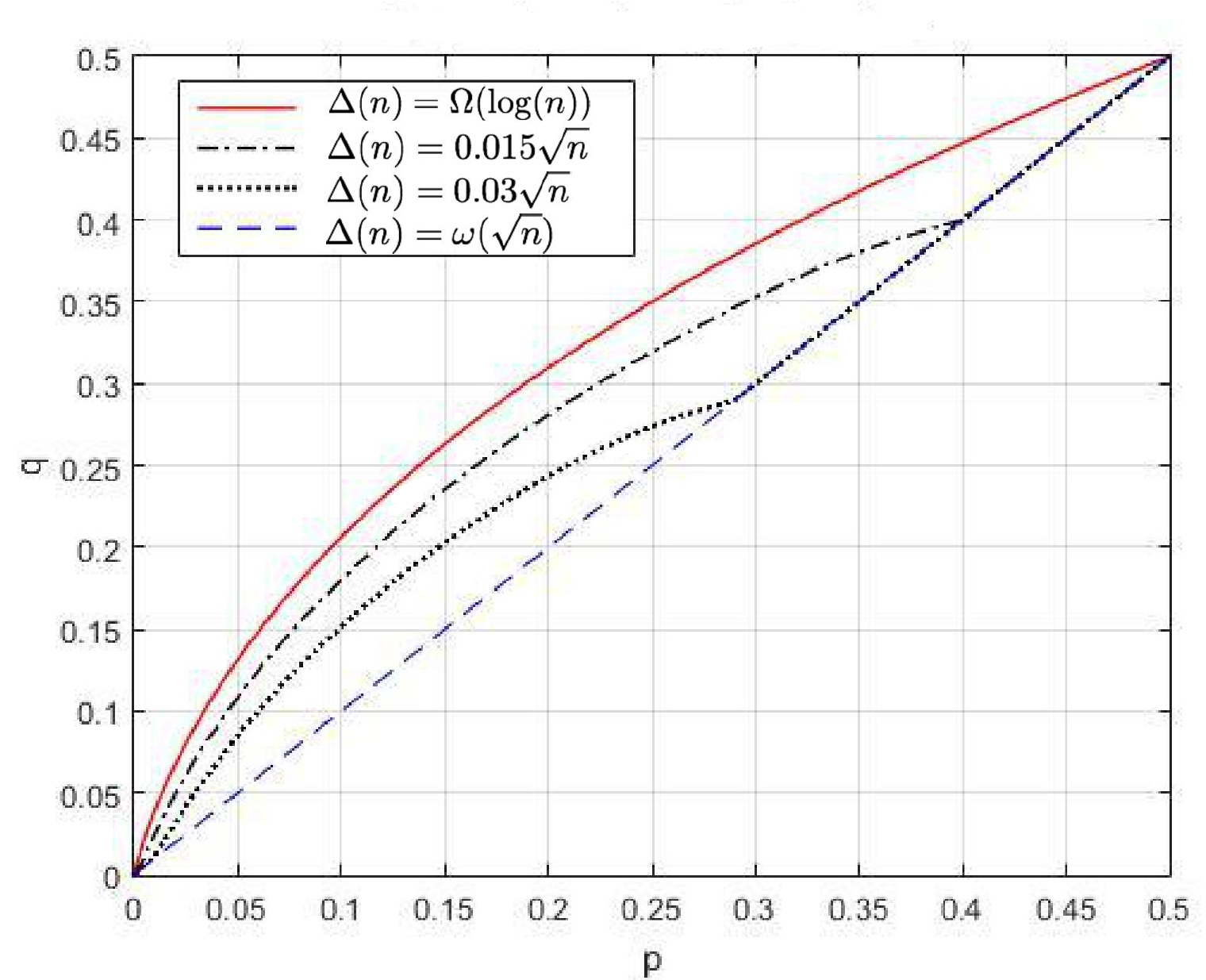}
		\caption{This figure shows the achievable positive throughput regions for different values of $\Delta(n)$. (1) For any $\Delta(n) \in (\Omega(\log(n)),o(\sqrt{n}))$, as shown via Theorem~\ref{thm:achievable}, covert communication is possible above the red curve. (2) When $\Delta(n) \in \Omega(\sqrt{n})$, as shown via Theorem~\ref{thm:achievable}, the achievable positive throughput region increases. The two black curves delineate the achievable positive throughput regions for $\Delta(n) = 0.015\sqrt{n}$ and $0.03 \sqrt{n}$ respectively. The achievable positive throughput regions for each corresponding $\Delta(n)$ are now above the respective black curves. (3) Regardless of the value of shared key $\Delta(n)$, no covert communication is possible below the blue dashed line corresponding to $p=q$. This is in contrast to ``classical'' covert communication~\cite{7407378} in the presence of a passive adversary (rather than an actively jamming adversary), wherein increasing amounts of shared key allow for covert communication even when $p>q$, i.e., the channel from Alice to Bob has more bit-flips than the channel from Alice to James. This is due to the fact that when $p>q$, among the classes of channels James can induce from Alice to Bob is one which has zero channel capacity even if $p<\frac{1}{2}$.} \label{fig:regime}
	\end{center}
\end{figure}

Next, we obtain an upper bound on the covert capacity that holds regardless of the amount of shared key available. Our strategy here is to bound the throughput of any simultaneously covert and reliable code by first showing that the average Hamming weight of codewords from such a code must be bounded from above by an appropriate function of the covertness parameter. Next, since the transmitted message must also be reliably decoded under all jamming strategies, this gives an upper bound on the number of distinct messages possible. To bound the number of codewords, we analyze Bob's reliability with respect to the mutual information $t(q,\epsilon_d)I_B(p,q)\sqrt{n}$ of the channel induced under James' myopic jamming strategy $W_{\mathbf{S}|\Z,\C}^{\text{(my)}}$. The proof of Theorem~\ref{thm:upperbound} can be found in Section~\ref{sec:upperbound}.
\begin{theorem}\label{thm:upperbound}
Let $\epsilon_d\in(0,1)$ and $p,q \in (0,\frac{1}{2})$. For every sequence $\{\Delta(n)\}$,
\begin{enumerate}
	\item if $q \le p$, then $r^{\ast}_{\Delta(n),\epsilon_d}(p,q) = 0$ (corresponds to the region below the blue dashed line in Fig.~\ref{fig:regime});
	\item if $p < q$, then $r^{\ast}_{\Delta(n),\epsilon_d}(p,q) \leq  t (q,\epsilon_d) I_B(p,q)$ (corresponds to the region above the blue dashed line in Fig.~\ref{fig:regime}). 
\end{enumerate}
\end{theorem}

\subsection{Achievability of covert communication with $\Delta(n)\ge 6\log(n)$}
Next, we give an achievability result based on low-weight random codes and list decoding. The crux of our proof is a novel {\it myopic list-decoding lemma} described in the introduction, and formally presented in Claim~\ref{claim:ratio2} in Section~\ref{sec:reliability}. This lemma first demonstrates that for the parameter regime under consideration, with high probability over the noise of the BSC($q$) from Alice to James, from James' perspective there are multiple (roughly $\mathcal{O}(\exp(\sqrt{n}))$) equally likely codewords (based on his observation $\Z$) --- hence James has a large ``uncertainty set'' of codewords. It then shows that, averaged over this uncertainty set, regardless of James' specific choice of jamming vector $\mathbf{S}$, very few codewords $\X$ in his uncertainty set are ``killed'' by $\mathbf{S}$, i.e., if Bob attempts to list-decode the corresponding $\Y = \X \oplus \mathbf{S}$, his list size is ``too large'' (larger than some polynomial --- say $n^2$). Hence, with high probability over the randomness in which $\X$ in the uncertainty set is instantiated, James is unable to force too large a list on Bob. To complete the argument we show that the dominant error event (among all joint distributions James can induced between $\Z$ and $\mathbf{S}$) corresponds to James behaving in the i.i.d. manner specific in the myopic jamming strategy. Bob is then able to use the ${\cal O}(\log(n))$-sized shared key to disambiguate the list down to a unique element via a hashing scheme. In the following, we present the {\it achievable positive throughput regions} ${\cal \underbar{R}}^+_{\Delta(n),\epsilon_d}(p,q)$ corresponding to the parameter regime where our codes have positive throughput. The achievable positive throughput regions $\underbar{{\cal R}}^+_{\Delta(n),\epsilon_d}(p,q)$ are subsets of the true positive throughput regions ${\cal R}^+_{\Delta(n),\epsilon_d}(p,q)$.

\begin{theorem}\label{thm:achievable}
	Let $\epsilon_d\in(0,1)$, $p,q \in (0,\frac{1}{2})$, and $\Delta(n) \ge 6\log(n)$. For three different regimes of $\D$, the achievable positive throughput regions $\underbar{{\cal R}}^+_{\Delta(n),\epsilon_d}(p,q)$ are given by 
	\begin{enumerate}
	\item {\bf small-sized key:} $\underbar{{\cal R}}^+_{\Delta(n),\epsilon_d}(p,q) \triangleq \left\{(p,q): p < q \mbox{ and } I_B(p,q) >I_J(q) \right\}$ if $\D \in (\Omega(\log(n)), o(\sqrt{n}))$. 
	\item {\bf moderate-sized key:} $\underbar{{\cal R}}^+_{\Delta(n),\epsilon_d}(p,q) \triangleq \left\{(p,q): p < q \mbox{ and } I_B(p,q)+\frac{\sigma}{t(q,\epsilon_d)} >I_J(q) \right\}$ if $\D = \sigma \sqrt{n}$ for a constant $\sigma > 0$.
	\item {\bf large-sized key:} $\underbar{{\cal R}}^+_{\Delta(n),\epsilon_d}(p,q) \triangleq \left\{(p,q): p < q \right\}$ if $\D \in \omega(\sqrt{n})$.
	\end{enumerate}
	For any $\D$ and $(p,q) \in \underbar{{\cal R}}^+_{\Delta(n),\epsilon_d}(p,q)$, the relative throughput 
	\begin{align}
	r_{\Delta(n),\epsilon_d}(p,q)= t(q,\epsilon_d) I_B(p,q) \label{eq:long}
	\end{align}
	is achievable, which implies that the covert capacity $r^{\ast}_{\Delta(n),\epsilon_d}(p,q) = t(q,\epsilon_d) I_B(p,q)$ (since~\eqref{eq:long} meets the upper bound derived in Theorem~\ref{thm:upperbound}). Both encoding and decoding may be performed with complexity $\exp(\mathcal{O}(\sqrt{n}))$.
 \end{theorem}
The proof of Theorem~\ref{thm:achievable} is included in Section~\ref{sec:achievability}. For any $0 < p < q < \frac{1}{2}$, to achieve relative throughput $t(q,\epsilon_d) I_B(p,q)$, the minimum size of the shared key is $\D = \mathcal{O}(\log(n))+ [t(q,\epsilon_d)(I_J(q)-I_B(p,q))]^+\sqrt{n}$, where $x^+ \triangleq \max(0,x)$. The intuition behind this scaling of $\Delta(n)$ is as follows: 
\begin{enumerate}
\item When the BSC$(q)$ from Alice to James is worse (has lower mutual information) than the worst channel he can instantiate from Alice to Bob, then ${\cal O}(\log(n))$ bits of shared key suffices for our scheme to work. 
\item Conversely, if James can make the channel from Alice to Bob to be worse than the channel to him, then Alice and Bob need a larger shared key (equaling at least the mutual information difference between the two channels) to cause James' uncertainty set to be large enough for the myopic list-decoding lemma (Claim~\ref{claim:ratio2}) to hold. Structurally this phenomenon in the presence of an active adversary James is intriguingly reminiscent of the phenomenon observed in~\cite{7407378} showing that covert communication in the presence of a  passive adversary is possible if and only if the key-rate exceeds the normalized mutual information difference between the main channel and the eavesdropped channel.
\end{enumerate}

Fig.~\ref{fig:regime} graphically represents the numerics of Theorems~\ref{thm:upperbound} and~\ref{thm:achievable}. Note that $\D = \omega(\sqrt{n})$ is the same as $\D = \infty$ (by comparing Theorems~\ref{thm:upperbound} and~\ref{thm:achievable}), since both the achievable positive throughput regime and the covert capacity are independent of $\D$ as long as it is larger than $\omega(\sqrt{n})$. Also, it is clear that the achievability and the converse may not match when $\Delta(n) \in (\Omega(\log(n)), o(\sqrt{n}))$ or $\D = \sigma \sqrt{n}$ (for some small $\sigma > 0$). 

\begin{figure*}
	\centering
	\begin{subfigure}{0.45\textwidth}
		\centering
		\includegraphics[width=0.9\textwidth]{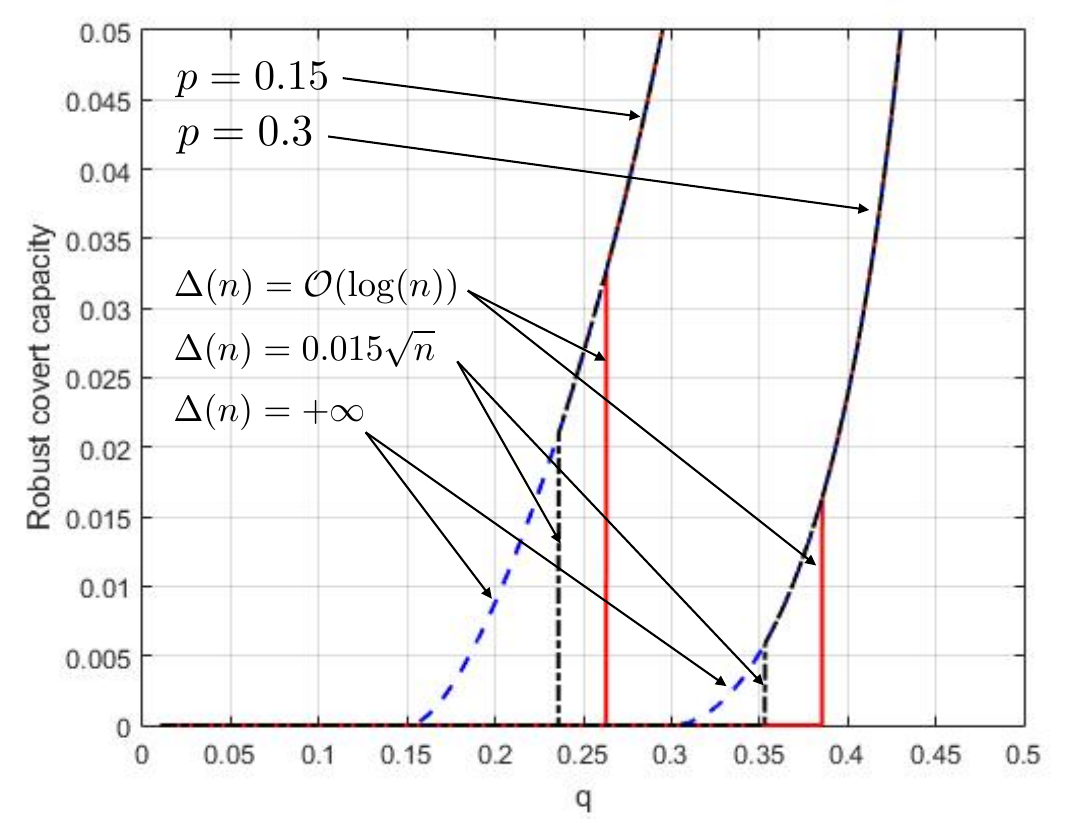} 
		\caption{The two sets of curves show covert capacities as functions of $q$ for two fixed values of $p$ ($p = 0.15$ and $p = 0.3$) and different amounts of shared key $\Delta(n)$ ($\Delta(n)=o(\sqrt{n}), \Delta(n)=0.015\sqrt{n},$ and $\Delta(n) = \infty$).}
		\label{fig:fixed_p_delta}
	\end{subfigure}
	\begin{subfigure}{0.45\textwidth}
		\centering
		\includegraphics[width=0.9\textwidth]{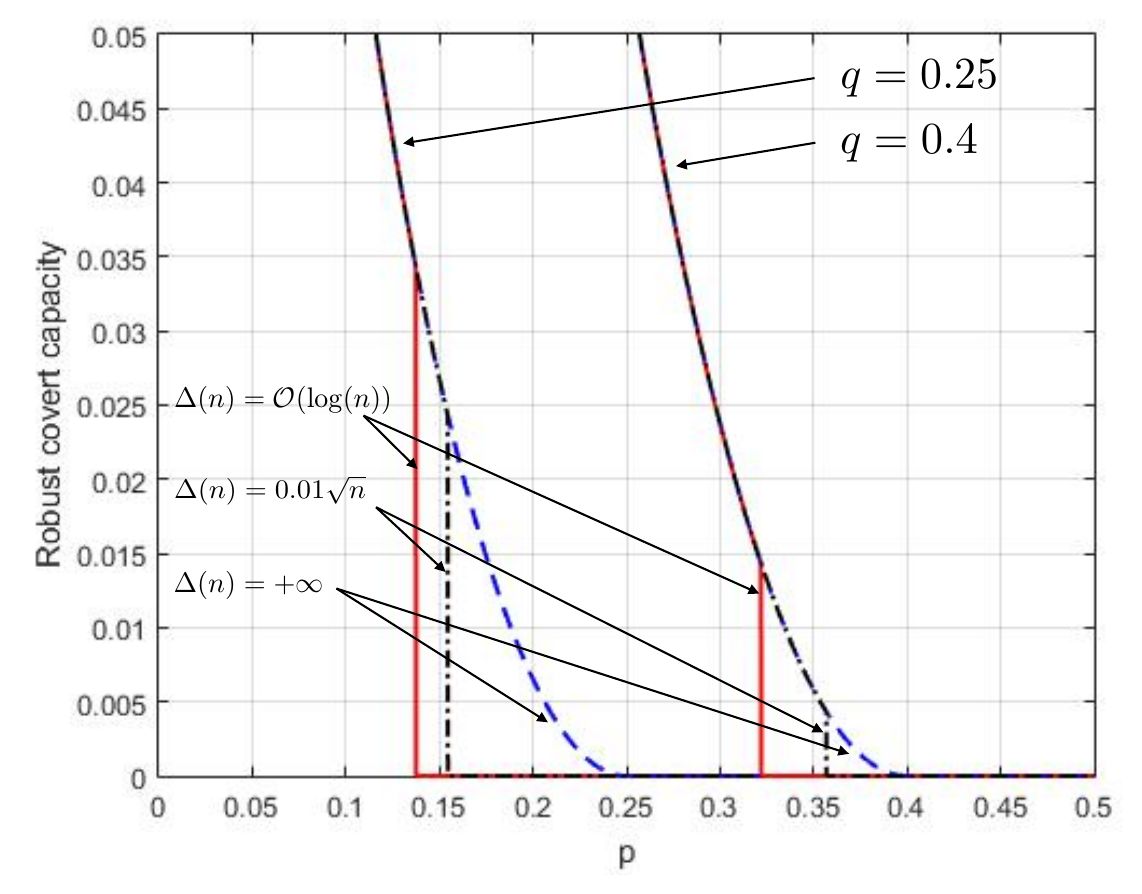} 
		\caption{The two sets of curves show covert capacities as functions of $p$ for two fixed values of $q$ ($q = 0.25$ and $q = 0.4$) and different amounts of shared key $\Delta(n)$ ($\Delta(n)=o(\sqrt{n}), \Delta(n)=0.01(\sqrt{n}),$ and $\Delta(n) = \infty$).}
		\label{fig:fixed_q_delta}
	\end{subfigure}
	\caption{Since the covert capacity region would require a three dimensional plot ($p$ and $q$ along the $x$ and $y$ axes respectively, and the relative throughput along the $z$ axis) that is hard to digest, we instead present here cross-sections of our partial characterization of the capacity region. The plot in Fig.~\ref{fig:fixed_p_delta} shows lower and upper bounds on the covert capacity for two values of $p$ ($p=0.15$ and $p=0.3$), and that in Fig.~\ref{fig:fixed_q_delta} shows the corresponding curves for two values of $q$ ($q=0.25$ and $q=0.4$), and the covertness parameter $\epsilon_d = 0.02$ for each of these curves. For each of these values of $p$ or $q$, the blue dashed curves indicate upper bounds on the covert capacity, and indeed, these are attainable via matching achievability schemes when $\Delta(n) \in \omega(\sqrt{n})$. As alluded to in the achievable positive throughput region plot in Fig.~\ref{fig:regime}, note the impact of increasing values of $\Delta(n)$ --- the achievable positive throughput region increases, and the corresponding throughput achievable by our coding scheme in Theorem~\ref{thm:achievable} tracks the blue curve corresponding to having unbounded shared keys. The red curve corresponds to the relative throughput attainable by our coding scheme for any value of $\Delta(n) \in (\Omega(\log(n)), o(\sqrt{n}))$, and the black curve corresponds to the attainable relative throughput for $\Delta(n)=0.015\sqrt{n}$ in Fig.~\ref{fig:fixed_p_delta} and $\Delta(n)=0.01\sqrt{n}$ in Fig.~\ref{fig:fixed_q_delta}.}
	\label{fig:fix_pq}
\end{figure*}

\begin{figure*}
	\centering
	\begin{subfigure}{0.45\textwidth}
		\centering
		\includegraphics[width=0.9\linewidth]{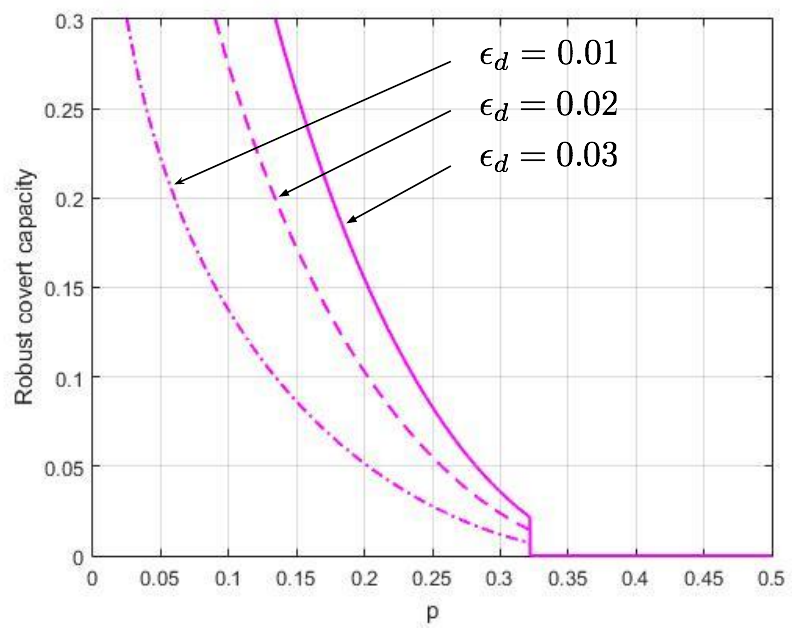} 
		\caption{Covert capacities as functions of $p$ for fixed $q$ ($q = 0.4$) and different covertness parameters $\epsilon_d$.}
		\label{fig:fixed_q_epsilon}
	\end{subfigure}
	\begin{subfigure}{0.45\textwidth}
		\centering
		\includegraphics[width=0.9\linewidth]{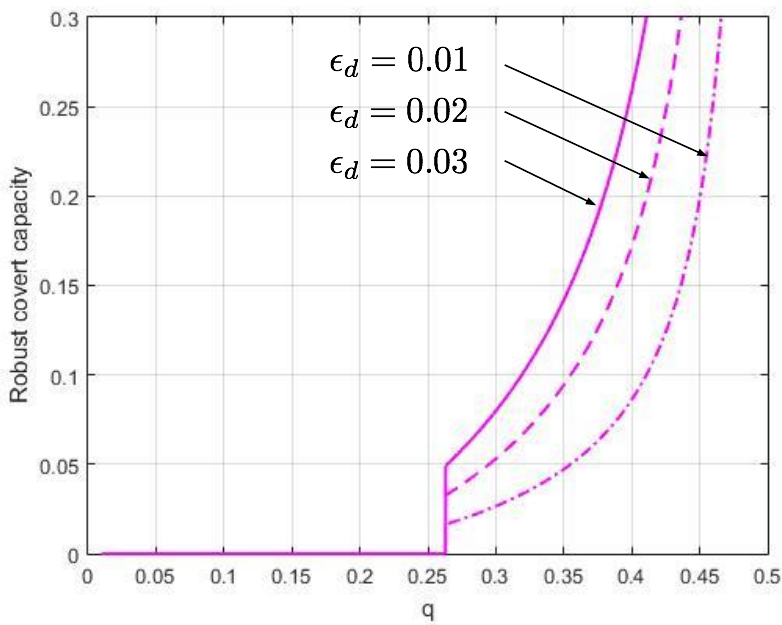} 
		\caption{Covert capacities as functions of $q$ for fixed $p$ ($p = 0.15$) and different covertness parameters $\epsilon_d$.}
		\label{fig:fixed_p_epsilon}
	\end{subfigure}
	\caption{The covertness parameter $\epsilon_d$ also has an impact on the covert capacity --- as shown in Fig.~\ref{fig:fixed_q_epsilon} and Fig.~\ref{fig:fixed_p_epsilon}, increasing $\epsilon_d$ increases the covert capacity, since Alice's codebook can comprise of somewhat ``heavier'' codewords.} 
	\label{fig:epsilon}
\end{figure*}

\begin{figure}
	\begin{center}
		\includegraphics[scale=0.35]{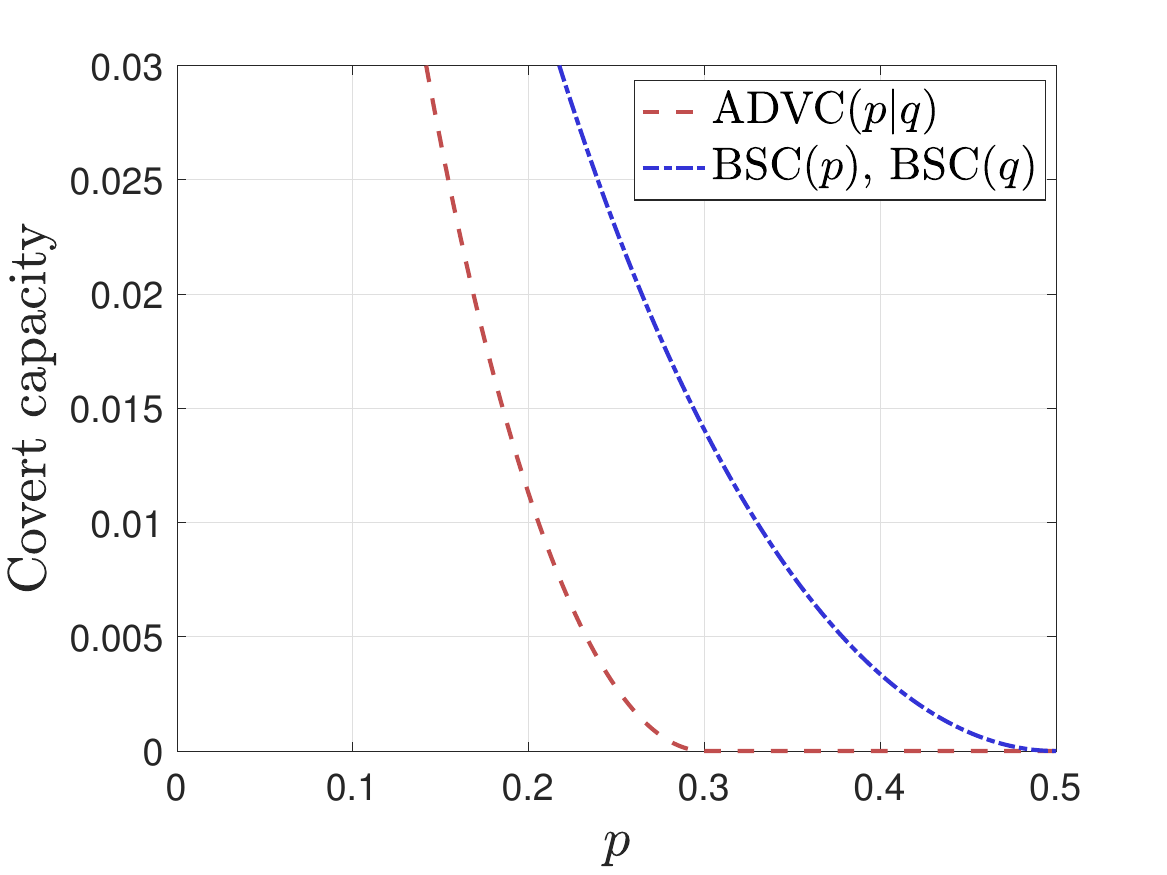}
		\caption{This figure plots the covert capacities (as a function of $p$ for fixed $q = 0.3$) of the adversarial channel model ADVC$(p|q)$ and the random noise setting with BSC($p$) and BSC($q$). We set the covertness parameter $\epsilon_d = 0.01$ and $\Delta(n) = \infty$. We note that (i) the covert capacity of the random noise setting is always larger than that of the ADVC($p|q$), and (ii) when $p \ge q$, the covert capacity of the random noise setting is positive, while that of the ADVC($p|q$) equals zero.} \label{fig:BSC}
	\end{center}
\end{figure}

\subsection{Graphical representations of covert capacity}\label{appendix:figure}

Fig.~\ref{fig:fix_pq}  provides a graphical representation of the lower and upper bounds on the covert capacities for different sizes of shared key $\Delta(n)$ --- specifically, Fig.~\ref{fig:fixed_p_delta} considers $q \in (0,0.5)$ and two fixed values of $p$, while Fig.~\ref{fig:fixed_q_delta} considers $p \in [0, 0.5]$ and two fixed values of $q$. Fig.~\ref{fig:epsilon} illustrates the impact of the covertness parameter $\epsilon_d$ on the covert capacity.

To better understand the covert capacity of the adversarial channel model ADVC($p|q$), we also compare it with the covert capacity of the well-studied \emph{random noise setting} in which the channel from Alice to Bob is a BSC($p$) and the channel from Alice to James is a BSC($q$). We assume the shared key $\Delta(n) = \infty$ such that the covert capacities of both settings are fully characterized. As shown in Theorems~\ref{thm:upperbound} and~\ref{thm:achievable}, the covert capacity of the ADVC($p|q$) is 
	\begin{align}
	r^{\ast}_{\Delta(n),\epsilon_d}(p,q) = \begin{cases} t(q,\epsilon_d) I_B(p,q), & \text{if } p < q; \\ 0, & \text{otherwise}. \end{cases} \label{eq:com1}
	\end{align}
	In contrast, the covert capacity of the random noise setting~\cite{tahmasbi2018first,zhang2019covert}, denoted by $\widetilde{r}_{\Delta(n),\epsilon_d}(p,q)$, is
	\begin{align}
	\widetilde{r}_{\Delta(n),\epsilon_d}(p,q) =  t(q,\epsilon_d) (1-2p)\log\left(\frac{1-p}{p}\right). \label{eq:com2}
	\end{align}
	Clearly, the most significant difference is that when $q \ge p$, the covert capacity of the ADVC($p|q$) is zero, while that of the random noise setting is positive. This is because in the former setting, if James employs the i.i.d. myopic jamming strategy, the induced channel from Alice to Bob is \emph{completely noisy} such that the corresponding mutual information is zero (for any input distribution). Additionally, we note that even when $q < p$, the covert capacity of the ADVC($p|q$) is also smaller than that of the random noise setting. This is illustrated in Fig.~\ref{fig:BSC} for $p \in [0, 0.5]$ and a fixed value of $q$ ($q = 0.3$).

\subsection{Computationally efficient coding schemes} \label{sec:result_efficient}
This subsection presents a computationally efficient encoding and decoding scheme when the amount of shared key is $\Omega(\sqrt{n}\log{n})$. Consider a binary asymmetric channel (BAC) with input alphabet $\mathcal{X}\in\{0,1\}$, output alphabet $\mathcal{Y}\in\{0,1\}$, and the bit flip probabilities $W_{Y|X}(1|0)=p$ and $W_{Y|X}(0|1)= \frac{(1-q)p}{q}$. This corresponds to the channel from Alice to Bob caused by the myopic jamming strategy. Let $C_{\text{BAC}}(p,q)\triangleq \max_{p(X)}I(X;Y)$ denote the channel capacity, and let Bern$(\rho^\ast)$ be the input distribution that achieves the maximum value of $I(X;Y)$.

\begin{theorem}\label{thm:efficient} Let $\epsilon_d \in(0,1)$, $0 < p < q < \frac{1}{2}$, and  $r< \frac{t(q,\epsilon_d)}{\rho^\ast}C_{\emph{BAC}}(p,q)$. There exists a sequence of codes $\{\C_n\}$ of blocklength $n$, relative throughput $r$, and $\Delta(n) = c\sqrt{n} \log(n)$ for some constant $c > 0$ such that for large enough $n$,
\begin{enumerate}
\item $\C_n$ is $(1-\epsilon_d)$-covert;
\item $\C_n$ ensures the probability of error $P_{\emph{err}} \le \epsilon_n$, where $\limsup_{n \to \infty} \epsilon_n = 0$; 
\item $\C_n$ can be encoded and decoded with ${\operatorfont{poly}}(n)$ complexity.
\end{enumerate}
\end{theorem}
The achievability scheme used for proving Theorem~\ref{thm:efficient} is presented in Section~\ref{sec:efficient}. It is based on a \emph{concatenated code}~\cite{forney1965concatenated} of blocklength $\mathcal{O}(\sqrt{n})$ and a permutation-based scheme described in the introduction. As argued in~\cite{7407378}, such a source-resolvability scheme results in covertness with respect to James. Also, as argued in~\cite{lipton1994new,langberg2004private} such codes also work well to scramble James' bit-flips, and make his actions behave in an i.i.d. manner.

{\bf Computationally bounded adversary:} Note that the above theorem does not place any computational restrictions on James. Following the standard cryptographic convention, we now consider the setting in which James' computational power is restricted to be polynomial in the blocklength $n$. Under this setting and under the assumption that \emph{one-way functions} exist, we show that the computationally efficient coding scheme used in Theorem~\ref{thm:efficient} can be implemented with much less amount of shared key, by introducing the notion of {\it pseudorandom generator (PRG)}. This leads to the following theorem.
\begin{theorem} \label{cor:prg}
Suppose James is computationally bounded.	Let $\epsilon_d \in(0,1)$, $0 < p < q < \frac{1}{2}$, and  $r< \frac{t(q,\epsilon_d)}{\rho^\ast}C_{\emph{BAC}}(p,q)$. There exists a sequence of codes $\{\C_n\}$ of blocklength $n$, relative throughput $r$, and $\Delta(n)=n^\xi$ (where $\xi > 0$ can be chosen arbitrarily small) such that for  large enough $n$,
\begin{enumerate}
\item For every polynomial-time estimator $\Phi$, $\liminf_{n \to \infty} P_{\emph{FA}}(\Phi)+P_{\emph{MD}}(\Phi)\ge 1-\epsilon_d$;
\item $\C_n$ ensures the probability of error $P_{\emph{err}} \le \epsilon_n$, where $\limsup_{n \to \infty} \epsilon_n = 0$;
\item $\C_n$ can be encoded and decoded with ${\operatorfont{poly}}(n)$ complexity.	
\end{enumerate}
\end{theorem}
Roughly speaking, a length-$u$ \emph{truly random shared key} can be used in conjunction with a PRG to generate a length-$\text{poly}(u)$ \emph{pseudorandom shared key}, without being detected by any polynomial-time algorithms~\cite[Theorems 7.6, 7.7]{katz2014introduction}. Substituting $u$ and $\text{poly}(u)$ with $n^{\xi}$ and $c\sqrt{n}\log n$, we have the following statement: there exist an efficiently computable function $g: \{0,1\}^{n^{\xi}}\rightarrow\{0,1\}^{c\sqrt{n}\log(n)}$ and a vanishing sequence $\nu_n$ (where $\nu_n < 1/p(n)$ for every polynomial $p(n)$ and sufficiently large $n$) such that if ${U}\sim \operatorname{Unif}(\{0,1\}^{n^\xi})$ and ${U'}\sim \operatorname{Unif}(\{0,1\}^{c\sqrt{n}\log(n)})$, then for all polynomial-time computable functions $D:\{0,1\}^{c\sqrt{n}\log(n)}\to\{0,1\}$, 
	\begin{align}
	\left|\PP_{{U}}(D(g({U}))=1)-\PP_{{U'}}(D({U'})=1)\right| \le \nu_n. \label{eq:video}
	\end{align}	 
That is, no polynomial-time computable function can distinguish the pseudorandom shared key $U$ from the truly random shared key $U'$ by a \emph{non-negligible} advantage (i.e., $\nu_n$ converges to 0 faster than any polynomial of $n$).

We say the shared key $K = g(U)$ if it comes from the output of a PRG with a seed $U \sim \text{Unif}(\{0,1\}^{n^{\xi}})$, and that $K = U'$ if it is truly uniformly distributed. Under the computational assumptions, we require our coding scheme to simultaneously satisfy
	\begin{align}
	\text{(Covertness)}\quad &\liminf_{n \to \infty} \PP(\widehat{T}=1|T=0)+\PP(\widehat{T}=0|T=1, K=g(U)) \ge 1 - \epsilon_d, \ \ \forall \text{ poly-time estimator }\Phi  \label{eq:phi},\\
	\text{(Reliability)}\quad &\limsup_{n \to \infty}  \max_{W_{\mathbf{S}|\Z,\C}} \left(
	\PP(\widehat{M} \neq 0 | T = 0) +
	\PP (\widehat{M} \neq M | T = 1, K = g(U)) 
	\right) = 0. \label{eq:rel}
	\end{align} 
	
To prove~\eqref{eq:phi}, it suffices to show that:
\begin{align}
\begin{cases}
\liminf_{n \to \infty} \PP(\widehat{T}=1|T=0)+\PP(\widehat{T}=0|T=1, K=U') \ge 1 - \epsilon_d, \quad &\forall \text{ poly-time estimator }\Phi, \\
\PP(\widehat{T}=1|T=1,K=g(U))-\PP(\widehat{T}=1|T=1, K=U') \le \nu_n, \quad &\forall \text{ poly-time estimator }\Phi.
 \label{eq:pseudo}
\end{cases}
\end{align} 
Note that the first condition in~\eqref{eq:pseudo} is satisfied since it has been shown in Theorem~\ref{thm:efficient} that the coding scheme with truly random shared key $K = U'$ is $(1-\epsilon_d)$-covert. The second condition in~\eqref{eq:pseudo} can be proved by contradiction --- if there were a polynomial-time estimator $\Phi$ satisfying $\PP(\widehat{T}=1|T=1,K=g(U))-\PP(\widehat{T}=1|T=1, K=U') > \nu_n$, then we would be able to construct a polynomial-time algorithm that can distinguish $g(U)$ and $U'$ with at least $\nu_n$ advantage, thus contradicting~\eqref{eq:video}. The detailed proof is provided in Appendix~\ref{app:bounded}.

To prove~\eqref{eq:rel}, it suffices to show that:
\begin{align}
\begin{cases}
\limsup_{n \to \infty}  \max_{W_{\mathbf{S}|\Z,\C}} \left(
\PP(\widehat{M} \neq 0 | T = 0) +
\PP (\widehat{M} \neq M | T = 1, K = U') 
\right) = 0,  \\
\PP(M \ne \widehat{M}|T=1, K=g(U))-\PP(M \ne \widehat{M}|T=1,K=U') \le \nu_n.
\label{eq:pseudo2}
\end{cases}
\end{align} 
The first condition in~\eqref{eq:pseudo2} is satisfied since it has been shown in Theorem~\ref{thm:efficient} that the coding scheme with truly random shared key $K = U'$ is reliable. The second condition in~\eqref{eq:pseudo2} can also be proved by contradiction --- if there were a polynomial-time decoder satisfying $\PP(M \ne \widehat{M}|T=1, K=g(U))-\PP(M \ne \widehat{M}|T=1,K=U') > \nu_n$, then we would be able to construct a polynomial-time algorithm that can distinguish $g(U)$ and $U'$ with at least $\nu_n$ advantage. Again, the detailed proof can be found in Appendix~\ref{app:bounded}.

\begin{table}[]
\small
\centering
\caption{Table of parameters}
\begin{tabular}{|l|l|l|l|}
\hline
\textbf{Symbol} & \textbf{Description} & \textbf{Equality/Range} & \textbf{Section} \\ \hline
$M$      & Message     & $M \in \{1,2,\ldots,N\}$    & Section~\ref{sec:model} \\ \hline
$T$       &  Transmission status   & $T \in \{0,1\}$   & Section~\ref{sec:model}  \\ \hline
$K$   & Shared key &  $K \in \{0,1\}^{\Delta(n)}$  & Section~\ref{sec:model}  \\ \hline
$p$  &  ADVC$(p)$ -- channel from Alice to Bob    &  $0 \le p \le 0.5$  & Section~\ref{sec:introduction} \\ \hline
$q$   &  BSC$(q)$ -- channel from Alice to James    &  $0 \le q < 0.5$     &Section~\ref{sec:introduction} \\ \hline
 $\epsilon_d$ &  Covertness parameter   & $\epsilon_d > 0$     &  Section~\ref{sec:model}       \\ \hline
  $\D$ &  Size of shared key   & N/A   &  Section~\ref{sec:model}       \\ \hline
$t(q,\epsilon_d)$ &  Code-weight design parameter  &   $t(q,\epsilon_d)  = \frac{2 \sqrt{q (1-q)}}{1-2q}\cdot Q^{-1}\left(\frac{1-\epsilon_d}{2}\right)$      & Section~\ref{sec:result}   \\ \hline
$\rho$ &  Normalized code-weight design parameter  &   $\rho = t(q,\epsilon_d)/\sqrt{n}$     &  Section~\ref{sec:result}       \\ \hline
$\X$   &  Codeword   &  $\X \in \{0,1\}^n$   &  Section~\ref{sec:model} \\ \hline
$\Z$  &   James' received vector  &  $\Z \in \{0,1\}^n$   &  Section~\ref{sec:model}   \\ \hline
$\mathbf{S}$  & James' jamming vector   &  $\mathbf{S} \in \{0,1\}^n$    &  Section~\ref{sec:model}    \\ \hline
 $\Y$  &  Bob's received vector  &   $\Y \in \{0,1\}^n$    & Section~\ref{sec:model}  \\ \hline
 $R$ &   Rate   &   $R = (\log N)/n$  &  Section~\ref{sec:model}       \\ \hline
 $r$  &  Relative throughput  &  $r = (\log N)/\sqrt{n}$   &   Section~\ref{sec:model}      \\ \hline
 $r^{\ast}_{\Delta(n),\epsilon_d}(p,q)$  &  Covert capacity     &   N/A   &   Section~\ref{sec:model}  \\ \hline
 $\mathcal{R}^+_{\Delta(n),\epsilon_d}(p,q)$  &  Positive throughput region    &   N/A    &  Section~\ref{sec:model}  \\ \hline
 $\underbar{\cal R}^+_{\Delta(n),\epsilon_d}(p,q)$  &  Achievable positive throughput region    &   N/A    &  Section~\ref{sec:model}  \\ \hline
 $Q_0(\Z)$   &  Innocent distribution of $\Z$ ($T = 0$)  &  N/A   &  Section~\ref{sec:model}  \\ \hline
 $Q_1(\Z)$  &   Active distribution of $\Z$ ($T = 1$)  &   N/A  & Section~\ref{sec:model} \\ \hline
 $I_J(q)$  &  Weight normalized mutual information   &   N/A  &  Section~\ref{sec:result}  \\ \hline
  $I_B(p,q)$ &   Weight normalized mutual information   &   N/A   &  Section~\ref{sec:result}   \\ \hline
\end{tabular}
\end{table}

\newpage

\section{Proof of Theorem~\ref{thm:converse1}} \label{sec:converse}
We now show that if $\D < \frac{1}{2}\log (n)$, the probability of error is bounded from below by $1-\max\left\{\frac{2}{N},\frac{2^{\D}\eta}{\sqrt{n}}\right\}$, for some constant $\eta$ independent of $n$. First note that due to the covertness constraint, most of the codewords have Hamming weight $\mathcal{O}(\sqrt{n})$, otherwise Alice's transmission status can be detected by James' weight-detector. Since James is able to flip $\mathcal{O}(n)$ bits, he can apply an oblivious jamming strategy --- generate his jamming vector by selecting $\mathcal{O}(\sqrt{n})$ codewords. Since the number of possible values of shared key is $2^{\D} < \mathcal{O}(\sqrt{n})$, he can select codewords in the following way (without loss of generality we assume $\x(m_0,k_0)$ --- the codeword corresponding to message $m_0$ and shared key $k_0$ --- is transmitted):
\begin{enumerate}
	\item For each value of $k \in \{ 1, \ldots, 2^{\Delta(n)}\}$, James randomly chooses $b = \min\left\{\frac{\mathcal{O}(\sqrt{n})}{2^{\D}}, \frac{N}{2}\right\}$ messages $m_1, m_2, \ldots, m_b$, and use Alice's encoding function to obtain codewords $\x(m_1,k), \x(m_2,k), \ldots, \x(m_b,k)$.
	\item Let $\mathcal{S}_k \triangleq \{ \x(m_1,k), \x(m_2,k), \ldots, \x(m_b,k)\}$. James' jamming vector $\s$ equals $\oplus_k \oplus_{\mathcal{S}_k} \x(m,k)$, i.e., the binary additions of all selected codewords. Bob's observation $\y$ equals $\x(m_0,k_0) \oplus \s$.    
\end{enumerate}
Now let's focus on the set $\mathcal{S}_{k_0}$. We define a modified set 
\begin{align}
\widehat{\mathcal{S}}_{k_0} \triangleq \begin{cases} \mathcal{S}_{k_0} \setminus \x(m_0,k_0), \text{ if } \ \x(m_0,k_0) \in \mathcal{S}_{k_0}, \\
\mathcal{S}_{k_0} \cup \x(m_0,k_0), \text{ if } \ \x(m_0,k_0) \notin \mathcal{S}_{k_0}.
\end{cases}
\end{align}
We assume there is an oracle who reveals to Bob the value of $k_0$, the set $\widehat{\mathcal{S}}_{k_0}$, and all $\mathcal{S}_{k}$ for $k \ne k_0$ selected by James, and the fact that whether or not Alice's true codeword in the set $\widehat{\mathcal{S}}_{k_0}$. Note that the oracle only strengthens Bob, since he can recover the received vector from the oracle revealed information. Thus, Bob's probability of decoding error with the knowledge of the oracle information is no larger than that without it. If $\x(m_0,k_0) \in \widehat{\mathcal{S}}_{k_0}$, from Bob's point of view the true message is uniformly distributed over the set $\widehat{\mathcal{S}}_{k_0}$, since he cannot distinguish the following $(b+1)$ equally likely events
\begin{itemize}
	\item $\mathcal{E}_{m_0}$: Alice transmits $\x(m_0,k_0)$ and James selects $\left\{\x(m_1,k_0), \x(m_2,k_0), \ldots, \x(m_b,k_0) \right\}$.
	\item $\mathcal{E}_{m_i}$ ($i \ne 0$): Alice transmits $\x(m_i,k_0)$ and James selects $\left\{\x(m_0,k_0), \x(m_1,k_0), \ldots,\x(m_b,k_0) \right\}\setminus \x(m_i,k_0)$.
\end{itemize}
Similarly, if $\x(m_0,k_0) \notin \widehat{\mathcal{S}}_{k_0}$, from Bob's point of view the true message is uniformly distributed over the set $\widehat{\mathcal{S}}_{k_0}^c$ (the complement of $\widehat{\mathcal{S}}_{k_0}$). These imply that the probability of decoding error (when $T = 1$) is bounded from below by $1 - \max\left\{\frac{2^{\D}\eta}{\sqrt{n}}, \frac{2}{N}\right\}$, for some $\eta > 0$.

\section{Proof of Theorem~\ref{thm:upperbound}} \label{sec:upperbound}
The upper bound in Theorem~\ref{thm:upperbound} is obtained by considering a specific myopic jamming strategy performed by James, as described in the following. This strategy leads to an artificial binary asymmetric channel (BAC) between Alice and Bob, and in turn limits the message size of any codes that simultaneously ensures $(1-\epsilon_d)$-covertness and a small probability of error $P_{\text{err}}$. 

\subsection{A myopic jamming strategy} 
Consider the jamming strategy $W^{\text{(my)}}_{\mathbf{S}|\Z,\C}$ described as follows. For each $i \in \{1, \ldots, n \}$, James does not flip bit $X_i$ if the corresponding $Z_i$ equals $0$, and flips with probability approximately $p/q$ if the corresponding $Z_i$ equals $1$. This ensures that his bit-flips are stochastically distributed in the support of the $\Z$ vector. Since the $\Z$ vector is correlated with Alice's transmission $\X$ via a BSC$(q)$, this ensures that James' jamming vector $\mathbf{S}$ is likelier to flip $1$'s in $\X$ to $0$'s, than it is to flip $0$'s in $\X$ to $1$'s.

More precisely, let $\nu = n^{-1/3}$ be a slackness parameter. For any $i \in \{1, \ldots, n \}$,
$$
S_i =
\begin{cases}
0, &\mbox{ with probability $1$ if $Z_i = 0$, }\\
0, &\mbox{ with probability $1-\frac{p(1 -\nu)}{q}$ if $Z_i = 1$, }\\
1, &\mbox{ with probability $\frac{p(1 -\nu)}{q}$ if $Z_i = 1$. }
\end{cases}
$$
Note that generating $\mathbf{S}$ in the i.i.d. manner specified above may in general result in James' exceeding his jamming budget $pn$. However, by setting the slackness parameter $\nu = n^{-1/3}$, we ensure with probability at least $1 - \exp(-\mathcal{O}(n^{\frac{1}{3}}))$, the Hamming weight of $\mathbf{S}$ is bounded from above by $pn$. 

By using this strategy, James induces a BAC from Alice to Bob with channel transition probabilities
\begin{align*}
&\wy(0|0) = 1-p(1-\nu), &\wy(1|0) &= p(1-\nu), \\
&\wy(0|1) = \frac{(1-q)p}{q}(1-\nu), &\wy(1|1) &= 1-\frac{(1-q)p}{q}(1-\nu).
\end{align*}
Note that when $q<\frac{1}{2}$, $\wy(0|1) > \wy(1|0)$, which means that the probability of a bit-flip is higher when $X_i = 1$, than when $X_i = 0$.

\subsection{Converse with respect to the BAC}
Though the error criterion of interest in this work is the average probability of error $P_{\text{err}}$ defined in~\eqref{eq:pe}, to prove the upper bound in Theorem~\ref{thm:upperbound}, we take a detour by introducing another error criterion --- the {\it max-average probability of error}
\begin{align}
\widetilde{P}_{\text{err}} \triangleq  \max_{k} \{ \mathbb{P}(M \ne \widehat{M}|K=k,T = 1)  + \mathbb{P}(\widehat{M} \ne 0 | T = 0)\},
\end{align} 
which is maximized over the shared key and averaged over the message. Lemma~\ref{lemma:reduction} below, which is adapted from~\cite[Lemma 6]{zhang2019covert}, establishes a nice connection between the two error criterions. For the benefit of the readers, we provide a detailed proof of Lemma~\ref{lemma:reduction} in the supplementary document~\cite{zhang2019supp}.

\begin{lemma}[Adapted from~\cite{zhang2019covert}] \label{lemma:reduction}
	Suppose a code $\C$, which contains $2^{\Delta(n)}$ sub-codes of size $\N$, guarantees $(1-\epsilon_d)$-covertness and $P_{\emph{err}} \le \epsilon_n$.
	Then, there exists another code $\C'$ containing $2^{\Delta'(n)}$ sub-codes of size $\N'$ guarantees $(1-\epsilon_d)$-covertness and $\widetilde{P}_{\emph{err}} \le \epsilon'_n$. In particular, 
	\begin{align*}
	\limsup_{n \to \infty} \epsilon_n = \limsup_{n \to \infty} \epsilon'_n = 0, \ \ \lim_{n \to \infty} \frac{\log \N}{\sqrt{n}} = \lim_{n \to \infty} \frac{\log \N'}{\sqrt{n}}, \ \ \lim_{n \to \infty} \frac{\Delta(n)}{\Delta'(n)} = 1.
	\end{align*}
\end{lemma}
First, we provide an upper bound on the number of bits that can be reliably and covertly transmitted, under max-average probability of error $\widetilde{P}_{\text{err}}$. Then, we use a reduction argument to show that the aforementioned upper bound is also valid under average probability of error $P_{\text{err}}$. The proof of converse leverages different techniques from~\cite{7407378, tahmasbi2017second, zhang2019covert}.

\subsubsection{Converse under $\widetilde{P}_{\text{err}}$} Consider any code $\C$ containing $2^{\Delta(n)}$ sub-codes of size $\N$ (indexed by $\{\C_i\}_{i=1}^{2^{\Delta(n)}}$) that ensures $(1-\epsilon_d)$-covertness and a vanishing max-average probability of error $\widetilde{P}_{\text{err}} \le \epsilon_n$, where $\limsup_{n \to \infty} \epsilon_n = 0$. We first find an upper bound on the maximum weight of codewords in a suitable sub-code. Specializing~\cite[Lemma 12]{tahmasbi2017second} to our setting, we obtain that for any $\gamma\in(0,1)$, there exists a subset $\C^\gamma$ of $\C$ such that
\begin{enumerate}
	\item $|\C^{\gamma}|\geq \gamma|\C|$
	\item there is a constant $c_0$ such that 
	\begin{align*}
	w(\C^\gamma)\triangleq\frac{\max_{\mathbf{x}\in\C^\gamma} \text{wt}_H(\mathbf{x})}{\sqrt{n}}\leq \frac{2\sqrt{q(1-q)}}{1-2q} Q^{-1}\left(\frac{1-\epsilon_d}{2}-\frac{c_0}{\sqrt{n}}-\gamma\right).
	\end{align*}
\end{enumerate}

For each sub-code $\C_i$ ($i \in \{ 1, \ldots, 2^{\Delta(n)}\}$), the intersection between $\C^{\gamma}$ and $\C_{i}$ is denoted by $\C_i^{\gamma}$. Note that there must exist a sub-code $\C_i$ such that the size of $\C_i^{\gamma}$ is at least $\gamma \N$, and we denote this sub-code by $\C_{i^*}$. 
Let $\gamma = \max \{\sqrt{\epsilon_n}, \exp(-n^{\frac{1}{2}-\varepsilon})\}$ for some small $\varepsilon > 0$. Since the average probability of error of $\C_{i^*}$ is at most $\epsilon_n$, the average probability of error of $\C_{i^*}^\gamma$, denoted by $\epsilon_n'$, is bounded from above as 
$$\epsilon_n' \le \epsilon_n/\gamma = \min\{\sqrt{\epsilon_n}, \epsilon_n \exp(n^{\frac{1}{2}-\varepsilon})\} \le \sqrt{\epsilon_n},$$ 
which is due to the fact that $|\C_{i^*}^\gamma|\cdot \epsilon_n' \le |\C_{i^*}|\cdot \epsilon_n$ if Bob simply employs the decoding rule for $\C_{i^*}$. Let $\widetilde{M}$ be the uniformly distributed random variable that corresponds to the message in $\C_{i^*}^\gamma$, $\bar{X}$ be the random variable distributed according to $\text{Bern}(w(\C^\gamma)/\sqrt{n})$, $\bar{Y}$ be the random variable corresponding to the output of the BAC $\wy$ with input $\bar{X}$. By standard information inequalities, we have
\begin{align}
\log \N +\log{\gamma}\le H(\widetilde{M}) &= I(\widetilde{M};\Y K) + H(M|\Y K) \label{eq:home} \\
& \le I(\widetilde{M};\Y K) + \epsilon_n'\cdot \log (\gamma \N) + 1  \label{eq:hom1} \\
& = I(\widetilde{M};\Y | K) + \epsilon_n'\cdot \log (\gamma \N) + 1 \label{eq:hom2}  \\
&\le I(\widetilde{M}K;\Y) + \epsilon_n'\cdot \log (\gamma \N) + 1  \label{eq:hom3}  \\ 
&\le I(\X;\Y) + \epsilon_n'\cdot \log (\gamma \N) + 1 \label{eq:home4} \\
&\le \sum_{i=1}^n I(X_i;Y_i) + \epsilon_n'\cdot \log (\gamma \N) + 1 \\
&\leq  n I(\bar{X};\bar{Y}) + \epsilon_n'\cdot \log (\gamma \N) + 1. \label{eq:home6}
\end{align} 
Inequality~\eqref{eq:hom1} follows from the Fano's inequality, equation~\eqref{eq:hom2} holds since $I(\widetilde{M};\Y K) = I(\widetilde{M};Y|K)+I(\widetilde{M};K)$ and $\widetilde{M}$ is independent of $K$. Similarly, inequality~\eqref{eq:hom3} holds since $I(\widetilde{M};\Y|K) = I(\widetilde{M}K;\Y)-I(K;\Y) \le I(\widetilde{M}K;\Y)$. Inequality~\eqref{eq:home4} is due to the data processing inequality, and~\eqref{eq:home6} is due to the concavity of mutual information with respect to the marginal distributions. Hence, we have
\begin{align}
\frac{\log \N}{\sqrt{n}} \le \frac{1}{1-\epsilon_n'}\sqrt{n}I(\bar{X};\bar{Y}) + \frac{1-(1-\epsilon'_n)\log \gamma}{(1-\epsilon_n')\sqrt{n}},
\end{align}
where $\frac{1-(1-\epsilon'_n)\log \gamma}{(1-\epsilon_n')\sqrt{n}}$ goes to zero for sufficiently large $n$, since $\gamma = \max \{\sqrt{\epsilon_n}, \exp(-n^{\frac{1}{2}-\varepsilon})\}$.
The mutual information $I(\bar{X};\bar{Y})$ of the BAC can be approximated as 
\begin{align*}
\lefteqn{\sqrt{n}I(\bar{X};\bar{Y})}\\
& = \sqrt{n}(H(\bar{Y}) - H(\bar{Y}|\bar{X})) \\
&\overset{\text{$n \to \infty$}}= \frac{w(\C_\gamma) p(q-1)}{q}\log \left(\frac{(1-p)(q-p+pq)}{p^2(1-q)}\right) + w(\C_\gamma) \log \left(\frac{q-p+pq}{pq}\right), \\
&\overset{\text{$n \to \infty$}}= t(q,\epsilon_d) I_B(p,q). 
\end{align*}
Therefore, we obtain that $\lim_{n \to \infty} \frac{\log \N}{\sqrt{n}} \le t(q,\epsilon_d) I_B(p,q)$.

\vspace{0.1cm}
\subsubsection{Converse under $P_{\text{err}}$ via reduction}
Now we use a reduction argument to show that the converse result also holds under $P_{\text{err}}$ (the probability of error of interest in this work), based on Lemma~\ref{lemma:reduction}. Suppose there exists a code $\C$, which contains $2^{\Delta(n)}$ sub-codes of size $\N$, ensures $(1-\epsilon_d)$-covertness and $P_{\text{err}} \le \epsilon_n$. If $\lim_{n \to \infty} \frac{\log \N}{\sqrt{n}} > t(q,\epsilon_d) I_B(p,q)$, then a contradiction arises since Lemma~\ref{lemma:reduction} says that there exists another code $\C'$ containing $2^{\Delta'(n)}$ sub-codes of size $\N'$ that ensures $(1-\epsilon_d)$-covertness, $\widetilde{P}_{\text{err}} \le \epsilon'_n$ (where $\limsup_{n \to \infty} \epsilon'_n = 0$), and 
\begin{align*}
\lim_{n \to \infty} \frac{\log \N'}{\sqrt{n}} = \lim_{n \to \infty} \frac{\log \N}{\sqrt{n}} > t(q,\epsilon_d) I_B(p,q).
\end{align*}     
Therefore, we conclude that any code $\C$ that ensures $(1-\epsilon_d)$-covertness and a vanishing average probability of error $P_{\text{err}}$ must satisfy $$\lim_{n \to \infty} \frac{\log \N}{\sqrt{n}} \leq t(q,\epsilon_d) I_B(p,q).$$
This completes the proof of Theorem~\ref{thm:upperbound}.

\section{Proof of theorem~\ref{thm:achievable}}\label{sec:achievability}

When the amount of shared key $\Delta(n) \in (\Omega(\log(n)),o(\sqrt{n}))$ (small-sized key regime\footnote{In fact, as we shall see, $\Delta(n) = 6\log(n)$ suffices.}), Theorem~\ref{thm:achievable} indicates that the optimal throughput $t(q,\epsilon_d) I_B(p,q)$ is achievable as long as $I_J(q)<I_B(p,q)$, and this is the main focus of this section. We introduce our coding scheme in Subsection~\ref{sec:code}, and sketch the proofs of reliability and covertness in Subsections~\ref{sec:reliability} and~\ref{sec:covertness}, respectively. After proving the above achievability results for small-sized key regime, it is then relatively straightforward to extend the achievability results to moderate-sized key regime and large-sized key regime, and we discuss such extensions in detail in Subsection~\ref{sec:large}.

Moreover, we also provide the detailed proofs of several technical lemmas (Lemmas~\ref{lemma:error1}-\ref{lemma:error3}, which are important for proving reliability) in Subsections~\ref{sec:reliability1}-\ref{sec:reliability3}, respectively. 

\subsection{Coding scheme} \label{sec:code}
\noindent{\textbf{Polynomial hash function:}} Let $\Delta(n) = 6\log(n)$. Alice and Bob partition the $6\log(n)$ bits of shared key $K$ into two equal parts, $K_1$ and $K_2$, and each part of the key contains $3\log(n)$ bits. Let
\begin{align}
L \triangleq n^3,
\end{align}
and both $K_1$ and $K_2$ can be viewed as elements of finite field $\mathbb{F}_{L}$. Let $l \triangleq r\sqrt{n}/(3\log(n))$. The message $M$ is partitioned into $3\log(n)$ sized small chunks $M_1, M_2, \ldots, M_l$. Likewise, each message chunk $M_i$ is also viewed as an element of $\mathbb{F}_{L}$. Alice uses the message $M$ and the shared key $K = (K_1,K_2)$ to compute a hash $G$ based on the {\it polynomial hash function}, which is defined as  
\begin{align}
G = G_K(M) \triangleq K_2 + \sum_{u=1}^l K_1^{u} M_{u}, \label{eq:hash}
\end{align}
where the additions and the multiplications are operated over $\mathbb{F}_{L}$. Note that this usage of shared key is distinct from the manner in which shared key is used in a wiretap secrecy setting~\cite{yamamoto1997rate, merhav2008shannon, kang2010wiretap},~\cite[Chapter 22.2]{el2011network}. In particular, in a wiretap secrecy setting, it is highly unlikely that a single codeword could correspond to many message-key pairs, while in our constructions, each codeword corresponds to multiple different message-key pairs. This property is critical since it ensures part of the shared key ($K_1$ in this work) is uniformly distributed from James' perspective even if he gains some information from his received vector, and this uniformity is critical in the list decoding argument.

\noindent{\textbf{Codebook generation:}} Let the relative throughput $r = t(q,\epsilon_d)I_B(p,q) - \delta$ (where $\delta > 0$ can be chosen arbitrarily small). For each message-hash pair $(i,j) \in \{0,1\}^{r\sqrt{n}}\times \{0,1\}^{3\log(n)}$, we generate a length-$n$ codeword $\x_{ij}$ according to $P_{\X} \triangleq \prod_{i=1}^n P_X$, where $P_X$ is a Bern$(\rho)$ distribution with 
\begin{align}
\rho \triangleq \frac{t(q,\epsilon_d)}{\sqrt{n}} =  \frac{2 \sqrt{q (1-q)}}{(1-2q)\sqrt{n}}\cdot Q^{-1}\left(\frac{1-\epsilon_d}{2}\right).
\end{align}
For different message-hash pairs, the codewords are generated independently. The codebook is a collection of $\x_{ij}, \forall (i,j)$.

\noindent{\textbf{Encoder:}} To encode a message $M=i$, Alice uses the shared key $K=k$ and the polynomial hash function to compute a hash $j = G_k(i)$. She then transmits the codeword $\x_{ij}$ to Bob. 
\begin{claim} \label{claim:remark}
For any message $M = i$ and hash $G=j$, the number of shared key $K=(K_1,K_2)$ that is consistent with $(i,j)$ equals $L = n^3$, i.e., 
\begin{align}
\sum_{k \in (\mathbb{F}_{L})^2} \mathbbm{1}\left\{j=G_k(i)\right\} = L. \label{eq:qinaide}
\end{align}
\end{claim}

As is common in the literature, we assume the message $M$ and the shared key $K$ are uniformly distributed, and $M$ and $K$ are generated independently. These assumptions together with~\eqref{eq:qinaide} imply that each of the codeword $\x_{ij}$ is equally likely to be transmitted, since
\begin{align}
\mathbb{P}(M=i,G=j) &= \sum_{k \in (\mathbb{F}_{L})^2} \mathbb{P}(M=i,K=k,G=j) \\
&= \sum_{k \in (\mathbb{F}_{L})^2} \mathbb{P}(M=i) \mathbb{P}(K=k) \mathbb{P}(G=j|M=i,K=k) \\
&=\frac{1}{N} \frac{1}{L^2}  \sum_{k \in (\mathbb{F}_{L})^2} \mathbbm{1}\left\{j=G_k(i)\right\} \\
&=\frac{1}{NL}.
\end{align} 
\qed

\noindent{\textbf{Decoding rule:}} Given a received vector $\y$, the list decoder $\mathcal{L}(\y)$ contains all the codewords satisfying the following constraints: 
\begin{align}
\mathcal{L}(\y) \triangleq \left\{ \x: 
\begin{array}{ll}
nf^{xy}_{10}(\x,\y) < \rho n\left(\frac{p(1-q)}{q}\right)(1+\varepsilon_1) \\
nf^{xy}_{11}(\x,\y) > \rho n\left(1-\frac{p(1-q)}{q}\right)(1-\varepsilon_2)	
\end{array}
\right\}. \label{eq:list_decoder}
\end{align}
In this work we set $\varepsilon_1 = \frac{1}{\log(n)}$ and $\varepsilon_2 =\frac{p-pq}{(q-p+pq)\log(n)}$, and explain the reason for such choices in Subsection~\ref{sec:reliability}. The decoding rule is as follows:
\begin{enumerate}
	\item Output all the codewords satisfying the list decoding rule~\eqref{eq:list_decoder} to $\mathcal{L}(\y)$;
	\item Decode $\widehat{M} = i$ if $\x_{ij}$ (for some $j$) is the unique codeword in $\mathcal{L}(\y)$ that is consistent with the shared key $k$ (i.e., $j = G_k(i)$). Decode $\widehat{M} =0$ if no codeword in $\mathcal{L}(\y)$ is consistent with $k$. Declare an error otherwise. 
\end{enumerate}

\noindent{\textbf{Decoding error events:}} When Alice is active $(T = 1)$, we suppose $M = i, K=k, G=j=G_k(i)$ without loss of generality. The decoding error $\mathcal{E}_1$ occurs if the transmitted codeword $\x_{ij}$ is not the unique codeword in the list $\mathcal{L}(\y)$ that is consistent with the shared key $k$, i.e.,
\begin{align}
\mathcal{E}_1:\big\{ \{\x_{ij} \notin \mathcal{L}(\y)\} \text{ or } \{ \exists (i',j')\ne(i,j): \x_{i'j'} \in \mathcal{L}(\y) \text{ and } j' = G_k(i') \} \big\}.
\end{align}
When Alice is silent $(T = 0)$, the decoding error $\mathcal{E}_0$ occurs if there exists a codeword $\x_{ij} \in \mathcal{L}(\y)$ such that $\x_{ij}$ is consistent with the shared key $k$, i.e., 
\begin{align}
\mathcal{E}_0: \big\{ \exists (i,j): \x_{ij} \in \mathcal{L}(\y) \text{ and } j = G_k(i) \big\}.
\end{align}

\subsection{Proof sketch of reliability} \label{sec:reliability}
\begin{figure}
	\begin{center}
		\includegraphics[scale=0.33]{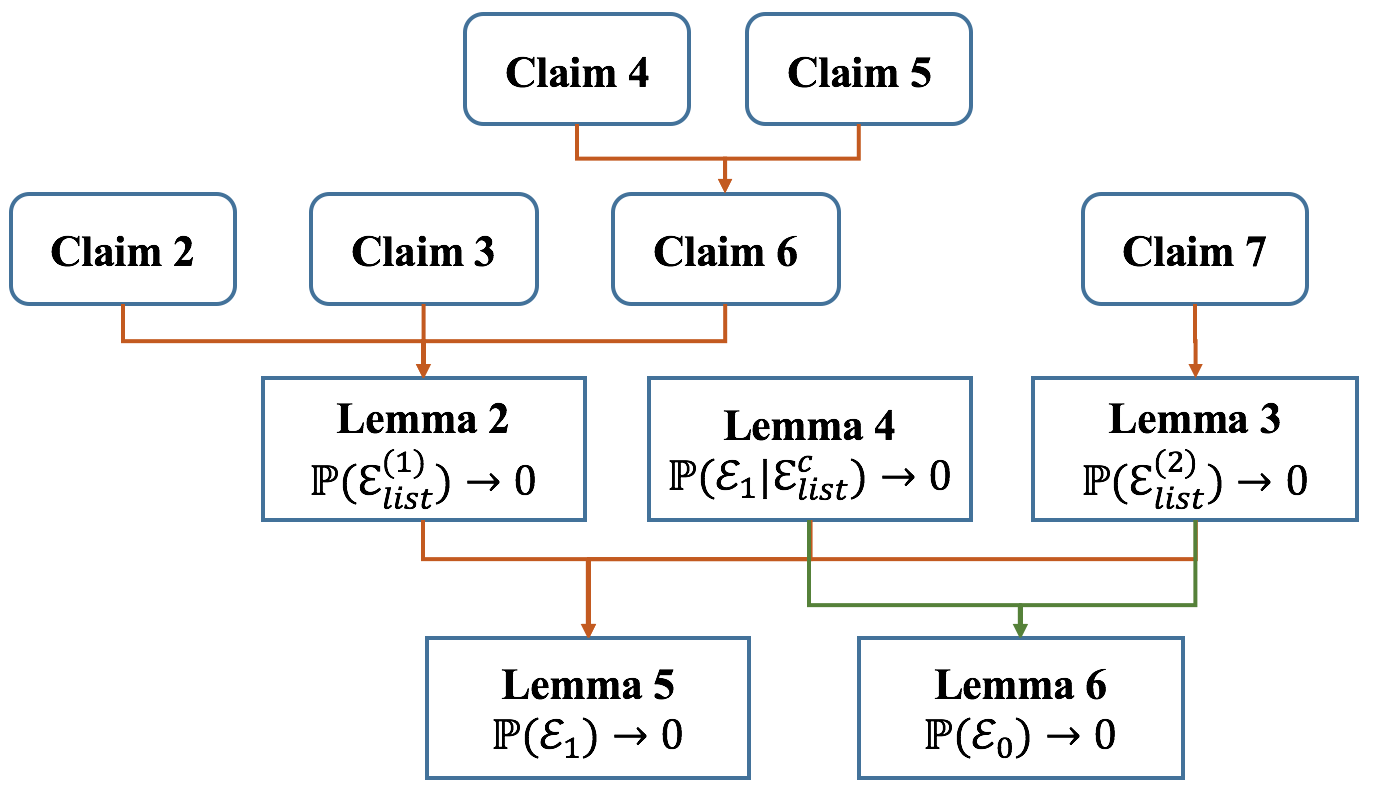}
		\caption{A road-map for the proof of reliability. Claims~\ref{claim:atypical1}-\ref{claim:n2} are presented in Subsections~\ref{sec:reliability1} and~\ref{sec:reliability2}.} \label{fig:map}
	\end{center}
\end{figure}

This subsection provides a proof sketch of reliability of our scheme. For readers' convenience, we also illustrate the road-map of our proof in Fig.~\ref{fig:map}.
\subsubsection{Transmission status $T = 1$} Let $\mathcal{E}_\text{list}$ be the error event corresponding to the list decoder, which occurs if one of the following two events occurs: 
\begin{itemize}
\item $\mathcal{E}_\text{list}^{(1)}$: the transmitted codeword $\x_{ij}$ does not belong to $\mathcal{L}(\y)$;
\vspace{2pt}
\item $\mathcal{E}_\text{list}^{(2)}$: the number of codewords $\x_{i'j'}$ (for $(i',j') \ne (i,j)$) falling into $\mathcal{L}(\y)$ is at least $n^2$.
\end{itemize}
Generally speaking, we hope that the list decoder contains the correct codeword $\x_{ij}$, and also keep the list size as small as possible (no larger than $n^2$). 
Lemmas~\ref{lemma:error1} and~\ref{lemma:error2} below respectively show that with high probability over the code design, a randomly chosen code $\C$ ensures that the probabilities of error events $\mathcal{E}_\text{list}^{(1)}$ and $\mathcal{E}_\text{list}^{(2)}$ go to zero as $n$ goes to infinity.
\begin{restatable}{lemma}{errorone}
	\label{lemma:error1}
	With probability at least $1 - \exp(-\mathcal{O}(n^{1/4}))$ over the code design, a randomly chosen code $\C$ ensures 
	\begin{align}
	\PP(\mathcal{E}_{\emph{list}}^{(1)}) \le 3\exp(-n^{1/8}). \notag
	\end{align}
\end{restatable}

\begin{restatable}{lemma}{errortwo}
	\label{lemma:error2}
	With probability at least $1 - \exp(-\mathcal{O}(\sqrt{n}))$ over the code design, a randomly chosen code $\C$ ensures 
	\begin{align}
	\PP(\mathcal{E}_{\emph{list}}^{(2)}) \le \exp(-n^{1/4}). \notag
	\end{align}
\end{restatable}

The proofs of Lemmas~\ref{lemma:error1} and~\ref{lemma:error2} are respectively provided in Subsections~\ref{sec:reliability1} and~\ref{sec:reliability2}. 
Combining Lemmas~\ref{lemma:error1} and~\ref{lemma:error2}, we have
\begin{align}
\PP(\mathcal{E}_\text{list}) = \PP(\mathcal{E}_\text{list}^{(1)} \cup \mathcal{E}_\text{list}^{(2)}) \le \PP(\mathcal{E}_\text{list}^{(1)}) + \PP(\mathcal{E}_\text{list}^{(2)}) \le 3\exp(-n^{1/8}) + \exp(-n^{1/4}). \label{eq:list}
\end{align}
Secondly, even if the list decoder does not make an error, one still need to worry the situation in which more than one codeword in $\mathcal{L}(\y)$ is consistent with the shared key $k$. In particular, the transmitted codeword $\x_{ij}$ is consistent with $k$ since $j = G_k(i)$ by the definition of the encoder, so we hope none of the other codewords $\x_{i'j'} \ne \x_{ij}$ are consistent with $k$. We denote the complement of $\mathcal{E}_{\text{list}}$ by $\mathcal{E}_{\text{list}}^c$, which means that the transmitted codeword $\x_{ij} \in \mathcal{L}(\y)$, and the number of codewords (other than $\x_{ij}$) falling into $\mathcal{L}(\y)$ is bounded from above by $n^2$. Lemma~\ref{lemma:error3} below shows that as long as the list decoder is ``well-behaved'' (i.e., $\mathcal{E}_{\text{list}}$ does not occur), the probability of decoding error $\PP(\mathcal{E}_1|\mathcal{E}_{\text{list}}^c)$ will be negligible. 

\begin{restatable}{lemma}{errorthree}
	\label{lemma:error3}
	Conditioned on $\mathcal{E}_{\text{list}}^c$, the error event $\mathcal{E}_1$ occurs with probability (over the shared key $K$) at most $\mathcal{O}\left(1/\sqrt{n}\log(n)\right)$.
\end{restatable}
We provide the detailed proofs of Lemmas~\ref{lemma:error1}-\ref{lemma:error3} in Subsections~\ref{sec:reliability1}-\ref{sec:reliability3}. Combining Lemmas~\ref{lemma:error1}-\ref{lemma:error3} and inequality~\eqref{eq:list}, we have the following lemma.
\begin{lemma} \label{lemma:e1}
	When Alice is active ($T = 1$), with probability at least $1 - \exp(-\mathcal{O}(n^{1/4}))$ over the code design, a randomly chosen code $\C$ ensures a vanishing probability of decoding error, i.e.,
	$$\PP(\mathcal{E}_1) = \max_{W_{\mathbf{S}|\Z,\C}}
	\PP(\widehat{M} \neq M | T = 1) \le \mathcal{O}\left(1/\sqrt{n}\log(n)\right).$$
\end{lemma}

\noindent{\it Proof:} By the total probability theorem, we have
\begin{align}
\PP(\mathcal{E}_1) &=  \PP(\mathcal{E}_\text{list})\PP(\mathcal{E}_1|\mathcal{E}_\text{list}) + \PP(\mathcal{E}^c_\text{list})\PP(\mathcal{E}_1|\mathcal{E}^c_\text{list}) \\
&\le \PP(\mathcal{E}_\text{list}) + \PP(\mathcal{E}_1|\mathcal{E}^c_\text{list}) \\
&\le 3\exp(-n^{1/8}) + \exp(-n^{1/4}) + \mathcal{O}\left(1/\sqrt{n}\log(n)\right) \\
&= \mathcal{O}\left(1/\sqrt{n}\log(n)\right). 
\end{align}
\qed

\subsubsection{Transmission status $T = 0$}
We provide an upper bound on the probability of error $\mathbb{P}(\mathcal{E}_0)$ as follows. 
\begin{lemma} \label{lemma:e0}
	With probability at least $1 - \exp(-\mathcal{O}(\sqrt{n}))$ over the code design, a randomly chosen code $\C$ ensures a vanishing probability of decoding error $\mathbb{P}(\mathcal{E}_0)$, i.e.,
	$$\mathbb{P}(\mathcal{E}_0) = \max_{W_{\mathbf{S}|\Z,\C}}
	\mathbb{P} (\widehat{M} \neq 0 | \T = 0) \le \mathcal{O}\left(1/\sqrt{n}\log(n)\right).$$
\end{lemma} 
The proof of Lemma~\ref{lemma:e0} is similar to that of Lemma~\ref{lemma:e1}. When $\T = 0$, no codeword is transmitted by Alice, and the list decoder makes an error $\mathcal{E}_{\text{list}}$ if and only if more than $n^2$ codewords falls into the list. Similar to Lemma~\ref{lemma:error2}, we argue that with probability at least $1 - \exp(-\mathcal{O}(\sqrt{n})$ over the code design, a randomly chosen code $\C$ ensures a vanishing probability of list-decoding error. This can be proved by simply reusing the proof of Lemma~\ref{lemma:error2} in Subsection~\ref{sec:reliability2}, by noting that a length-$n$ zero vector can be view as a typical codeword, as defined in~\eqref{eq:typical_x}. Secondly, conditioned on $\mathcal{E}_{\text{list}}^c$, the probability (over the shared key) that more than one codeword in the list satisfies the polynomial hash function is at most $\mathcal{O}\left(1/\sqrt{n}\log(n)\right)$. This completes the proof sketch of Lemma~\ref{lemma:e0}.
Therefore,
\begin{align}
P_{\text{err}} \le \mathbb{P}(\mathcal{E}_0) + \mathbb{P}(\mathcal{E}_1) \le \mathcal{O}\left(1/\sqrt{n}\log(n)\right).
\end{align}

\subsection{Proof of covertness} \label{sec:covertness}
Note that the proof of covertness directly follows from prior work on covert communication over probabilistic channels, since James' observation $\Z$ (which is used to estimate Alice's transmission status) depends only on the probabilistic wiretap channel BSC($q$), and is independent of the adversarial jamming stucture. Hence, we only provide a high-level proof sketch, and refer the interested readers to~\cite{CheBJ:13,7407378,7447769,TahmasbiB17} for detailed proofs. 

The proof of covertness essentially connects to the analysis of the distributions of James' channel outputs $\Z$. Let $Q_0(\Z)$ be the $n$-letter innocent distribution of James' channel output $\Z$ when Alice is silent ($\T = 0$), and $Q_1(\Z)$ be the $n$-letter active distribution of James' channel output $\Z$ when Alice is transmitting ($\T = 1$). A standard statistical arguments~\cite{lehmann2006testing} shows that the optimal estimator $\widehat{\Phi}$ satisfies $P_{\text{FA}}(\widehat{\Phi}) + P_{\text{MD}}(\widehat{\Phi}) = 1 - \mathbb{V}(Q_0(\Z), Q_1(\Z))$, where $\mathbb{V}(Q_0(\Z), Q_1(\Z))$ is the {\it variational distance} between the two distributions. Therefore, to prove $(1-\epsilon_d)$-covertness, it suffice to show 
\begin{align}
\limsup_{n \to \infty} \mathbb{V}(Q_0(\Z), Q_1(\Z))  \le \epsilon_d.
\end{align}
The $n$-letter innocent distribution $Q_0(\Z)$ is a Binomial$(n,q)$ distribution, with 
\begin{align}
Q_0(\z) \triangleq W_{\Z|\X}(\z|\mathbf{0}) =  q^{\text{wt}_H(\z)}(1-q)^{(n-\text{wt}_H(\z))},\ \forall \z \in \{0,1\}^n.
\end{align}
The $n$-letter active distribution $Q_1(\Z)$ depends on the specific codebook, and is given by
\begin{align}
Q_1(\z) \triangleq \sum_{i=1}^N \sum_{j=1}^L \frac{1}{NL} W_{\Z|\X}(\z|\x_{ij}), \ \forall \z\in \{0,1\}^n. \label{eq:p1}
\end{align}
For the purpose of analysis, we also define an \emph{$n$-letter ensemble-averaged active distribution} $\mathbb{E}_{\C}\left(Q_1(\Z)\right)$, which is essentially the active distribution $Q_1(\Z)$ averaged over all the possible codebooks, as 
\begin{align}
\mathbb{E}_{\C}\left(Q_1(\z)\right) \triangleq \mathbb{E}_{\C}\left(\sum_{i=1}^N \sum_{j=1}^L \frac{1}{NL} W_{\Z|\X}(\z|\x_{ij})\right) &= \sum_{i=1}^N \sum_{j=1}^L \frac{1}{NL} \sum_{\x_{ij} \in \{0,1\}^n} P_{\X}(\x_{ij}) W_{\Z|\X}(\z|\x_{ij}) \\
&=\sum_{\x \in \{0,1\}^n} P_{\X}(\x) W_{\Z|\X}(\z|\x).
\end{align}

To prove $\mathbb{V}(Q_0, Q_1) \le \epsilon_d$, we first note that $\mathbb{V}(Q_0, Q_1)$ is no larger than $\mathbb{V}(Q_0, \mathbb{E}_{\C}(Q_1)) + \mathbb{V}(\mathbb{E}_{\C}(Q_1),  Q_1)$ by the {\it triangle inequality}, and then bound the two terms from above separately.
\begin{itemize}
	\item Following the lead of~\cite{tahmasbi2017second} (based on the {\it Berry-Esseen theorem}), it has been proved that by setting the code-weight parameter $t(q,\epsilon_d)  = \frac{2 \sqrt{q (1-q)}}{1-2q}\cdot Q^{-1}\left(\frac{1-\epsilon_d}{2}\right)$, 
	$\limsup_{n \to \infty} \mathbb{V}(Q_0, \mathbb{E}_{\C}(Q_1)) \le \epsilon_d.$
	\item As long as $r\sqrt{n} + \Delta(n)$ (the normalized size of the code) is greater than $t(q,\epsilon_d) I_J(q)\sqrt{n}$ (the mutual information from Alice to James), with high probability over the code design, the output distribution $Q_1$ induced by the randomly chosen code $\C$ is indistinguishable from the ensemble-averaged active distribution $\mathbb{E}_{\C}(Q_1)$, i.e., $\limsup_{n \to \infty} \mathbb{V}(\mathbb{E}_{\C}(Q_1),  Q_1) = 0$. This result was discovered independently by~\cite{CheBJ:13} based on the type class decompositions, and by~\cite{7407378} based on the {\it channel resolvability}.
\end{itemize}

For any values of $(p,q)$ such that $I_B(p,q) > I_J(q) > 0$, the coding scheme described above then ensures $(1-\epsilon_d)$-covertness, since the relative throughput $r = t(q,\epsilon_d) I_B(p,q) - \delta$ and $\delta > 0$ can be chosen arbitrarily small. This completes the proof sketch of covertness, as well as the proof of the achievability result for small-sized key regime in Theorem~\ref{thm:achievable}.

\subsection{Achievability scheme with moderate-sized and large-sized key} \label{sec:large}

We now provide a modified coding scheme when the amount of shared key is moderate or large. First, let $\Delta(n) = \sigma \sqrt{n} + 6\log (n)$ for some constant $\sigma > 0$ (which asymptotically equals $\sigma \sqrt{n}$ when $n$ is sufficiently large). Alice and Bob generate a public code $\C$ that contains $2^{\sigma \sqrt{n}}$ sub-codes, and each sub-code (containing $r\sqrt{n}$ message bits and $6\log(n)$ bits of shared key) is generated independently according to the codebook generation process described in Subsection~\ref{sec:code}. Again, the relative throughput $r = t(q,\epsilon_d)I_B(p,q) - \delta$ for some arbitrarily small $\delta > 0$. The extra $\sigma \sqrt{n}$ bits of shared key is used for Alice and Bob to select which sub-code to use during transmission, and the selected one is kept secret from James. It is worth noting that each sub-code also contains $6\log(n)$ bits of shared key and it is critical for list decoding. 

From Bob's perspective, the size of the selected sub-code is small enough so that he can reliably decode (the proof follows from Subsection~\ref{sec:reliability}). From James' perspective, the size of the public code is sufficiently large, since he does not know the shared key and the sub-code used by Alice and Bob. In particular, the normalized size of the public code roughly equals $$(r+\sigma)\sqrt{n} = (I_B(p,q)t(q,\epsilon_d)+\sigma)\sqrt{n},$$ which is greater than $t(q,\epsilon_d) I_J(q)\sqrt{n}$ (the criterion for achieving covertness provided in Subsection~\ref{sec:covertness}) as long as $I_B(p,q) + \frac{\sigma}{t(q,\epsilon_d)} > I_J(q)$. This implies the achievability result for the moderated-sized key regime in Theorem~\ref{thm:achievable}. 

Further, we note that in the regime $\Delta(n) \in \omega(\sqrt{n})$ (large-sized key regime), the criterion for achieving covertness is always satisfied since $\sigma = \omega(1)$ is larger than any constant. Therefore, the optimal throughput $t(q,\epsilon_d)I_B(p,q)$ is achievable for any values of $(p,q)$ such that $p < q$.

\subsection{Definitions of typical sets and type classes}\label{sec:def} 

The proof of Lemmas~\ref{lemma:error1} and~\ref{lemma:error2}  relies critically on the {\it type class decompositions}, hence we first define the concepts of typical sets and type classes as follows. The fractional Hamming weight of $\x$ and $\z$, and the fraction of pair $(u,v)$ in $(\x, \z)$ (where $u,v \in \{0,1\}$), are respectively denoted by 
\begin{align}
f^x_1(\x) \triangleq \frac{\text{wt}_H(\x)}{n}, \ \ f^z_1(\z) \triangleq \frac{\text{wt}_H(\z)}{n}, \ \ f^{xz}_{uv}(\x,\z) \triangleq \frac{\left|i \in \{1,\ldots,n\}:(x_i,z_i)=(u,v)\right|}{n}.
\end{align}
The $n${\it -letter typical set} of $\X$ and $\Z$ are respectively defined as 
\begin{align}
&\ax \triangleq \left\{ \x \in \{0,1\}^n: f^x_1(\x) \le 2 \rho \right\}, \label{eq:typical_x}\\
&\az \triangleq \left\{ \z \in \{0,1\}^n: (\rho * q) (1 - n^{-1/4}) \le f^z_1(\z) \le (\rho * q) (1 + n^{-1/4})  \right\}. \label{eq:typical_z}
\end{align}
Given a fixed $\z$, the $n${\it -letter conditional type class of $\X$} (with type $(f^{xz}_{10}, f^{xz}_{11})$) is defined as 
\begin{align}
\txz \triangleq \left\{ \x\in \{0,1\}^{n}: 
\begin{array}{ll}
\big|i:(x_i,z_i)=(1,0)\big| = n f^{xz}_{10} \\
\big|i:(x_i,z_i)=(1,1)\big| = n f^{xz}_{11}
\end{array}
\right\},
\end{align}
and the $n${\it -letter conditionally typical set of $\X$} is defined as 
\begin{align}
\axz \triangleq \left\{ \x \in \{0,1\}^{n}: 
\begin{array}{ll}
\rho q(1 - n^{-1/8}) \le f^{xz}_{10}(\x,\z) \le \rho q(1 + n^{-1/8}) \\
\rho (1-q)(1 - n^{-1/8}) \le f^{xz}_{11}(\x,\z) \le \rho (1-q)(1 + n^{-1/8})	
\end{array}
\right\}, \label{eq:typical_xz}
\end{align}
Note that the conditionally typical set can be represented as the union of typical conditional type classes, i.e.,
\begin{align}
\axz = \bigcup_{(f^{xz}_{10}, f^{xz}_{11})\in \mathcal{F}^{xz}_{n}} \txz,
\end{align}
where $\mathcal{F}^{xz}_{n}$ is the set of typical fractional Hamming weight, and is given by 
\begin{align}
\mathcal{F}^{xz}_{n} \triangleq \left\{ (f^{xz}_{10}, f^{xz}_{11}) :
\begin{array}{ll}
\rho q(1 - n^{-1/8}) \le f^{xz}_{10}(\x,\z) \le \rho q(1 + n^{-1/8}) \\
\rho (1-q)(1 - n^{-1/8}) \le f^{xz}_{11}(\x,\z) \le \rho (1-q)(1 + n^{-1/8})	 \\
nf^{xz}_{10} \in \mathbb{Z}^{\ast}, \ nf^{xz}_{11} \in \mathbb{Z}^{\ast}
\end{array}
\right\}. 
\end{align}

\noindent{\bf Oracle argument:} Before stating the formal proof, we first introduce the {\it oracle argument} that is frequently used in the myopic adversarial setting. When Alice transmits a codeword $\x_{ij}$ and James receives a vector $\z$, the only knowledge that James has is the received vector $\z$ and the public code $\C$. We now assume that there is an oracle which helps James by revealing the type class $\tau = \txz$ that the transmitted codeword $\x_{ij}$ lies in. Note that this extra information $\tau$ strengthens James in the sense that it reduces James' uncertainty about which codeword is transmitted by Alice (since only the codewords in $\tau$ are likely to be the transmitted codewords). If our coding scheme is proven to be reliable against the stronger adversary, it will also succeed against the original adversary. With the extra information $\tau$, James' jamming strategy may depend on the received vector $\z$, the public code $\C$, as well as $\tau$. Hence, in the following analysis, we denote James' jamming function by the $n$-letter conditional distribution $W_{\mathbf{S}|\Z,\C,\tau}$, instead of $W_{\mathbf{S}|\Z,\C}$ defined in Section~\ref{sec:model}. The main purpose of introducing an oracle is to simplify the analysis, and such oracle is by no means necessary for analysis.  Note that James has the flexibility to optimize his jamming function $W_{\mathbf{S}|\Z,\C,\tau}$; however, our proofs show that with high probability, a randomly generated code guarantees a small probability of error regardless of James' choices of $W_{\mathbf{S}|\Z,\C,\tau}$.

\subsection{Proof of Lemma~\ref{lemma:error1}}\label{sec:reliability1} 

Recall that the error event $\mathcal{E}_\text{list}^{(1)}$ occurs if the transmitted codeword $\x_{ij}$ does not belong to $\mathcal{L}(\y)$. For a fixed code $\C$, the probability of $\mathcal{E}_\text{list}^{(1)}$ is given by 
\begin{align}
\mathbb{P}(\mathcal{E}_\text{list}^{(1)}) &= \max_{W_{\mathbf{S}|\Z,\C,\tau}}\left\{ \sum_{i=1}^N \sum_{j=1}^L \frac{1}{NL} \sum_{\z}W_{\Z|\X}(\z|\x_{ij}) \sum_{\s} W_{\mathbf{S}|\Z,\C, \tau}(\s|\z,\C,\tau)\cdot  \mathbbm{1}\{\x_{ij} \notin \mathcal{L}(\x_{ij}+\s)\} \right\} \notag \\
&\le \max_{W_{\mathbf{S}|\Z,\C,\tau}}\left\{\frac{1}{NL}\sum_{\z \in \az} \sum_{(i,j): \x_{ij} \in \axz}  W_{\Z|\X}(\z|\x_{ij})  \sum_{\s} W_{\mathbf{S}|\Z,\C,\tau}(\s|\z,\C,\tau) \cdot  \mathbbm{1}\{\x_{ij} \notin \mathcal{L}(\x_{ij}+\s)\}\right\} \notag \\
& + \max_{W_{\mathbf{S}|\Z,\C,\tau}}\left\{ \frac{1}{NL}\sum_{\z \in \az} \sum_{(i,j): \x_{ij} \notin \axz} W_{\Z|\X}(\z|\x_{ij})  \sum_{\s} W_{\mathbf{S}|\Z,\C,\tau}(\s|\z,\C,\tau)\cdot  \mathbbm{1}\{\x_{ij} \notin \mathcal{L}(\x_{ij}+\s)\} \right\}   \notag \\
&+\max_{W_{\mathbf{S}|\Z,\C,\tau}}\left\{ \frac{1}{NL} \sum_{i=1}^N \sum_{j=1}^L \sum_{\z \notin \az} W_{\Z|\X}(\z|\x_{ij})  \sum_{\s}W_{\mathbf{S}|\Z,\C,\tau}(\s|\z,\C,\tau)\cdot  \mathbbm{1}\{\x_{ij} \notin \mathcal{L}(\x_{ij}+\s)\} \right\}. \label{eq:xi1}
\end{align}
In~\eqref{eq:xi1}, we partition James' received vector $\z$ into typical and atypical sets; for any typical $\z$, we further partition all the codewords into conditionally typical and atypical sets. Since the indicator function $\mathbbm{1}(\cdot)$ is always upper-bounded by one, the two atypical terms in~\eqref{eq:xi1} can be respectively upper-bounded as 
\begin{align}
\frac{1}{\N L}\sum_{\z \in \az} \sum_{(i,j): \x_{ij} \notin \axz}  W_{\Z|\X}(\z|\x_{ij}) + \frac{1}{\N L} \sum_{i=1}^{\N} \sum_{j=1}^L \sum_{\z \notin \az} W_{\Z|\X}(\z|\x_{ij}). \label{eq:vanish2}
\end{align}
The following two claims state that the probabilities of error caused by the two atypical events are vanishing, and the detailed proofs are deferred to Appendix~\ref{appendix:reliability}.

\begin{restatable}{claim}{claimfour}
 \label{claim:atypical1}
With probability at least $1 - \exp(-\mathcal{O}(n^{1/4}))$ over the code design, 
$$\frac{1}{NL}\sum_{\z \in \az} \sum_{(i,j): \x_{ij} \notin \axz}  W_{\Z|\X}(\z|\x_{ij}) < \exp(-n^{1/8}).$$
\end{restatable} 

\begin{restatable}{claim}{claimfive}
 \label{claim:atypical2}
With probability at least $1 - \exp(-\mathcal{O}(\sqrt{n}))$ over the code design, 
$$\frac{1}{NL} \sum_{i=1}^N \sum_{j=1}^L \sum_{\z \notin \az} W_{\Z|\X}(\z|\x_{ij}) <\exp(-n^{1/4}).$$
\end{restatable}

From now on we consider the the typical event in~\eqref{eq:xi1} --- a typical $\z$ is received and a conditionally typical codeword is transmitted. One critical step in our proof is to decompose the conditionally typical set $\axz$ into the conditionally typical type classes $\txz$ (where $(f^{xz}_{10}, f^{xz}_{11})\in \mathcal{F}^{xz}_{n}$) that comprise it. Let 
\begin{align}
c \triangleq r-t(q,\epsilon_d)\cdot I_J(q) > 0.
\end{align} 
Claim~\ref{claim:ratio1} below shows that for any typical $\z$ and conditionally typical type class $\txz$, with probability super-exponentially\footnote{Note that this super-exponential concentration result is critical since we need to take a union bound over exponentially many typical $\z$ and type classes $\txz$.} close to one (over the code design), the number of codewords falling into $\txz$ is tightly concentrated around $2^{c\sqrt{n}}$.  

\begin{claim} \label{claim:ratio1}
For any typical $\z$ and any conditionally typical type class $\txz$, the expected number of codewords falling into $\txz$ is super-polynomially large, i.e., 
$$\mathbb{E}_{\C}\left(\sum_{i=1}^N \sum_{j=1}^L \mathbbm{1}\{\x_{ij} \in \txz \} \right) = 2^{c\sqrt{n}}.$$ 
Further, with probability at least $1-\exp(-2^{\mathcal{O}(\sqrt{n})})$ over the code design, a randomly chosen code $\C$ satisfies
\begin{align*}
&\sum_{i=1}^N \sum_{j=1}^L \mathbbm{1}\{\x_{ij} \in \txz \} > \left(1-\exp(-n^{\frac{1}{4}})\right) \cdot 2^{c\sqrt{n}}. \notag
\end{align*}
\end{claim}
The first part of Claim~\ref{claim:ratio1} is due to the fact that the relative throughput $r = t(q,\epsilon_d)I_B(p,q) - \delta$ is larger than the normalized mutual information $t(q,\epsilon_d)I_J(q)$ of the BSC($q$) from Alice to James, while the second part follows from the Chernoff bound. Thus, Claim~\ref{claim:ratio1} relies critically on the fact that the channel from Alice to James is sufficiently noisy. We provide the detailed proof of Claim~\ref{claim:ratio1} in Appendix~\ref{appendix:ratio1}. 
\begin{remark}
With the help of the oracle revealed information $\tau =\txz$, James knows the transmitted codeword must belong to $\txz$, but each of the codeword in $\txz$ is equally likely from his perspective. Thus, the number of codewords in $\txz$ essentially reflects James' uncertainty about the transmitted codeword. It is critical in our proof that James' uncertainty should be large enough, and this is exactly what Claim~\ref{claim:ratio1} shows. 
\end{remark}

\begin{figure}
	\begin{center}
	\includegraphics[scale=0.5]{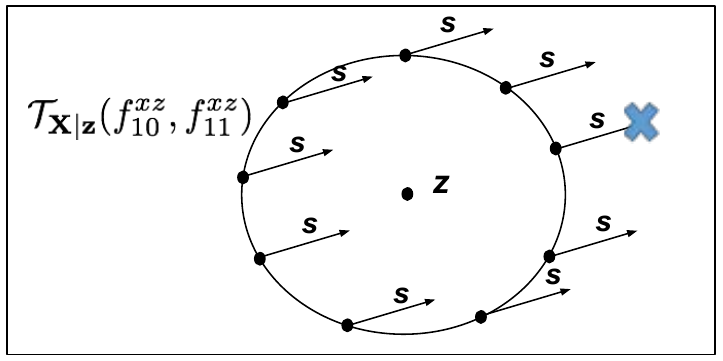}
	\caption{We consider a typical $\z$ and a conditionally typical type class $\txz$ with respect to $\z$. We prove that the number of codewords falling into $\txz$ is super-polynomially large, and no matter which jamming vector $\s$ is chosen, only a small fraction of codewords that belong to $\txz$ are killed.} \label{fig:kill}
	\end{center}
\end{figure}

We say a codeword $\x$ is \emph{killed} by a jamming vector $\s$ if $\x$ is pushed out of the list decoder by $\s$, i.e., $\x \notin \mathcal{L}(\x+\s)$. 
If magically James is able to find a jamming vector $\s$ such that each of the codeword in $\txz$ is killed by $\s$, then the jamming vector must result in a decoding error --- this is because the true transmitted codeword $\x$ belongs to $\txz$ and is also killed by $\s$. Fortunately, Claim~\ref{claim:ratio2} below shows that for typical $\z$ and $\txz$, with probability super-exponentially close to one (over the code design), no matter which $\s \in \{0,1\}^n$ James chooses, only a decaying fraction of codewords in $\txz$ are killed by $\s$ (as illustrated in Fig.~\ref{fig:kill}).

\begin{claim}[{\bf Myopic list-decoding lemma}] \label{claim:ratio2}
For any typical $\z$ and any conditionally typical type class $\txz$, with probability at least $1 - \exp\left(-2^{\mathcal{O}(\sqrt{n})}\right)$  over the code design,
$$\sum_{i=1}^N \sum_{j=1}^L \mathbbm{1}\left\{\left[\x_{ij} \in \txz\right] \cap \left[\x_{ij} \notin \mathcal{L}(\x+\s)\right] \right\} < \exp(-n^{1/4}) \cdot 2^{c\sqrt{n}}, \ \ \forall \s \in \{0,1\}^n.$$
\end{claim}

The proof of Claim~\ref{claim:ratio2} is deferred to Appendix~\ref{appendix:ratio2}. Based on Claims~\ref{claim:ratio1} and~\ref{claim:ratio2}, we obtain Claim~\ref{claim:ratio3} which shows that for typical $\z$ and $\txz$, a decaying fraction of codewords $\x \in \txz$ being killed (regardless of $\s$) implies a vanishing probability of error. Finally, we also needs to take a union bound over all typical $\z$ and conditionally typical type class $\txz$. 

\begin{claim}[First term in~\eqref{eq:xi1}] \label{claim:ratio3}
With probability at least $1-\exp(-2^{\mathcal{O}(\sqrt{n})})$ over the code design, a randomly chosen code $\C$ satisfies
\begin{align*}
&\max_{W_{\mathbf{S}|\Z,\C,\tau}}\left\{ \frac{1}{NL}\sum_{\z \in \az} \sum_{(i,j): \x_{ij} \in \axz}  W_{\Z|\X}(\z|\x_{ij})  \sum_{\s} W_{\mathbf{S}|\Z,\C,\tau}(\s|\z,\C,\tau)\cdot  \mathbbm{1}\{\x_{ij} \notin \mathcal{L}(\x_{ij}+\s)\}\right\} \le \exp(-n^{\frac{1}{4}}+1).
\end{align*}
\end{claim}
\noindent{{\it Proof:}} For any typical $\z$ and conditionally typical type class $\txz$, Claim~\ref{claim:ratio1} and Claim~\ref{claim:ratio2} guarantee that a randomly chosen code $\C$ satisfies 
\begin{align}
\frac{\sum_{i=1}^N \sum_{j=1}^L \mathbbm{1}\left\{\left[\x_{ij} \in \txz\right] \cap \left[\x_{ij} \notin \mathcal{L}(\x+\s)\right] \right\}}{\sum_{i=1}^N \sum_{j=1}^L \mathbbm{1}\{\x_{ij} \in \txz \}} &< \frac{\exp(-n^{1/4})\cdot 2^{c\sqrt{n}}}{(1-\exp(-n^{1/4})) \cdot 2^{c\sqrt{n}}} \label{eq:live}\\
&\le \exp(-n^{1/4}+1), \ \ \forall \s \in \{0,1\}^n,
\end{align}
with probability at least $1-\exp(-2^{\mathcal{O}(\sqrt{n})})$ over the code design. We now turn to analyze the first term in~\eqref{eq:xi1}. For any jamming strategy $W_{\mathbf{S}|\Z,\C,\tau}$, we have 

\begin{align}
&\frac{1}{NL}\sum_{\z \in \az} \sum_{(i,j): \x_{ij} \in \axz}  W_{\Z|\X}(\z|\x_{ij})  \sum_{\s} W_{\mathbf{S}|\Z,\C,\tau}(\s|\z,\C,\tau)\cdot  \mathbbm{1}\{\x_{ij} \notin \mathcal{L}(\x_{ij}+\s)\} \label{eq:start}\\
&=\frac{1}{NL} \sum_{\z \in \az} \sum_{(\fa, \fb) \in \mathcal{F}_n^{xz}} \ \sum_{(i,j): \x_{ij} \in \txz} W_{\Z|\X}(\z|\x_{ij})  \sum_{\s} W_{\mathbf{S}|\Z,\C,\tau}(\s|\z,\C,\tau)\cdot  \mathbbm{1}\{\x_{ij} \notin \mathcal{L}(\x_{ij}+\s)\} \label{eq:start2} \\
&= \frac{1}{NL}\sum_{\z \in \az} \sum_{(\fa, \fb) \in \mathcal{F}_n^{xz}} W_{\Z|\X}(\z|\txz)  \sum_{\s} W_{\mathbf{S}|\Z,\C,\tau}(\s|\z,\C,\tau) \notag \\
& \qquad\qquad\qquad\qquad\qquad\qquad\qquad\qquad\qquad  \sum_{i=1}^N \sum_{j=1}^L \mathbbm{1}\left\{\left[\x_{ij} \in \txz\right] \cap \left[\x_{ij} \notin \mathcal{L}(\x+\s)\right] \right\} \\
&\overset{\text{w.h.p.}}\le{} \frac{\exp(-n^{\frac{1}{4}}+1)}{NL} \sum_{\z \in \az} \sum_{(\fa, \fb) \in \mathcal{F}_n^{xz}} W_{\Z|\X}(\z|\txz) \sum_{\s} W_{\mathbf{S}|\Z,\C,\tau}(\s|\z,\C,\tau) \notag \\
&\qquad\qquad\qquad\qquad\qquad\qquad\qquad\qquad\qquad   \sum_{i=1}^N \sum_{j=1}^L \mathbbm{1}\left\{\x_{ij} \in \txz \right\} \label{eq:tutorial} \\
&= \exp(-n^{\frac{1}{4}}+1) \cdot \frac{1}{NL} \sum_{\z \in \az} \sum_{(\fa, \fb) \in \mathcal{F}_n^{xz}} \ \sum_{(i,j):\x_{ij} \in \txz} W_{\Z|\X}(\z|\x_{ij}) \label{eq:tutorial2} \\
&\le \exp(-n^{\frac{1}{4}}+1) \cdot \frac{1}{NL} \sum_{\z} \sum_{i=1}^N \sum_{j=1}^L W_{\Z|\X}(\z|\x_{ij}) \label{eq:tutorial3} \\
&= \exp(-n^{\frac{1}{4}}+1). \label{eq:error1-1}
\end{align}
In~\eqref{eq:start2}, we decompose the conditionally typical set $\axz$ into the union of all conditionally typical type classes $\txz$.
Inequality~\eqref{eq:tutorial} follows from~\eqref{eq:live}, and holds with probability at least $1-\exp(-2^{\mathcal{O}(\sqrt{n})})$ over the code design. Note that in~\eqref{eq:tutorial} we need to take a union bound over exponentially many $\z, \s$ and $(\fa, \fb)$, which is valid since $1-\exp(-2^{\mathcal{O}(\sqrt{n})})$ is super-exponentially large. Equation~\eqref{eq:tutorial2} follows since $\sum_{s}W_{\mathbf{S}|\Z,\C,\tau}(\s|\z,\C,\tau) = 1$, and inequality~\eqref{eq:tutorial3} is obtained by relaxing the constraints on $\z$. Note that equations~\eqref{eq:start}-\eqref{eq:error1-1} holds for arbitrary jamming strategy $W_{\mathbf{S}|\Z,\C,\tau}$, hence Claim~\ref{claim:ratio3} is proved. \qed 

By combining Claims~\ref{claim:atypical1},~\ref{claim:atypical2}, and~\ref{claim:ratio3}, we finally prove that with probability at least $1 - \exp(-\mathcal{O}(n^{1/4}))$ over the code design, a randomly chosen code $\C$ ensures the probability of the error event $\mathcal{E}_\text{list}^{(1)}$ is bounded from above as 
\begin{align}
\mathbb{P}(\mathcal{E}_\text{list}^{(1)}) \le \exp(-n^{1/8}) + \exp(-n^{1/4}) + \exp(-n^{\frac{1}{4}}+1) \le 3\exp(-n^{1/8}). 
\end{align}
This completes the proof of Lemma~\ref{lemma:error1}.

\subsection{Proof of Lemma~\ref{lemma:error2}} \label{sec:reliability2}

Recall that the error event $\mathcal{E}_\text{list}^{(2)}$ occurs if more than $n^2$ codewords (other than the transmitted codeword) fall into the list $\mathcal{L}(\y)$. 

\begin{claim} \label{claim:n2}
Fix a typical transmitted codeword $\x_{ij}$ and a jamming vector $\s$ satisfying $\emph{wt}_H(\s) \le pn$. With probability at least $1- \exp(-\mathcal{O}(n^{5/2}))$ over the code design, the number of codewords $\x_{i'j'}$ (where $(i',j') \ne (i,j)$) falling into the list $\mathcal{L}(\x_{ij}+\s)$ is bounded from above by $n^2$.  
\end{claim}
\noindent{\it Proof:} The Hamming weight of Bob's received vector $\y = \x_{ij} + \s$ satisfies
\begin{align}
\text{wt}_H(\y) = \text{wt}_H\left(\x_{ij} + \s\right) \le \text{wt}_H(\x_{ij}) + \text{wt}_H(\s) \le 2\rho n + pn,
\end{align} 
since $\text{wt}_H(\s) \le pn$, $\text{wt}_H(\x_{ij}) \le 2\rho n$ for typical $\x_{ij}$, and the intersection between $\x_{ij}$ and $\s$ is greater than zero. For any codeword $\x_{i'j'}$ such that $(i',j') \ne (i,j)$, $\x_{i'j'} \in \mathcal{L}(\y)$ if and only if 
\begin{align}
\begin{cases}
nf^{xy}_{10}(\x_{i'j'}, \y) < \rho n\left(\frac{p(1-q)}{q}\right)(1+\varepsilon_1),\\
nf^{xy}_{11}(\x_{i'j'},\y) > \rho n\left(1-\frac{p(1-q)}{q}\right)(1-\varepsilon_2).
\end{cases}
\end{align}
Note that the complement of the support of $\y$ has size greater than $(1-p)n - 2\rho n$, hence we have 
$$\mathbb{E}\left(nf^{xy}_{10}(\X_{i'j'}, \y)\right) \ge \rho \left((1-p)n - 2\rho n\right),$$
since each bit of $\X_{i'j'}$ is generated i.i.d. according to Bern$(\rho)$. Let $\kappa_1 = 1 - \frac{(p-pq)(1+\varepsilon_1)}{q-pq+2\rho q}$. By the {\it Chernoff–Hoeffding Theorem}~\cite{hoeffding1994probability}, we have 
\begin{align}
&\mathbb{P} \left(nf^{xy}_{10}(\X_{i'j'}, \y) < \rho n\left(\frac{p(1-q)}{q}\right)(1+\varepsilon_1) \right)  \\
&= \mathbb{P} \left(nf^{xy}_{10}(\X_{i'j'}, \y) < (1-\kappa_1)\mathbb{E}_{\C}\left(nf^{xy}_{10}(\X_{i'j'}, \y)\right) \right) \\
&\le 2^{-\mathbb{D}(\rho(1-\kappa_1) \Vert \rho)((1-p)n-2\rho n) \log e}. \label{eq:kappa1}
\end{align} 
Similarly, since $\text{wt}_H(\y) \le 2\rho n + pn$, we have  
$$\mathbb{E}\left(nf^{xy}_{11}(\X_{i'j'}, \y)\right) \le \rho \left(2\rho n + pn\right).$$
Let $\kappa_2 = \frac{(q-p+pq)(1-\varepsilon_2)}{q(2\rho + p)} - 1$. By the Chernoff–Hoeffding Theorem, we have 
\begin{align}
&\mathbb{P} \left(nf^{xy}_{11}(\X_{i'j'}, \y) > \rho n\left(1 - \frac{p(1-q)}{q}\right)(1-\varepsilon_2) \right)  \\
&= \mathbb{P} \left(nf^{xy}_{11}(\X_{i'j'}, \y) > (1+\kappa_2)\mathbb{E}_{\C}\left(nf^{xy}_{11}(\X_{i'j'}, \y)\right) \right) \\
&\le 2^{-\mathbb{D}(\rho(1+\kappa_2) \parallel \rho)(pn+2\rho n) \log e}. \label{eq:kappa2}
\end{align}  
Combining inequalities~\eqref{eq:kappa1} and~\eqref{eq:kappa2}, we have
\begin{align}
&\mathbb{P} \left(\X_{i'j'} \in \mathcal{L}(\y) \right) \notag \\
&= \mathbb{P} \left(nf^{xy}_{10}(\X_{i'j'}, \y) < \rho n\left(\frac{p(1-q)}{q}\right)(1+\varepsilon_1) \right) \cdot \mathbb{P}_{\X_{i'j'}} \left(nf^{xy}_{11}(\X_{i'j'}, \y) > \rho n\left(1 - \frac{p(1-q)}{q}\right)(1-\varepsilon_2) \right) \notag \\
&\le 2^{-\mathbb{D}(\rho(1-\kappa_1) \parallel \rho)((1-p)n-2\rho n) \log e} \cdot 2^{-\mathbb{D}(\rho(1+\kappa_2) \parallel \rho)(pn+2\rho n) \log e} \\
& \overset{\text{$n \to \infty$}}={} 2^{-t(q,\epsilon_d) I_B(p,q)\sqrt{n}},
\end{align}
since the two events (the number of ones of $\X_{i'j'}$ inside the support of $\y$ and outside the support of $\y$) are independent. On expectation, the total number of codewords (other than the transmitted codeword $\x_{ij}$) falling into the list $\mathcal{L}(\y)$ is given by
\begin{align}
\mathbb{E}\left(\sum_{(i'j')\ne(i,j)} \mathbbm{1}\left\{\X_{i'j'} \in \mathcal{L}(\y)\right\} \right) \le  2^{r\sqrt{n} + 3\log n} \cdot 2^{-t(q,\epsilon_d)I_B(p,q)\sqrt{n}} = 2^{(r-t(q,\epsilon_d) I_B(p,q))\sqrt{n} + 3\log n},
\end{align}
which is super-polynomially small since $r < t(q,\epsilon_d) I_B(p,q)$. 
Therefore, we use a counting argument to characterize the probability that more than $n^2$ codewords falling into the list $\mathcal{L}(\y)$. As long as $r < t(q,\epsilon_d) I_B(p,q)$, we have 
\begin{align}
&\PP_{\C \setminus \x_{ij}} \left(\sum_{(i'j')\ne(i,j)} \mathbbm{1}\left\{\X_{i'j'} \in \mathcal{L}(\y)\right\} \ge n^2 \right) \\
& = \sum_{\theta=n^2}^{2^{r\sqrt{n}}} \mathbb{P}_{\C \setminus \x_{ij}} \left(\sum_{(i'j')\ne(i,j)} \mathbbm{1}\left\{\X_{i'j'} \in \mathcal{L}(\y)\right\} = \theta \right) \label{eq:eye1}\\
& = \sum_{\theta=n^2}^{2^{r\sqrt{n}}} \binom{2^{r\sqrt{n}}}{\theta}\left(2^{-t(q,\epsilon_d) I_B(p,q)\sqrt{n}}\right)^\theta \left(1- 2^{-t(q,\epsilon_d) I_B(p,q)\sqrt{n}}\right)^{(2^{r\sqrt{n}}-\theta)} \\
&\le 2^{r\sqrt{n}} \binom{2^{r\sqrt{n}}}{n^2}\left(2^{-t(q,\epsilon_d) I_B(p,q)\sqrt{n}}\right)^{n^2} \label{eq:eye2} \\
& \le 2^{r\sqrt{n}} \left(\frac{e\cdot 2^{r\sqrt{n}}}{n^2}\right)^{n^2}\left(2^{-t(q,\epsilon_d) I_B(p,q)\sqrt{n}}\right)^{n^2} \label{eq:eye3}\\
& = 2^{r\sqrt{n}}\left(\frac{e\cdot 2^{(r-t(q,\epsilon_d) I_B(p,q))\sqrt{n}}}{n^2}\right)^{n^2} \\
&= \exp(-\mathcal{O}(n^{5/2})). \label{eq:eye4}
\end{align}
Inequality~\eqref{eq:eye2} follows since $\theta=n^2$ maximizes the probability in~\eqref{eq:eye1}, and we bound the number of summations from above by $2^{r\sqrt{n}}$. Inequality~\eqref{eq:eye3} follows from the inequality 
\begin{align}
\binom{n}{k} \le \left(\frac{en}{k}\right)^k.
\end{align} 
Finally, we obtain~\eqref{eq:eye4} by using the fact that $r < t(q,\epsilon_d) I_B(p,q)$.
\qed

In the following we also consider the atypical events, and prove that with high probability over the code design, a randomly chosen code $\C$ ensures that the probability of the error events $\mathcal{E}_\text{list}^{(2)}$ goes to zero. Note that
\begin{align}
\mathbb{P}(\mathcal{E}_\text{list}^{(2)}) = \max_{W_{\mathbf{S}|\Z,\C,\tau}}\left\{ \frac{1}{NL}\sum_{i=1}^N \sum_{j=1}^L \sum_{\z}W_{\Z|\X}(\z|\x_{ij}) \sum_{\s}W_{\mathbf{S}|\Z,\C,\tau}(\s|\z, \C,\tau) \cdot \mathbbm{1}\bigg\{\sum_{(i',j')\ne(i,j)}\mathbbm{1}\left\{\x_{i'j'} \in \mathcal{L}(\y) \right\} \ge n^2\bigg\} \right\} \notag,
\end{align}
and regardless of James' jamming strategy $W_{\mathbf{S}|\Z,\C,\tau}$, 
\begin{align}
&\mathbb{E}_{\C}\left[ \frac{1}{NL}\sum_{i=1}^N \sum_{j=1}^L \sum_{\z}W_{\Z|\X}(\z|\x_{ij}) \sum_{\s}W_{\mathbf{S}|\Z,\C,\tau}(\s|\z, \C,\tau) \cdot \mathbbm{1}\bigg\{\sum_{(i',j')\ne(i,j)}\mathbbm{1}\left\{\x_{i'j'} \in \mathcal{L}(\y) \right\} \ge n^2\bigg\} \right] \\
&=\mathbb{E}_{\C}\left[ \sum_{\z}W_{\Z|\X}(\z|\x_{11}) \sum_{\s}W_{\mathbf{S}|\Z,\C,\tau}(\s|\z, \C,\tau) \cdot \mathbbm{1}\bigg\{\sum_{(i',j')\ne(1,1)}\mathbbm{1}\left\{\x_{i'j'} \in \mathcal{L}(\y) \right\} \ge n^2\bigg\} \right] \label{eq:atl1} \\
&= \sum_{\x_{11} \in \{0,1\}^n} P_{\X}(\x_{11}) \sum_{\C \setminus \x_{11}} P_{\C \setminus \X_{11}}(\C \setminus \x_{11}) \sum_{\z}W_{\Z|\X}(\z|\x_{11}) \notag \\
&\qquad\qquad\qquad\qquad\qquad\qquad\qquad\qquad  \sum_{\s}W_{\mathbf{S}|\Z,\C,\tau}(\s|\z, \C,\tau) \cdot \mathbbm{1}\bigg\{\sum_{(i',j')\ne(1,1)}\mathbbm{1}\left\{\x_{i'j'} \in \mathcal{L}(\y) \right\} \ge n^2\bigg\}\label{eq:jide1} \\
& \le \sum_{\x_{11} \in \mathcal{A}_{\X}}  P_{\X}(\x_{11}) \sum_{\s} \sum_{\C \setminus \x_{11}}P_{\C \setminus \X_{11}}(\C \setminus \x_{11}) \cdot \mathbbm{1}\bigg\{\sum_{(i',j')\ne(1,1)}\mathbbm{1}\left\{\x_{i'j'} \in \mathcal{L}(\y) \right\} \ge n^2\bigg\} + \sum_{\x_{11} \notin \mathcal{A}_{\X}}  P_{\X}(\x_{11})  \label{eq:jide3}\\
& \le \sum_{\x_{11} \in \mathcal{A}_{\X}}  P_{\X}(\x_{11}) \sum_{\s} P_{\C \setminus \x_{11}} \left(\sum_{(i'j')\ne(1,1)} \mathbbm{1}\left\{\X_{i'j'} \in \mathcal{L}(\y)\right\} \ge n^2 \right) + \sum_{\x_{11} \notin \mathcal{A}_{\X}}  P_{\X}(\x_{11})\\
& \le \left(\sum_{\s} \exp(-\mathcal{O}(n^{5/2}))\right) + \exp \left(-\frac{1}{3} t(q,\epsilon_d) \sqrt{n}\right)  \label{eq:jide4} \\
& = \exp(-\mathcal{O}(\sqrt{n})). \label{eq:atl2}
\end{align}
Equation~\eqref{eq:atl1} is obtained by noting that for each codeword $\x_{ij}$, the averaged probability of error (over the code design) is the same. Hence, without loss of generality, we consider the average probability of error corresponding to the codeword $\x_{11}$ being transmitted. The notation $P_{\C \setminus \X_{11}}(\C \setminus \x_{11})$ in~\eqref{eq:jide1} represents the probability of generating a code $\C$ excluding the transmitted codeword $\x_{11}$. In~\eqref{eq:jide3}, we again consider the transmitted codeword $\x_{11}$ to be either typical or atypical, and 
\begin{itemize}
	\item When $\x_{11}$ is atypical, we simply bound the indicator function $\mathbbm{1}\{\sum_{(i',j')\ne(1,1)}\mathbbm{1}\{\x_{i'j'} \in \mathcal{L}(\y)\} \ge n^2 \}$ from above by one. 
	\item When $\x_{11}$ is typical, we bound the probability $W_{\mathbf{S}|\Z,\C,\tau}(\s|\z,\C,\tau)$ from above by one, and then interchange the order of summations. Note that if we keep the term $W_{\mathbf{S}|\Z,\C,\tau}(\s|\z,\C,\tau)$, the order of summations cannot be changed since James' jamming strategy $W_{\mathbf{S}|\Z,\C,\tau}(\s|\z,\C,\tau)$ depends on the realization of the code $\C$.
\end{itemize}
Inequality~\eqref{eq:jide4} follows from Claim~\ref{claim:n2} (which is valid for all typical transmitted codewords) and the Chernoff bound.
Finally, by noting that~\eqref{eq:atl2} holds for every possible jamming strategy $W_{\mathbf{S}|\Z,\C,\tau}$, the Markov's inequality yields 
\begin{align}
\mathbb{P}_{\C}\left( \mathbb{P}(\mathcal{E}_\text{list}^{(2)}) \ge \exp(-n^{1/4})\right)  \le \exp(-\mathcal{O}(\sqrt{n})).
\end{align}
This completes the proof of Lemma~\ref{lemma:error2}.
\qed

\subsection{Proof of Lemma~\ref{lemma:error3}} \label{sec:reliability3}

If $\mathcal{E}_\text{list}$ does not occur, the transmitted codeword $\x_{ij}$ belong to the list $\mathcal{L}(\y)$, and the number of codewords (other than $\x_{ij}$) falling into $\mathcal{L}(\y)$ is bounded from above by $n^2$. Let 
\begin{align}
V \triangleq \{\x_{i',j'}: (i',j')\ne (i,j) \text{ and } \x_{i'j'} \in \mathcal{L}(\y) \}
\end{align}
be the set of codewords that belong to $\mathcal{L}(\y)$, where $|V| \le n^2$. In the following, we show that with high probability (over the shared key $K$), none of the codewords in $V$ is consistent with $K$.

Recall that the polynomial hash function, first defined in~\eqref{eq:hash}, is given by 
\begin{align}
G = G_K(M) \triangleq K_2 + \sum_{u=1}^{w} K_1^{u} M_{u},
\end{align}
where the additions and multiplications are over $\mathbb{F}_{n^3}$, and $w \triangleq  \frac{r\sqrt{n}}{3\log(n)}$.
Though the shared key $K$ is {\it a priori} uniformly distributed, it may not necessarily be uniform from James' perspective, since his observations $\z$ may reveal some information about $K$. Nevertheless, we argue that the first part of the key, $K_1$, is still uniformly distributed from James' perspective, even if James knows $\z$ as well as the message-hash pair $(i,j)$ transmitted by Alice. Note that the above argument also holds without the extra assumption that the transmitted message-hash pair is revealed, since this assumption only strengthens James, and the purpose of introducing it is merely to simplify the analysis.

As James knows $M = i$ ($M_u = i_u, \forall u \in \{1,2,\ldots,w\}$) and $G = j$, he certainly knows that the shared key $K = (K_1,K_2)$ satisfies 
\begin{align}
j =  K_2 + \sum_{u=1}^w K_1^{u} i_{u}. \label{eq:85}
\end{align} 
For each value of $K_1 \in \mathbb{F}_{n^3}$, there exists a unique $K_2$ such that the $(K_1,K_2)$ pair satisfies equation~\eqref{eq:85}. Saying differently, the total number of $(K_1,K_2)$ pairs satisfying equation~\eqref{eq:85} is $n^3$, and each pair contains a distinct $K_1$. Thus, from James' perspective, $K_1$ is uniformly distributed, while $K_2$ may or may not be uniformly distributed.  
For any $(i', j') \ne (i,j)$, by Schwartz–Zippel lemma and the uniformity of $K_1$, the probability that $(i', j')$ is consistent with $K$ is given by 
\begin{align}
&\PP_{K} \left( j' = K_2 + \sum_{u=1}^{w}K_1^{u}i'_u \Big| j = K_2 + \sum_{u=1}^{w}K_1^{u}i_u\right) = \PP_{K} \left( j'-j= \sum_{u=1}^{w}K_1^{u}(i'_u-i_u) \right)  \le \frac{w}{n^3} = \frac{rn^{-5/2}}{3\log(n)}.
\end{align}
By taking a union bound over all the codewords in $V$ (where $|V| \le n^2$), one can prove that with probability at least 
\begin{align}
1 - n^2 \cdot \frac{rn^{-5/2}}{3\log(n)} = 1- \mathcal{O}\left(\frac{1}{\sqrt{n}\log(n)}\right)
\end{align}
over $K$, none of the codewords in $|V|$ is consistent with $K$.\qed

\begin{remark}
Note that the error probability essentially depends on the amount of shared key between Alice and Bob, and there is a fundamental tradeoff between shared key and error probability. As our goal is to use minimum amount of shared key to ensure reliable communication, we show that $6\log(n)$ bits of shared key suffices, while the error probability is relatively high (of order $\mathcal{O}(1/\sqrt{n} \log n)$). In contrast, if our goal were to minimize the error probability to achieve a better performance in the finite blocklength regime, we would need a larger amount of shared key. By a close inspection of the proof above, we have that if the amount of shared key were $(6+\chi)\log(n)$ bits for some $\chi > 0$, the error probability would scale as $\mathcal{O}(1/n^{\frac{1+\chi}{2}}\log(n))$ and thus decay faster.
\end{remark}

\section{Permutation-based concatenated codes} \label{sec:efficient}
Recall that covert communication requires the average Hamming weights of codewords to be at most $t(q,\epsilon_d)\sqrt{n}$. Instead of generating a low-weight code of blocklength $n$ directly, here we first generate a code $\widetilde{\mathcal{C}}$ of blocklength $\mathcal{O}(\sqrt{n})$, and then expand the codewords to length-$n$ transmitted vectors. The code $\widetilde{\mathcal{C}}$ adopted here is a capacity-achieving concantenated code for the binary asymmetric channel (BAC) described in Section~\ref{sec:result_efficient}, and a favorable feature is that it can be encoded and decoded in polynomial time.

\noindent{\underline{Concatenated codes:}} Let $l \triangleq \sqrt{n}/\log n$ and $d \triangleq t(q,\epsilon_d)/\rho^*$, where $t(q,\epsilon_d)$ is  defined in~\eqref{eq:q} and $\text{Bern}(\rho^*)$ is the capacity-achieving input distribution of the BAC. The blocklength of the code $\widetilde{\mathcal{C}}$ is $d\sqrt{n}$, and the rate is $R = C_{\text{BAC}}(p,q) - \epsilon$, where $\epsilon > 0$ can be made arbitrarily small. Each message $M = m$ contains $dR\sqrt{n}$ bits, and can further be partitioned into $l$ equal-sized chunks $[m^{(1)}, m^{(2)}, \ldots, m^{(l)}]$, where each $m^{(i)}$ ($i \in \{1,2,\ldots, l \}$) contains $dR\log{n}$ bits.  Let $l' \triangleq (1+\frac{1}{\log n})l$.  

The concantenated code $\widetilde{\mathcal{C}}$ consists of an outer code $\widetilde{\mathcal{C}}_{\mathrm{out}}$ and $l'$ inner codes $\{\widetilde{\mathcal{C}}_{\mathrm{in}}^{(1)}, \ldots, \widetilde{\mathcal{C}}_{\mathrm{in}}^{(l')}\}$. 
\begin{itemize}
	\item The outer code $\widetilde{\mathcal{C}}_{\mathrm{out}}$ is chosen to be a $(l',l)$-Reed-Solomon (RS) code over finite field $\mathbb{F}_{2^{dR\log{n}}}$. 
	\item Inner codes $\{\widetilde{\mathcal{C}}_{\mathrm{in}}^{(1)}, \ldots, \widetilde{\mathcal{C}}_{\mathrm{in}}^{(l')}\}$ are generated randomly and independently. The $i$-th inner code $\widetilde{\mathcal{C}}_{\mathrm{in}}^{(i)}$ contains $2^{dR\log{n}}$ inner-codewords  $\{\x^{(i)}_{1}, \x^{(i)}_{2}, \ldots, \x^{(i)}_{2^{dR\log{n}}}\}$, where each inner-codeword is of length $d\sqrt{n}/l'$ and is generated according to the product distribution of $\text{Bern}(\rho^*)$. The decoder of $\widetilde{\mathcal{C}}_{\mathrm{in}}^{(i)}$ follows from the principle of \emph{typicality decoding} --- it outputs $j$ if the $j$-th codeword $\x^{(i)}_j$ is the unique inner-codeword in $\widetilde{\mathcal{C}}_{\mathrm{in}}^{(i)}$ such that the fractions of $(1,0)$ and $(0,1)$ pairs in $(\x_j^{(i)}, \y^{(i)})$ satisfy $f_{10}^{xy}(\x_j^{(i)}, \y^{(i)}) \le \frac{(1-q)p(1+\epsilon)}{q}$ and $f_{10}^{xy}(\x_j^{(i)}, \y^{(i)}) \le p(1+\epsilon)$ respectively, and it declares an error otherwise. 
\end{itemize}

\begin{figure}
	\begin{center}
		\includegraphics[scale=0.4]{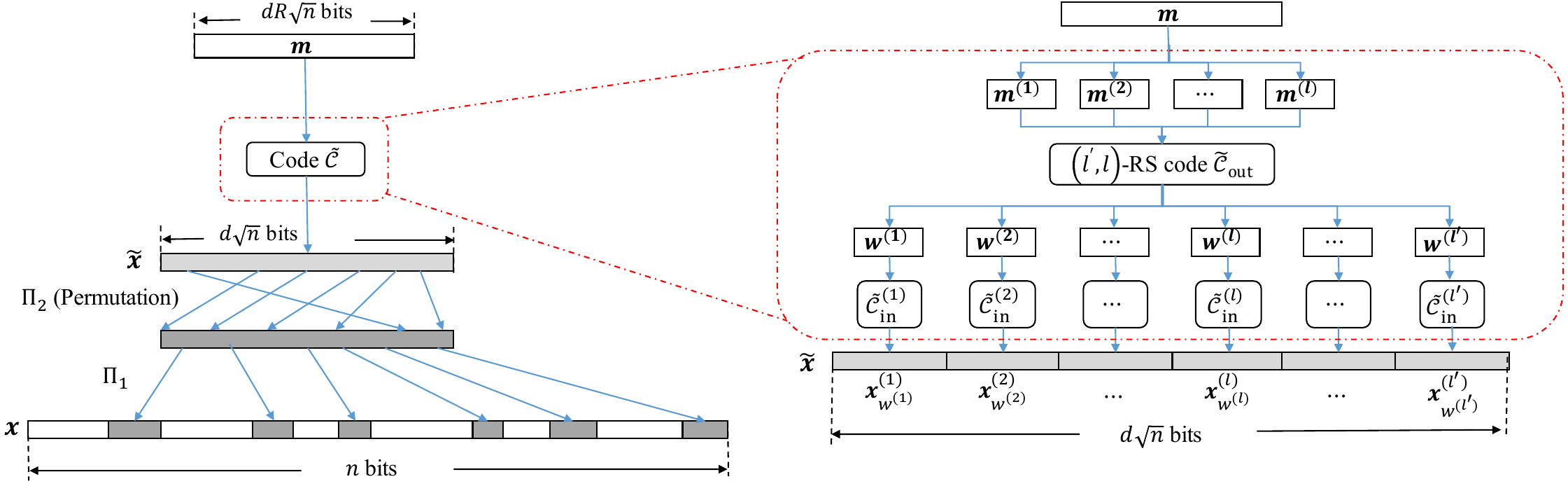}
		\caption{The procedure of encoding a message $m$ to a length-$n$ transmitted vector $\x$.} \label{fig:CC_code}
	\end{center}
\end{figure}

\noindent{\underline{Encoder:}} To send a message $M = m$, Alice adopts the following procedure to produce a length-$n$ transmitted vector $\x$. The encoding procedure is illustrated in Fig.~\ref{fig:CC_code}. 

\begin{itemize}
	\item Alice first uses the outer code $\widetilde{\mathcal{C}}_{\mathrm{out}}$ to encode the partitioned message $m = [m^{(1)}, m^{(2)}, \ldots, m^{(l)}]$ to $l'$ coded-chunks $[w^{(1)}, w^{(2)}, \ldots, w^{(l')}]$, where $w^{(i)}$ is referred to as the \emph{$i$-th inner-message}. 

	\item For $i \in \{1,2,\ldots, l'\}$, Alice uses the inner code $\widetilde{\mathcal{C}}_{\mathrm{in}}^{(i)}$ to encode the inner-message $w^{(i)}$ (where $w^{(i)}$ is assumed to take values in $\{1,2,\ldots, 2^{dR\log{n}} \}$) to the inner-codeword $\x^{(i)}_{w^{(i)}}$. The codeword  $\widetilde{\x} = [\x^{(1)}_{w^{(1)}}, \x^{(2)}_{w^{(2)}}, \ldots, \x^{(l')}_{w^{(l')}}]$ is a concatenation of $l'$ inner-codewords, and is of length $d\sqrt{n}$ bits.

	\item Alice generates a uniformly distributed shared key $\Pi_1$ to select $d\sqrt{n}$ slots (out of $n$ slots) to carry the codeword $\widetilde{\x}$, and generates another uniformly distributed shared key $\Pi_2$ to select a permutation on the $d\sqrt{n}$ slots to permute the codeword $\widetilde{\x}$. 	The length-$n$ transmitted vector $\x$ is obtained by first permuting the $d\sqrt{n}$ bits of the codeword $\widetilde{\x}$, and then inserting the permuted codeword into the $d\sqrt{n}$ selected slots (while all the remaining $n - d\sqrt{n}$ slots comprise entirely of zeros). Note that the key $\Pi_1$ is of length $\log \left(\binom{n}{d\sqrt{n}}\right)= \mathcal{O}(\sqrt{n}\log(n))$ bits, the key $\Pi_2$ is of length $\log\left((d\sqrt{n})!\right) = \mathcal{O}(\sqrt{n}\log(n))$ bits. $\Pi_1$ and $\Pi_2$ are only known to Alice and Bob.
\end{itemize}

\noindent{\underline{Decoder:}} Upon receiving a length-$n$ vector $\y$, Bob adopts the following procedure to reconstruct the message.

\begin{itemize}
	
\item Based on his knowledge of the shared key $\Pi_1$ and $\Pi_2$, Bob first performs inverse operations of permutation and insertion to extract $\widetilde{\y}$ --- the noisy version of the codeword  $\widetilde{\x}$. He then partitions $\widetilde{\y}$ into $l'$ equal-sized chunks $[\y^{(1)}, \y^{(2)}, \ldots, \y^{(l')}]$.

\item  For $i \in \{1,2,\ldots, l'\}$, Bob uses the decoder of the $i$-th inner code $\widetilde{\mathcal{C}}_{\mathrm{in}}^{(i)}$ to obtain an estimate $\widehat{w}^{(i)}$ of the $i$-th inner-message, based on $\y^{(i)}$.

\item Having obtained $[\widehat{w}^{(1)}, \widehat{w}^{(2)}, \ldots, \widehat{w}^{(l')}]$, Bob uses the decoder of the RS code $\widetilde{\mathcal{C}}_{\mathrm{out}}$ to obtain an estimate $\widehat{m} = [\widehat{m}^{(1)}, \widehat{m}^{(2)}, \ldots, \widehat{m}^{(l)}]$ of the transmitted message. 
\end{itemize}	

\noindent{\underline{Encoding and decoding complexities:}} The encoding complexity is dominated by the complexity of the RS encoder, which requires $\mathcal{O}(l'\log(l')\log(|\mathbb{F}_{2^{dR\log n}}|)) = \mathcal{O}(\sqrt{n}\log n)$ binary operations, by applying a \emph{fast Fourier transform}~\cite{preparata1977computational} over the finite field $\mathbb{F}_{2^{dR\log n}}$. The decoding complexity of the best known RS decoder requires $\mathcal{O}((l')^2\log(l')\log(|\mathbb{F}_{2^{dR\log n}}|)) = \mathcal{O}(n)$ binary
operations~\cite{wicker1995error}. For each inner code $\widetilde{\mathcal{C}}_{\mathrm{in}}^{(i)}$, the decoder needs to compare $\y^{(i)}$ with each of the inner-codeword in $\widetilde{\mathcal{C}}_{\mathrm{in}}^{(i)}$, which requires $2^{dR \log n}\times (d\sqrt{n}/l')$ binary operations. Thus, the overall decoding complexity (including the decoders for the RS code and $l'$ inner codes) is $\mathcal{O}\big(\max\{n^{\frac{t(q,\epsilon_d)C_{\text{BAC}}}{\rho^\ast} + \frac{1}{2}}, n \} \big)$.

\noindent{\underline{Analysis:}} First note that James can adopt the myopic jamming strategy (described in Section~\ref{sec:upperbound}) to corrupt the length-$n$ transmitted vector $\X$, i.e., flip $X_i$ with probability approximately $p/q$ if $Z_i=1$, and does not flip $X_i$ otherwise. This jamming strategy roughly flips $p$ fraction of zeros and $(1-q)p/q$ fraction of ones inside the length-$d\sqrt{n}$ codeword $\widetilde{\X}$, hence it effectively induces a BAC on the $d\sqrt{n}$ slots that carry the codeword $\widetilde{\X}$, with bit flip probabilities $W_{Y|X}(1|0)=p$ and $W_{Y|X}(0|1)= \frac{(1-q)p}{q}$.

Next, one can show that the myopic jamming strategy is almost as good as the best jamming strategy that James can adopt --- this is equivalent to saying that the instantiated channel from Alice to Bob (on the selected $d\sqrt{n}$ slots) cannot be significantly noisier than the aforementioned BAC. The rationale behind is that James is unable to gain much information about $\Pi_1$ and $\Pi_2$ based on his observation $\z$, and a careful inspection (based on an oracle argument  similar to that described in Section~\ref{sec:def}) shows that super-polynomially many pairs of $(\Pi_1,\Pi_2)$ are equally likely from James' perspective. Over the randomness of these equally likely pairs of $(\Pi_1,\Pi_2)$, it can be shown that with high probability, no matter which jamming  strategy James uses, he cannot flip more than $p(1+\varepsilon)$ fraction of zeros or more than $(1-q)p(1+\varepsilon)/q$ fraction of ones inside the codeword $\widetilde{\X}$ (where $\varepsilon > 0$ can be made arbitrarily small). This implies that the instantiated channel is at most as noisy as the aforementioned BAC (by ignoring the slackness parameter $\varepsilon$). Therefore, Bob is able to correctly decode the message with high probability since the concatenated code $\widetilde{\mathcal{C}}$ is able to tolerate the noise induced by the BAC.

Finally, note that the average Hamming weight of the length-$n$ transmitted vector $\X$ is $d\sqrt{n}\cdot \rho^{\ast} = t(q,\epsilon_d) \sqrt{n}$, and the support of $\X$ is uniformly distributed over the $n$ transmitted slots (due to the use of $\Pi_1$ and $\Pi_2$). This ensures the covertness of our scheme.


\section{Concluding Remarks} \label{sec:conclusion}
This work considers the problem of covert communication against an active adversary who is able to maliciously jam the communication based on what he eavesdrops. We first show that a shared key of size $\Omega(\log n)$ is necessary for legitimate parties to communicate covertly and reliably, and then provide lower and upper bounds (which match for a wide range of parameter regime) on the covert capacity as a function of the amount of shared key. We also develop a computationally efficient concatenated coding scheme when the amount of shared key available is $\Omega(\sqrt{n} \log(n))$, and further show that this scheme can be implemented with much less amount of shared key when the adversary is assumed to be computationally bounded.

Finally, we put forth two directions that are worthy exploring for future research.

{\it (1) Extensions to general channel models:} One would expect to extend our results for binary channels to a general setting in which the channel from Alice to James is an arbitrary DMC and the channel from Alice to Bob is a general AVC. This is essentially a \emph{myopic AVC problem} with input and state constraints, where ``state'' is the terminology for the jamming vector in the AVC literature, and by ``myopic'' we mean the state is determined by
James as a function of his noisy observation. This problem is conceivably more challenging due to the following reasons. Recall that our achievability scheme relies critically on list-decoding. However, to the best of our konwledge, the list-decoding problem for AVCs with input and state constraints is only partially understood~\cite{sarwate2012list} when the state is chosen by James in an \emph{oblivious} manner (i.e., James has no knowledge of Alice's transmission), and remains open when the state is chosen by James in a \emph{myopic} manner (i.e., James has noisy observations of Alice's transmission, as considered in this work). Thus, the bottleneck in list-decoding makes it challenging to extend to general channel models. On the contrary, any progress on the proposed problem would also be helpful to understand the list-decoding problem for AVCs (which is perhaps more fundamental).

{\it (2) Second-order asymptotics:} As the main objective of this work is to characterize the capacity of covert communication over adversarially jammed channels, our first proof-of-concept coding scheme (presented in Theorem~\ref{thm:achievable}) only guarantees the error probability to be vanishing, without characterizing the speed of decay. Meanwhile, we note that finite blocklength results for some related problems have been derived --- Ref.~\cite{tahmasbi2018first} provided the second-order results for covert communication over DMCs,  while~\cite{kosut2018finite} provided the second-order results for the non-symmetrizable AVC. However, both results (or a combination of these results) do not apply to our setting directly since our adversarial channel model with a stringent covertness constraint can  essentially be regarded a symmetrizable AVC, which differs from the model studied in~\cite{kosut2018finite}. Thus, characterizing the second-order results for our problem may be challenging, but would also be a fruitful endeavour.

\begin{appendices}

\section{Proofs of Claims~\ref{claim:atypical1} and~\ref{claim:atypical2}} \label{appendix:reliability}
\claimfour*
\noindent{\it Proof:} 
First note that 
\begin{align}
&\mathbb{E}_{\C}\left(\frac{1}{NL}\sum_{\z \in \az} \sum_{(i,j):\x_{ij} \notin \axz}  W_{\Z|\X}(\z|\x_{ij})\right) \\
& = \frac{1}{NL} \sum_{i=1}^N \sum_{j=1}^L \E_{\C}\left(\sum_{\z} \pzx(\z|\x_{ij}) \mathbbm{1}\Big\{(\z \in \az) \cap (\x_{ij} \in \axz) \Big\}\right) \\
& = \E_{\C}\left(\sum_{\z} \pzx(\z|\x) \mathbbm{1}\Big\{(\z \in \az) \cap (\x \in \axz) \Big\}\right) \label{eq:ball1}\\
&= \sum_{\x \in\{0,1\}^n} \sum_{\z} P_{\X}(\x) \pzx(\z|\x) \mathbbm{1}\Big\{(\z \in \az) \cap (\x \in \axz) \Big\} \label{eq:cui}\\
&= \PP_{\X\Z}\left(\Z \in \az \cap \X \in \axz \right) \\
&\le \PP_{\X\Z}\left(\X \in \axz \right)  \\
&= \PP_{\X\Z}\left( f_{10}^{xz}(\X,\Z) \notin \rho q(1\pm n^{-\frac{1}{8}}) \text{ or } f_{11}^{xz}(\X,\Z) \notin \rho (1-q)(1\pm n^{-\frac{1}{8}})\right)  \label{eq:ball2}\\
& \le \exp\left(-\frac{q\cdot t(q,\epsilon_d)}{3} n^{\frac{1}{4}}\right) + \exp\left(-\frac{(1-q)\cdot t(q,\epsilon_d)}{3} n^{\frac{1}{4}}\right).\label{eq:ball3}
\end{align}
We simplify the notation in~\eqref{eq:ball1} since the expectation $\E_{\C}(.)$ for each codeword $\x_{ij}$ is exactly the same. Equation~\eqref{eq:ball2} follows from the definition of the conditionally typical set $\axz$, while~\eqref{eq:ball3} is due to the Chernoff bound. Finally, by applying the Markov's inequality, we have 
\begin{align}
\PP_{\C}\left(\frac{1}{NL}\sum_{\z \in \az} \sum_{(i,j): \x_{ij} \notin \axz}  W_{\Z|\X}(\z|\x_{ij}) \ge \exp(-n^{1/8})\right) \le \exp(-\mathcal{O}(n^{1/4})). \label{eq:error1-2}
\end{align}  
\qed

\claimfive*
\noindent{\it Proof:} Note that
\begin{align}
\mathbb{E}_{\C}\left(\frac{1}{NL}\sum_{i=1}^N \sum_{j=1}^L\sum_{\z \notin \az} W_{\Z|\X}(\z|\x_{ij})\right) &= \mathbb{E}_{\C}\left(\sum_{\z \notin \az} W_{\Z|\X}(\z|\x)\right) \\ 
&= \sum_{\z \notin \az} \sum_{\x \in \{0,1\}^n} P_{\X}(\x) W_{\Z|\X}(\z|\x) \\
&= \PP\left(f_1^z(\Z) \notin (\rho *q)\cdot (1\pm n^{-\frac{1}{4}}) \right) \\
&\le \exp\left(-\frac{1}{3}(\rho * q) \sqrt{n}\right),
\end{align}
where the last step is due to the Chernoff bound.
By the Markov's inequality, we have 
\begin{align}
\PP_{\C}\left(\frac{1}{NL}\sum_{i=1}^N \sum_{j=1}^L\sum_{\z \notin \az} W_{\Z|\X}(\z|\x_{ij}) \ge \exp(-n^{1/4}) \right) \le \exp(-\mathcal{O}(\sqrt{n})). \label{eq:error1-3}
\end{align} 
\qed

\section{Proof of Claim~\ref{claim:ratio1}} \label{appendix:ratio1}
As first shown in~\cite{CheBJ:13}, the expected number of codewords falling into the a type class is given by
\begin{align}
&\mathbb{E}_{\C}\left(\sum_{i=1}^N \sum_{j=1}^L \mathbbm{1}\{\x_{ij} \in \txz \} \right) \\
&= \sum_{i=1}^N \sum_{j=1}^L \PP_{\C}\left(\X_{ij} \in \txz \right)  \label{eq:temp1} \\
&= \sum_{i=1}^N \sum_{j=1}^L \binom{n(f^{xz}_{01}+f^{xz}_{11})}{nf^{xz}_{11}} \rho^{nf^{xz}_{11}} (1-\rho)^{nf^{xz}_{01}} \cdot \binom{n(f^{xz}_{00}+f^{xz}_{10})}{nf^{xz}_{10}} \rho^{nf^{xz}_{10}} (1-\rho)^{nf^{xz}_{00}} \\
&=2^{r\sqrt{n} + 3 \log n} \cdot 2^{-n[\mathbb{I}(\x;\z)+ \mathbb{D}(\x \parallel \rho)] - 2\log \left(n+1\right)},
\end{align}
and for any typical $\z$ and any conditionally typical $\x$, i.e., $(f^{xz}_{10}, f^{xz}_{11}) \in \mathcal{F}_n^{xz}$,
\begin{align}
\mathbb{I}(\x;\z) = \rho (1-2q) \log \left(\frac{1-q}{q}\right) + \mathcal{O}(n^{-3/4}), \ \ \mathbb{D}(\x \parallel \rho) = \mathcal{O}(1).
\end{align}
Hence, for sufficiently large $n$, we have
\begin{align}
\mathbb{E}_{\C}\left(\sum_{i=1}^N \sum_{j=1}^L \mathbbm{1}\{\x_{ij} \in \txz \} \right) &\ge 2^{r\sqrt{n}+3\log n} \cdot 2^{-t(q,\epsilon_d)(1-2q)\log((1-q)/q) \sqrt{n}+o(\sqrt{n})} \\
&\overset{\text{$n \to \infty$}}={} 2^{(r-t(q,\epsilon_d)\cdot I_J(q))\sqrt{n}} \label{eq:temp2} \\
& = 2^{c\sqrt{n}}.
\end{align}
Finally, the Chernoff bound ensures that with probability at least $1-\exp(-2^{\mathcal{O}(\sqrt{n})})$ over the code design, a randomly chosen code $\C$ satisfies
\begin{align}
&\sum_{i=1}^N \sum_{j=1}^L \mathbbm{1}\{\x_{ij} \in \txz \} > \left(1-\exp(-n^{\frac{1}{4}})\right) \cdot 2^{c\sqrt{n}}. \notag
\end{align}

\section{Proof of Claim~\ref{claim:ratio2}} \label{appendix:ratio2}
\begin{figure}
  \begin{center}
    \includegraphics[scale=0.45]{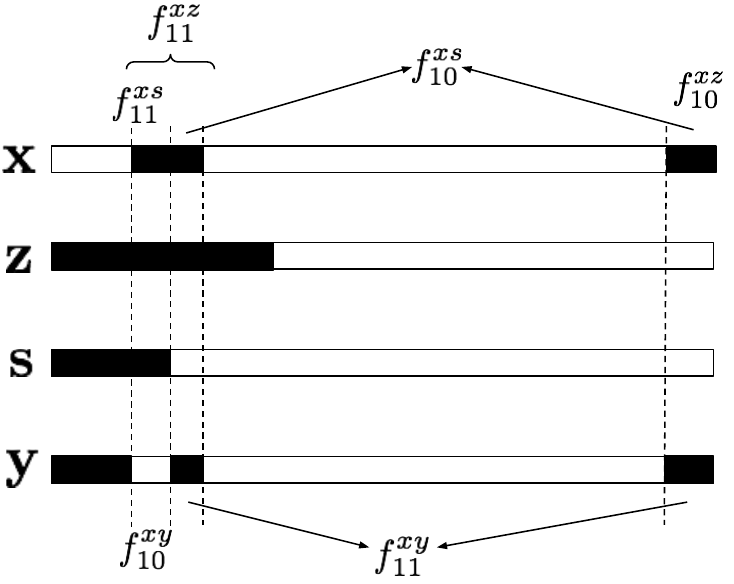}
    \caption{The black region represents ones in the vector while the white region represents zeros in the vector. We denote the joint type classes between $\X$ and $\Z$,$\mathbf{S}$,$\Y$ respectively by $f_{ij}^{xz}$, $f_{ij}^{xs}$, $f_{ij}^{xy}$, for $(i,j) \in \{0,1\} \times \{0,1\}$.} \label{fig:fraction}
  \end{center}
\end{figure}

The key step is to calculate the probability that a randomly generated codeword $\X$ falls into the the type class $\txz$ and is simultaneously killed by a jamming vector $\s$. The probability is maximized when the support of $\s$ is entirely inside the support of $\z$. We now fix a typical $\z$ and a worst-case jamming vector $\s$ satisfying $\big| \text{supp}(\z) \cap \text{supp}(\s)\big| = pn$. By the list decoding rule, a codeword $\x$ is included in the list $\mathcal{L}(\y)$ (or $\mathcal{L}(\x + \s)$) if
\begin{align}
\begin{cases} nf^{xy}_{10}(\x,\y) < \rho n\left(\frac{p(1-q)}{q}\right)(1+\varepsilon_1), \\
nf^{xy}_{11}(\x,\y) > \rho n\left(1-\frac{p(1-q)}{q}\right)(1-\varepsilon_2).
\end{cases} \label{eq:constraint1}
\end{align}
Note that $f^{xy}_{10}(\x,\y) = f^{xs}_{11}(\x,\s)$ and $f^{xy}_{11}(\x,\y) = f^{xs}_{10}(\x,\s)$ (as illustrated in Fig.~\ref{fig:fraction}), hence the constraint in~\eqref{eq:constraint1} is equivalent to 
\begin{align}
\begin{cases} nf^{xs}_{11}(\x,\s) < \rho n\left(\frac{p(1-q)}{q}\right)(1+\varepsilon_1), \\
nf^{xs}_{10}(\x,\s) > \rho n\left(1-\frac{p(1-q)}{q}\right)(1-\varepsilon_2).
\end{cases} \label{eq:constraint2}
\end{align}

\noindent{We further notice that $f^{xs}_{10}(\x,\s) = f^{xz}_{11}(\x,\z) - f^{xs}_{11}(\x,\s) + f^{xz}_{10}(\x,\z)$, and $f^{xz}_{10}(\x,\z)$, $f^{xz}_{11}(\x,\z)$ are tightly concentrated since $\z \in \az$ and $\x \in \axz$. By setting $\varepsilon_1 = \frac{1}{\log(n)}$ and $\varepsilon_2 =\frac{p-pq}{(q-p+pq)\log(n)}$, the constraints in~\eqref{eq:constraint2} is also equivalent to}
\begin{align}
\begin{cases} nf^{xs}_{11}(\x,\s) < \rho n\left(\frac{p(1-q)}{q}\right)(1+\frac{1}{\log(n)}), \\
 nf^{xs}_{11}(\x,\s) < \rho n\left(\frac{p(1-q)}{q}\right)(1+\frac{1}{\log(n)}) + \mathcal{O}(n^{-1/8}).
\end{cases} \label{eq:constraint3}
\end{align}
Without loss of correctness, we ignore the lower order term $\mathcal{O}(n^{-1/8})$ in~\eqref{eq:constraint3} in the following analysis. Let $i_0 = \rho n \left(\frac{p(1-q)}{q}\right)(1+ \frac{1}{\log(n)})$ be the minimum amount of intersections between $\x$ and $\s$ such that $\x$ is killed by $\s$. A codeword $\x$ does not fall into the list $\mathcal{L}(\x+\s)$ if $i_0 \le nf^{xs}_{11}(\x,\s) \le nf^{xz}_{11}$. The probability that a randomly generated codeword $\X$ falls into the type class $\txz$ and does not fall into the list $\mathcal{L}(\X+\s)$ is bounded from above as 
\begin{align}
&\mathbb{P}_{\X}\left(\left[\X \in \txz\right] \cap \left[\X \notin \mathcal{L}(\X+\s)\right] \right) \\
&= \mathbb{P}_{\X}\left(\X \in \txz \right) \cdot \mathbb{P}_{\X} \left(\X \notin \mathcal{L}(\X+\s) \big| \X \in \txz \right) \\
& = 2^{-t(q,\epsilon_d) \cdot I_J(q) \sqrt{n}+\mathcal{O}(n^{1/4})} \cdot \mathbb{P}_{\X} \left(\X \notin \mathcal{L}(\X+\s) \big| \X \in \txz \right) \label{eq:movie1}\\
& = 2^{-t(q,\epsilon_d) \cdot I_J(q) \sqrt{n}+\mathcal{O}(n^{1/4})} \cdot \frac{\sum_{i=i_0}^{nf^{xz}_{11}} \binom{pn}{i}\binom{n\left(f_{01}^{xz}+f_{11}^{xz}\right)-pn}{nf_{11}^{xz}-i}\binom{n\left(f_{00}^{xz}+f_{10}^{xz}\right)}{nf_{10}^{xz}}}{\binom{n\left(f_{01}^{xz}+f_{11}^{xz}\right)}{nf_{11}^{xz}} \binom{n\left(f_{00}^{xz}+f_{10}^{xz}\right)}{nf_{10}^{xz}}} \label{eq:movie2}\\
&=2^{-t(q,\epsilon_d) \cdot I_J(q) \sqrt{n}+\mathcal{O}(n^{1/4})} \cdot \frac{\sum_{i=i_0}^{nf^{xz}_{11}} \binom{pn}{i}\binom{n\left(f_{01}^{xz}+f_{11}^{xz}\right)-pn}{nf_{11}^{xz}-i}}{\binom{n\left(f_{01}^{xz}+f_{11}^{xz}\right)}{nf_{11}^{xz}} } \\
&= 2^{-t(q,\epsilon_d) \cdot I_J(q) \sqrt{n}+\mathcal{O}(n^{1/4})} \cdot \frac{\sum_{i=i_0}^{nf^{xz}_{11}} \binom{pn}{i}\binom{n\left(f_{01}^{xz}+f_{11}^{xz}\right)-pn}{nf_{11}^{xz}-i}}{\sum_{j=0}^{nf^{xz}_{11}} \binom{pn}{j}\binom{n\left(f_{01}^{xz}+f_{11}^{xz}\right)-pn}{nf_{11}^{xz}-j}} \label{eq:g(i)}.
\end{align}
The calculation in~\eqref{eq:movie1} follows from equations~\eqref{eq:temp1}-\ref{eq:temp2}. In equation~\eqref{eq:movie2}, the denominator is the total number of $(\x,\z)$ pairs that belong to $\txz$, while the numerator is the number of $(\x,\z)$ pairs that belong to $\txz$ and are simultaneously killed by $\s$. In equation~\eqref{eq:g(i)}, we reformulate the denominator such that it has similar structure to the numerator.
We define the auxiliary function $g(i)$ as 
\begin{align}
 g(i) = \binom{pn}{i}\binom{n\left(f_{01}^{xz}+f_{11}^{xz}\right)-pn}{nf_{11}^{xz}-i}.
 \end{align}
To find the maximum value of $g(i)$ when $0 \le i \le nf^{xz}_{11}$, we calculate the ratio between the two successive terms in the following:
\begin{align}
\frac{g(i+1)}{g(i)} &= \frac{\binom{pn}{i+1}\binom{n\left(f_{01}^{xz}+f_{11}^{xz}\right)-pn}{nf_{11}^{xz}-i-1}}{\binom{pn}{i}\binom{n\left(f_{01}^{xz}+f_{11}^{xz}\right)-pn}{nf_{11}^{xz}-i}} = \frac{(pn-i)(nf^{xz}_{11})}{(i+1)(f^{xz}_{01}-pn+i+1)}.
\end{align}
Let $\phi \triangleq \frac{pf^{xz}_{11}n^2-nf^{xz}_{01}+pn-1}{nf^{xz}_{01}+nf^{xz}_{11}+2}$. It turns out that $g(i+1)/g(i) > 1$ when $i < \phi$, and $g(i+1)/g(i) < 1$ when $i > \phi$, which means the function $g(i)$ achieves its maximum when $i = \left \lceil{\phi}\right \rceil $. Note that the parameter $\phi$ itself depends on $f^{xz}_{01}$ and $f^{xz}_{11}$, i.e., the particular type class. One can prove that for typical $\z$ and conditionally typical type class $\txz$, the maximum value of $\phi$ is always bounded from above as 
\begin{align}
\phi \le \frac{\rho n p(1-q)}{q}\left(1+n^{-1/8}\right) \triangleq \phi_{\max}.
\end{align}
Note that as $n$ grows without bound, $i_0$ is larger than $\phi_{\max}$, hence $g(i_0)$ is always smaller than $g(\phi_{\max})$. On the other hand, $g({i_0})$ is always greater than $g(\tilde{i})$, for any $\tilde{i} > i_0$. We now bound the second term in~\eqref{eq:g(i)} as 
\begin{align}
\frac{\sum_{i=i_0}^{nf^{xz}_{11}} \binom{pn}{i}\binom{n\left(f_{01}^{xz}+f_{11}^{xz}\right)-pn}{nf_{11}^{xz}-i}}{\sum_{j=0}^{nf^{xz}_{11}} \binom{pn}{j}\binom{n\left(f_{01}^{xz}+f_{11}^{xz}\right)-pn}{nf_{11}^{xz}-j}} \le \frac{\sum_{i=i_0}^{nf^{xz}_{11}}g(i)}{g(\phi)} \le  \frac{\sum_{i=i_0}^{nf^{xz}_{11}}g(i)}{g(\phi_{\max})} \le \frac{g(i_0)\cdot \log(n)}{g(\phi_{\max})}. \label{eq:shu}
\end{align} 
The last step follows from the geometric sequence 
\begin{align}
\sum_{i=i_0}^{nf^{xz}_{11}}g(i) = \sum_{i=i_0}^{\infty}g(i) & = g(i_0) + g(i_0) \frac{g(i_0+1)}{g(i_0)} + g(i_0) \frac{g(i_0+1)}{g(i_0)}\frac{g(i_0+2)}{g(i_0+1)} + \cdots \cdots \\
& \le g(i_0) + g(i_0) \frac{g(i_0+1)}{g(i_0)} + g(i_0) \left(\frac{g(i_0+1)}{g(i_0)}\right)^2 + \cdots \cdots \label{eq:jun1} \\
&= g(i_0) \cdot \frac{1- \left(\frac{g(i_0+1)}{g(i_0)}\right)^{\infty}}{1-\frac{g(i_0+1)}{g(i_0)}} \\
&\le g(i_0)\cdot \log(n), \label{eq:jun2}
\end{align}
where inequality~\eqref{eq:jun1} holds since $g(i+1)/g(i)$ is monotonically decreasing, and inequality~\eqref{eq:jun2} follows from the fact $g(i_0+1)/g(i_0) \le 1 - 1/\log(n)$.

To calculate the ratio between $g(i_0)$ and $g(\phi_{\max})$, we introduce an interpolation point $\phi' \triangleq \frac{\rho n p(1-q)}{q}\left(1 + \frac{1}{(\log(n))^2}\right)$. Note that $g(\phi_{\max}) \ge g(\phi')$ since $\phi' \ge \phi_{\max}$ and $g(i)$ is monotonically decreasing when $i \ge \phi_{\max}$. Now we consider the ratio between $g(i_0)$ and $g(\phi_{\max})$ as follows:
\begin{align}
\frac{g(i_0)}{g(\phi_{\max})} \le \frac{g(i_0)}{g(\phi')} &= \frac{g(\phi'+1)}{g(\phi')} \frac{g(\phi'+2)}{g(\phi'+1)}  \frac{g(\phi'+3)}{g(\phi'+2)} \cdots \frac{g(i_0)}{g(i_0-1)} \\
&\le \left(\frac{g(\phi'+1)}{g(\phi')}\right)^{i_0 - \phi'} \\
&= \left(1 - \frac{1}{(\log(n))^2}\right)^{\frac{t(q,\epsilon_d) p(1-q)\sqrt{n}}{q}\left(\frac{1}{\log(n)} - \frac{1}{(\log(n))^2}\right)} \label{eq:anjing}\\
& \le \left(1 - \frac{1}{(\log(n))^2}\right)^{c_1\sqrt{n}/\log(n)},
\end{align}
for some constant $c_1 > 0$. Inequality~\eqref{eq:anjing} follows since $g(\phi'+1)/g(\phi') \le 1 - 1/(\log(n))^2$. Using the approximation $\lim_{n \to \infty}(1+1/n)^n = 1/e$, as $n$ grows without bound, we obtain 
\begin{align}
\frac{g(i_0)}{g(\phi_{\max})} \le e^{-c_1\sqrt{n}/(\log(n))^3}. \label{eq:shu2}
\end{align}
By combining~\eqref{eq:g(i)},~\eqref{eq:shu} and~\eqref{eq:shu2}, we finally show that 
\begin{align}
\mathbb{P}_{\X}\left(\left[\X \in \txz\right] \cap \left[\X \notin \mathcal{L}(\X+\s)\right] \right) \le 2^{-(t(q,\epsilon_d) \cdot I_J(q)+c_2/(\log(n))^3)\sqrt{n} + \mathcal{O}(n^{1/4})},
\end{align}
where $c_2 = c_1 \ln 2$. Without loss of correctness, we ignore the lower order terms to simplify the following analysis.
The expected number of codewords falling into the type class $\txz$ and is simultaneously killed by $\s$ equals 
\begin{align}
\mu_2 &\triangleq \mathbb{E}_{\C}\left(\sum_{i=1}^N \sum_{j=1}^L \mathbbm{1}\left\{\left[\x_{ij} \in \txz\right] \cap \left[\x_{ij} \notin \mathcal{L}(\x+\s)\right] \right\}\right) \\
&= 2^{r\sqrt{n}}\cdot \mathbb{P}_{\X}\left(\left[\X \in \txz\right] \cap \left[\X \notin \mathcal{L}(\X+\s)\right] \right) \\
&\le 2^{(c-c_2/(\log(n))^3)\sqrt{n}},
\end{align}
where $c = r - t(q,\epsilon_d)\cdot I_J(q) > 0$.
The probability that more than $\epsilon_1 \cdot 2^{c\sqrt{n}}$ messages that falls into the type class $\txz$ as well as being killed by $\s$ is bounded from above as 
\begin{align}
&\mathbb{P}_{\C}\left(\sum_{i=1}^N \sum_{j=1}^L \mathbbm{1}\left\{\left[\x_{ij} \in \txz\right] \cap \left[\x_{ij} \notin \mathcal{L}(\x+\s)\right] \right\} \ge \epsilon_1 \cdot 2^{c\sqrt{n}}\right) \\
&=\mathbb{P}_{\C}\left(\sum_{i=1}^N \sum_{j=1}^L \mathbbm{1}\left\{\left[\x_{ij} \in \txz\right] \cap \left[\x_{ij} \notin \mathcal{L}(\x+\s)\right] \right\} \ge \left(1+\frac{\epsilon_1 \cdot 2^{c\sqrt{n}}}{\mu_2}-1\right)\mu_2\right) \\
&\le \exp\left(-\frac{1}{3}\left(\frac{\epsilon_1 \cdot 2^{c\sqrt{n}}}{\mu_2}-1\right)\mu_2 \right) \\
&\le \exp\left(-\frac{2^{c\sqrt{n}}}{3}\left(\epsilon_1 - 2^{-c_2\sqrt{n}/(\log(n))^3}\right) \right). 
\end{align}
By setting $\epsilon_1 = \exp(-n^{1/4})$, we have 
\begin{align*}
&\mathbb{P}_{\C}\left(\sum_{i=1}^N \sum_{j=1}^L \mathbbm{1}\left\{\left[\x_{ij} \in \txz\right] \cap \left[\x_{ij} \notin \mathcal{L}(\x+\s)\right] \right\} < \exp(-n^{1/4}) \cdot 2^{c\sqrt{n}}\right) \ge 1 - \exp\left(-2^{\mathcal{O}(\sqrt{n})}\right).
\end{align*} \qed

\section{Proof of Theorem~\ref{cor:prg}} \label{app:bounded}
\subsubsection{Covertness}

\begin{figure}
	\begin{center}
		\includegraphics[scale=0.66]{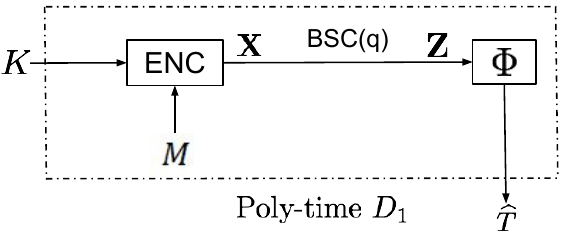}
		\caption{A polynomial-time algorithm $D_1$ that can distinguish $K = g(U)$ and $K = U'$ based on the assumptions on the estimator $\Phi$.} \label{fig:pseudo_covert}
	\end{center}
\end{figure}

\begin{figure}
	\begin{center}
		\includegraphics[scale=0.6]{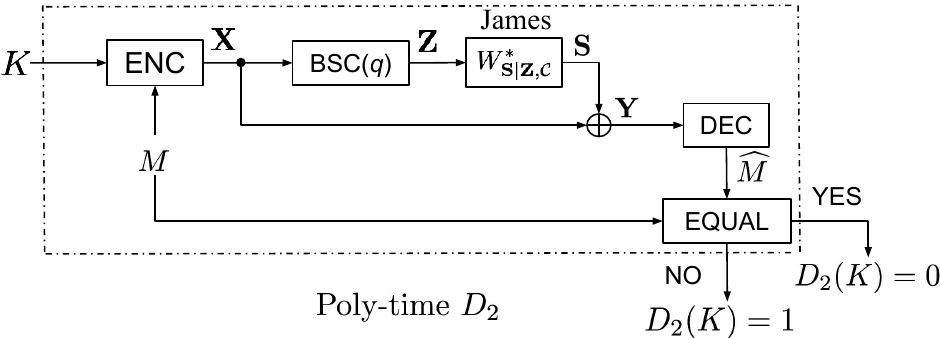}
		\caption{A polynomial-time algorithm $D_2$ that can distinguish $K = g(U)$ and $K = U'$ based on the assumptions on the decoder.} \label{fig:pseudo_decoding}
	\end{center}
\end{figure}

Assume there is a polynomial-time estimator $\Phi$ such that
\begin{align} \PP(\widehat{T}=1|T=1,K=g(U))-\PP(\widehat{T}=1|T=1, K=U') > \nu_n. \label{eq:yu}
\end{align} 
James is then able to design a probabilistic polynomial-time algorithm $D_1$ by generating an artificial system that contains the message, encoder, channel, and estimator $\Phi$ (as shown in Fig.~\ref{fig:pseudo_covert}). The algorithm $D_1$ takes $K$ as input, and outputs $D_1(K) = \widehat{T}$. Substituting $\widehat{T}$ with $D_1(K)$ in~\eqref{eq:yu}, we have 
$$\PP(D_1(K)=1|T=1,K=g(U))-\PP(D_1(K)=1|T=1, K=U') > \nu_n,$$
which means that $D_1$ can distinguish $g(U)$ from $U'$ with at least $\nu_n$ advantage. Also, note that $D_1$ runs in polynomial time, since both the encoder and the estimator $\Phi$ run in polynomial time.
The existence of such $D_1$ contradicts with~\eqref{eq:video}, thus the second condition in~\eqref{eq:pseudo} holds.  

\subsubsection{Reliability}

Suppose there is a polynomial-time decoder satisfying
$$\PP(M \ne \widehat{M}|T=1, K=g(U))-\PP(M \ne \widehat{M}|T=1,K=U') > \nu_n,$$
under some jamming strategy $W^*_{\mathbf{S}|\Z,\C}$.
Bob is then able to use this decoder to design a probabilistic polynomial-time algorithm $D_2$ by generating an artificial system as illustrated in Fig.~\ref{fig:pseudo_decoding}, wherein James' jamming strategy is $W^*_{\mathbf{S}|\Z,\C}$. Let $D_2$ take $K$ as input, and output $D_2(K) = \mathbbm{1}\{\widehat{M} \ne M \}$. By assumption, $D_2$ can distinguish $g(U)$ from $U'$ with at least $\nu_n$ advantage, i.e.,
$$\PP(D_2(K)=1|T=1, K=g(U))-\PP(D_2(K)=1|T=1, K=U') > \nu_n,$$ 
and runs in polynomial time (since both the encoder and decoder run in polynomial time).  The existence of such $D_2$ contradicts with~\eqref{eq:video}, thus any polynomial-time decoder satisfies 
\begin{align}
\PP(M \ne \widehat{M}|T=1,K=g(U))-\PP(M \ne \widehat{M}|T=1,K=U') \le \nu_n, \quad \text{for all } W_{\mathbf{S}|\Z,\C}.  \label{eq:v1}
\end{align}

\section*{Acknowledgement}
The authors would like to thank Pascal O. Vontobel for his valuable comments. 

\end{appendices}

\bibliographystyle{IEEEtran}
\bibliography{steg_mayank,mainfunction}
\end{document}